\documentclass[twocolumn]{aastex6}
\usepackage{graphicx}
\usepackage[flushleft]{threeparttable}

\shorttitle{ASPECS: molecular gas in high-$z$ galaxies}
\shortauthors{Decarli et al.}

\def\Lsun{L$_\odot$}
\def\Msun{M$_\odot$}

\def\Oii{[O\,{\sc ii}]}

\def\Cii{[C\,{\sc ii}]}

\def\kms{km\,s$^{-1}$}

\def\Kkmspc{K~km\,s$^{-1}$\,pc$^2$}

\def\lsim{\mathrel{\rlap{\lower 3pt \hbox{$\sim$}} \raise 2.0pt \hbox{$<$}}}
\def\gsim{\mathrel{\rlap{\lower 3pt \hbox{$\sim$}} \raise 2.0pt \hbox{$>$}}}

\begin{document}

\title{ALMA Spectroscopic Survey in the Hubble Ultra Deep Field: Molecular gas reservoirs in high--redshift galaxies}

\author{
Roberto Decarli\altaffilmark{1}, 
Fabian Walter\altaffilmark{1,2,3}, 
Manuel Aravena\altaffilmark{4}, 
Chris Carilli\altaffilmark{3,5}, 
Rychard Bouwens\altaffilmark{6}, 
Elisabete da Cunha\altaffilmark{7,8}, 
Emanuele Daddi\altaffilmark{9}, 
David Elbaz\altaffilmark{9}, 
Dominik Riechers\altaffilmark{10}, 
Ian Smail\altaffilmark{11}, 
Mark Swinbank\altaffilmark{11}, 
Axel Weiss\altaffilmark{12}, 
Roland Bacon\altaffilmark{13},
Franz Bauer\altaffilmark{14,15,16}, 
Eric F.~Bell\altaffilmark{17}, 
Frank Bertoldi\altaffilmark{18}, 
Scott Chapman\altaffilmark{19}, 
Luis Colina\altaffilmark{20}, 
Paulo C.~Cortes\altaffilmark{21,22}, 
Pierre Cox\altaffilmark{21}, 
Jorge G\'onzalez-L\'opez\altaffilmark{23}, 
Hanae Inami\altaffilmark{13},
Rob Ivison\altaffilmark{24,25}, 
Jacqueline Hodge\altaffilmark{6}, 
Alex Karim\altaffilmark{18}, 
Benjamin Magnelli\altaffilmark{18}, 
Kazuaki Ota\altaffilmark{26,5}, 
Gerg\"{o} Popping\altaffilmark{24}, 
Hans--Walter Rix\altaffilmark{1}, 
Mark Sargent\altaffilmark{27}, 
Arjen van der Wel\altaffilmark{1},
Paul van der Werf\altaffilmark{6}
%
}
\altaffiltext{1}{Max-Planck Institut f\"{u}r Astronomie, K\"{o}nigstuhl 17, D-69117, Heidelberg, Germany. E-mail: {\sf decarli@mpia.de}}
\altaffiltext{2}{Astronomy Department, California Institute of Technology, MC105-24, Pasadena, California 91125, USA}
\altaffiltext{3}{National Radio Astronomy Observatory, Pete V.\,Domenici Array Science Center, P.O.\, Box O, Socorro, NM, 87801, USA}
\altaffiltext{4}{N\'{u}cleo de Astronom\'{\i}a, Facultad de Ingenier\'{\i}a, Universidad Diego Portales, Av. Ej\'{e}rcito 441, Santiago, Chile}
\altaffiltext{5}{Cavendish Laboratory, University of Cambridge, 19 J J Thomson Avenue, Cambridge CB3 0HE, UK}
\altaffiltext{6}{Leiden Observatory, Leiden University, PO Box 9513, NL2300 RA Leiden, The Netherland}
\altaffiltext{7}{Centre for Astrophysics and Supercomputing, Swinburne University of Technology, Hawthorn, Victoria 3122, Australia}
\altaffiltext{8}{Research School of Astronomy and Astrophysics, Australian National University, Canberra, ACT 2611, Australia}
\altaffiltext{9}{Laboratoire AIM, CEA/DSM-CNRS-Universite Paris Diderot, Irfu/Service d'Astrophysique, CEA Saclay, Orme des Merisiers, 91191 Gif-sur-Yvette cedex, France}
\altaffiltext{10}{Cornell University, 220 Space Sciences Building, Ithaca, NY 14853, USA}
\altaffiltext{11}{6 Centre for Extragalactic Astronomy, Department of Physics, Durham University, South Road, Durham, DH1 3LE, UK}
\altaffiltext{12}{Max-Planck-Institut f\"ur Radioastronomie, Auf dem H\"ugel 69, 53121 Bonn, Germany}
\altaffiltext{13}{Universit\'{e} Lyon 1, 9 Avenue Charles Andr\'{e}, 69561 Saint Genis Laval, France}
\altaffiltext{14}{Instituto de Astrof\'{\i}sica, Facultad de F\'{\i}sica, Pontificia Universidad Cat\'olica de Chile Av. Vicu\~na Mackenna 4860, 782-0436 Macul, Santiago, Chile}
\altaffiltext{15}{Millennium Institute of Astrophysics (MAS), Nuncio Monse{\~{n}}or S{\'{o}}tero Sanz 100, Providencia, Santiago, Chile}
\altaffiltext{16}{Space Science Institute, 4750 Walnut Street, Suite 205, Boulder, CO 80301, USA}
\altaffiltext{17}{Department of Astronomy, University of Michigan, 1085 South University Ave., Ann Arbor, MI 48109, USA}
\altaffiltext{18}{Argelander Institute for Astronomy, University of Bonn, Auf dem H\"{u}gel 71, 53121 Bonn, Germany}
\altaffiltext{19}{Dalhousie University, Halifax, Nova Scotia, Canada}
\altaffiltext{20}{ASTRO-UAM, UAM, Unidad Asociada CSIC, Spain}
\altaffiltext{21}{Joint ALMA Observatory - ESO, Av. Alonso de C\'ordova, 3104, Santiago, Chile}
\altaffiltext{22}{National Radio Astronomy Observatory, 520 Edgemont Rd, Charlottesville, VA, 22903, USA}
\altaffiltext{23}{Instituto de Astrof\'{\i}sica, Facultad de F\'{\i}sica, Pontificia Universidad Cat\'olica de Chile Av. Vicu\~na Mackenna 4860, 782-0436 Macul, Santiago, Chile}
\altaffiltext{24}{European Southern Observatory, Karl-Schwarzschild-Strasse 2, 85748, Garching, Germany}
\altaffiltext{25}{Institute for Astronomy, University of Edinburgh, Royal Observatory, Blackford Hill, Edinburgh EH9 3HJ}
\altaffiltext{26}{Kavli Institute for Cosmology, University of Cambridge, Madingley Road, Cambridge CB3 0HA, UK}
\altaffiltext{27}{Astronomy Centre, Department of Physics and Astronomy, University of Sussex, Brighton, BN1 9QH, UK}

\begin{abstract}
We study the molecular gas properties of high--$z$ galaxies observed in the ALMA Spectroscopic Survey (ASPECS) that targets a $\sim1$\,arcmin$^2$ region in the Hubble Ultra Deep Field (UDF), a blind survey of CO emission (tracing molecular gas) in the 3\,mm and 1\,mm bands. Of a total of 1302 galaxies in the field, 56 have spectroscopic redshifts and correspondingly well--defined physical properties. Among these, 11 have infrared luminosities $L_{\rm{}IR}>10^{11}$\,\Lsun{}, i.e. a detection in CO emission was expected. Out these, 7 are detected at various significance in CO, and 4 are undetected in CO emission. In the CO--detected sources, we find CO excitation conditions that are lower than typically found in starburst/SMG/QSO environments. 
We use the CO luminosities (including limits for non-detections) to derive molecular gas masses. We discuss our findings in context of previous molecular gas observations at high redshift (star--formation law, gas depletion times, gas fractions): The CO--detected galaxies in the UDF tend to reside on the low-$L_{\rm{}IR}$ envelope of the scatter in the $L_{\rm{}IR}-L'_{\rm{}CO}$ relation, but exceptions exist. For the CO--detected sources, we find an average depletion time of $\sim$\,1\,Gyr, with significant scatter. The average molecular--to--stellar mass ratio ($M_{\rm{}H2}$/$M_*$) is consistent with earlier measurements of main sequence galaxies at these redshifts, and again shows large variations among sources. In some cases, we also measure dust continuum emission.  On average, the dust--based estimates of the molecular gas are a factor $\sim$2--5$\times$ smaller than those based on CO. Accounting for detections as well as non--detections, we find large diversity in the molecular gas properties of the high--redshift galaxies covered by ASPECS.
\end{abstract}


\section{Introduction}

Molecular gas observations of galaxies throughout cosmic time are fundamental to understand the cosmic history of formation and evolution of galaxies \citep[see reviews by][]{kennicutt12,carilli13}. The molecular gas provides the fuel for star formation, thus by characterizing its properties we place quantitative constraints on the physical processes that lead to the stellar mass growth of galaxies. This has been a demanding task in terms of telescope time. To date, only a couple hundred sources at $z>1$ have been detected in a molecular gas tracer \citep[typically the rotational transitions of the carbon monoxide $^{12}$CO molecule; e.g.,][]{carilli13}. This sample is dominated by `extreme' sources, such as QSO host galaxies \citep[e.g.,][]{walter03,bertoldi03,weiss07,wang13} or sub-mm galaxies \citep[e.g.,][]{frayer98,neri03,greve05,bothwell13,riechers13,aravena16}, with IR luminosities $L_{\rm IR}\gg 10^{12}$\,\Lsun{} and Star Formation Rates (SFR) $\gg 100$\,\Msun{}\,yr$^{-1}$. These extreme sources  might contribute significantly to the star formation budget in the Universe at $z>4$, but their role declines with cosmic time \citep{casey14}. Indeed, the bulk of star formation up to $z\sim2$ is observed in galaxies along the so-called `main sequence' \citep{noeske07, elbaz07, elbaz11, daddi10a,daddi10b, genzel10, wuyts11, whitaker12}, a tight (scatter rms $\sim 0.3$\,dex) relation between the SFR and the stellar mass, $M_*$. Addressing the molecular gas content of main sequence galaxies beyond the local universe has become feasible only in recent years. 

The first step in the characterization of the molecular gas content of galaxies is the measure of the molecular gas mass, $M_{\rm H2}$. The $^{12}$CO molecule (hereafter, CO) is the second most abundant molecule in the Universe, and it is relatively easy to target thanks to its bright rotational transitions. The use of CO as a tracer for the molecular gas mass requires assuming a conversion factor, $\alpha_{\rm CO}$, to pass from CO(1-0) luminosities to H$_2$ masses. At $z\sim0$, the conversion factor that is typically used is $\sim4$\,\Msun{}(\Kkmspc)$^{-1}$ for ``normal'' $M_*>10^9$\,\Msun{} star-forming galaxies with metallicities close to solar \citep[see][for a recent review]{bolatto13}. If other CO transitions are observed instead of the J=1$\rightarrow$0 ground state one, a further factor is required to account for the CO excitation \citep[see, e.g.,][]{weiss07,daddi15}. \citet{tacconi10} and \citet{daddi10a} investigated the molecular gas content of highly star-forming galaxies at $z\sim1.2$ and $z\sim2.3$ via the CO(3-2) transition. They found large reservoirs of gas, yielding molecular--to--stellar mass ratios $M_{\rm H2}/M_*\sim1$. These values are significantly higher than those observed in local galaxies \citep[$\sim 0.1$, see e.g.][]{leroy08}, suggesting a strong evolution of $M_{\rm H2}/M_*$ with redshift \citep[see also][]{riechers10, geach11, casey11, magnelli12, aravena12, aravena16, bothwell13, tacconi13, saintonge13, chapman15a, genzel15}. 

An alternative approach to estimate gas masses is via dust emission. The dust mass in a galaxy can be retrieved via the study of its rest-frame sub-mm spectral energy distribution (SED) \citep[e.g.,][]{magdis11, magdis12, magnelli12, santini14, bethermin15, berta16}, in particular via the Rayleigh-Jeans tail, which is less sensitive to the dust temperature \citep[see, e.g.,][]{scoville14,groves15}. Using the dust as a proxy of the molecular gas does not require assumptions on CO excitation and on $\alpha_{\rm CO}$. However, this approach relies on the assumption of a dust-to-gas mass ratio (DGR), which typically depends on the gas metallicity \citep{sandstrom13,bolatto13,genzel15}. Recent ALMA results report substantially lower values of $M_{\rm gas}$ than typically obtained in CO--based studies \citep{scoville14,scoville15}.

In the present paper, we study the molecular gas properties of galaxies in ASPECS, the ALMA Spectroscopic Survey in the Hubble Ultra Deep Field (UDF). This is a blind search for CO emission using the Atacama Large Millimeter/sub-millimeter Array (ALMA). The goal is to constrain the molecular gas content of an unbiased sample of galaxies. The targeted region is one of the best studied areas of the sky, with exquisitely deep photometry in $>25$ X-ray--to--far-infrared (IR) bands, photometric redshifts, and dozens of spectroscopic redshifts. This provides us with an exquisite wealth of ancillary data, which is instrumental to put our CO measurements in the context of galaxy properties. Thanks to the deep field nature of our approach, we avoid potential biases related to the pre-selection of targets, and include both detections and non-detections in our analysis. Our dataset combines 3mm and 1mm observations of the same galaxies, thus providing constraints on the CO excitation. Furthermore, the combination of the spectral line survey and the 1mm continuum image allows us to compare CO- and dust-based estimates of the gas mass. In other papers of this series, we present the dataset and the catalog of blindly-selected CO emitters (Paper I, Walter et al.), we study the properties of 1.2mm-detected sources (Paper II, Aravena et al.), we discuss the inferred constraints on the luminosity functions of CO (Paper III, Decarli et al.) and we search for \Cii{} emission in $z$=6--8 galaxies (Paper V, Aravena et al.). Paper VI (Bouwens et al.) places our findings in the context of the dust extinction law for $z>2$ galaxies, and Paper VII (Carilli et al.) uses ASPECS to put first direct constraints on intensity mapping experiments. Here we put the CO detections in the context of the properties of the associated galaxies.  In sec.~\ref{sec_obs} we summarize the observational dataset; in sec.~\ref{sec_sample} we describe our sample; in sec.~\ref{sec_results} we present CO-based measurements, which are discussed in the context of galaxy properties in sec.~\ref{sec_discussion}. We summarize our findings in sec.~\ref{sec_conclusions}.

Throughout the paper we assume a standard $\Lambda$CDM cosmology with $H_0=70$ km s$^{-1}$ Mpc$^{-1}$, $\Omega_{\rm m}=0.3$ and $\Omega_{\Lambda}=0.7$ \citep[broadly consistent with the measurements in][]{planck15}, and a \citet{chabrier03} stellar initial mass function. Magnitudes are expressed in the AB photometric system \citep{oke83}.

\section{Observations}\label{sec_obs}

\subsection{The ALMA dataset}

The details of the ALMA dataset (observations and data reduction) are presented in Paper~I of this series. Here we briefly summarize the observational details that are relevant for the present study. The dataset consists of two frequency scans in ALMA band 3 (3\,mm, 84--115 GHz) and in band~6 (1\,mm, 212--272 GHz). In the case of the 3mm observations, we obtained a single pointing centered at RA = 03:32:37.900, Dec = --27:46:25.00 (J2000.0), close to the Northern corner of the Hubble eXtremely Deep Field \citep[XDF,][]{illingworth13}. The primary beam has a diameter of $\approx 65''$ at the central frequency of the band ($99.5$\,GHz). The typical noise rms is $\sim 0.18$\,mJy\,beam$^{-1}$ per 50\,\kms{} channel. The 1\,mm observations consist of a 7-pointing mosaic covering approximately the same area as the 3\,mm observations. The typical noise rms is 0.44\,mJy\,beam$^{-1}$ per 50\,\kms{} channel. The resulting 1mm continuum image reaches a noise rms of $12.7$\,$\mu$Jy\,beam$^{-1}$ at the center of the mosaic (see Paper II). 



\subsection{Ancillary Data}

We complement the ALMA data with X-ray--to--far-IR photometry from public catalogs of this field, as well as optical/near-IR spectroscopic information where available. The main sources for the photometry are the compilations by \citet{coe06} and \citet{skelton14}. The former includes optical photometry based on the original {\em HST} Advanced Camera for Survey (ACS) images of the {\em Hubble} Ultra Deep Field \citep[UDF,][]{beckwith06} and near-IR images obtained with {\em HST} NICMOS. The latter compiles also optical/near-IR observations with {\em HST} ACS and Wide Field Camera 3 (WFC3) from the {\em Hubble} XDF \citep{illingworth13}, {\em Spitzer} IRAC, as well as a wealth of ground-based optical/near-IR observations. {\em Spitzer} MIPS data at 24$\mu$m and 70$\mu$m, as well as {\em Herschel} PACS and SPIRE data come from the work by \citet{elbaz11}. X-ray data are taken from the Extended {\em Chandra} Deep Field South Survey \citep{lehmer05} and from the {\em Chandra} Source Catalogue \citep{evans10}.

Photometric redshifts ($z_{\rm photo}$) are available for all of the optically-selected sources in the field. At a limiting magnitude $i$=28\,mag, the median uncertainty is $\delta z_{\rm photo}\sim 0.5$, and it reaches $\delta z_{\rm photo}\sim 1$ at $i\approx 30$ mag \citep{coe06}. The compilation of \citet{skelton14} provides even more robust photometric redshifts, thanks to the expanded photometric dataset. The agreement with available spectroscopic redshifts is typically very good in these cases, with a standard deviation on $\Delta z/(1+z)$ of $\approx0.01$ for \citep{skelton14}. In addition, the 3D--HST survey provides {\em HST} ACS and WFC3 grism observations of the field, yielding grism redshifts for tens of sources in our pointing \citep{momcheva16}. Slit spectroscopy for 74 (mostly bright) galaxies in the field is also available \citep{lefevre05,skelton14,morris15}. Finally, integral field spectroscopy of this field has been secured with ESO VLT/MUSE. These data are part of a Guaranteed Time observing program targeting the UDF. In particular, a single (1 arcmin$^2$) deep (21\,hr on source) pointing overlaps with $\sim70$\% of the ASPECS coverage. The cubes have been processed and analyzed with the improved MUSE GTO pipeline. These observations will be presented in Bacon et al. (in prep). 

\begin{figure}\begin{center}
\includegraphics[width=0.99\columnwidth]{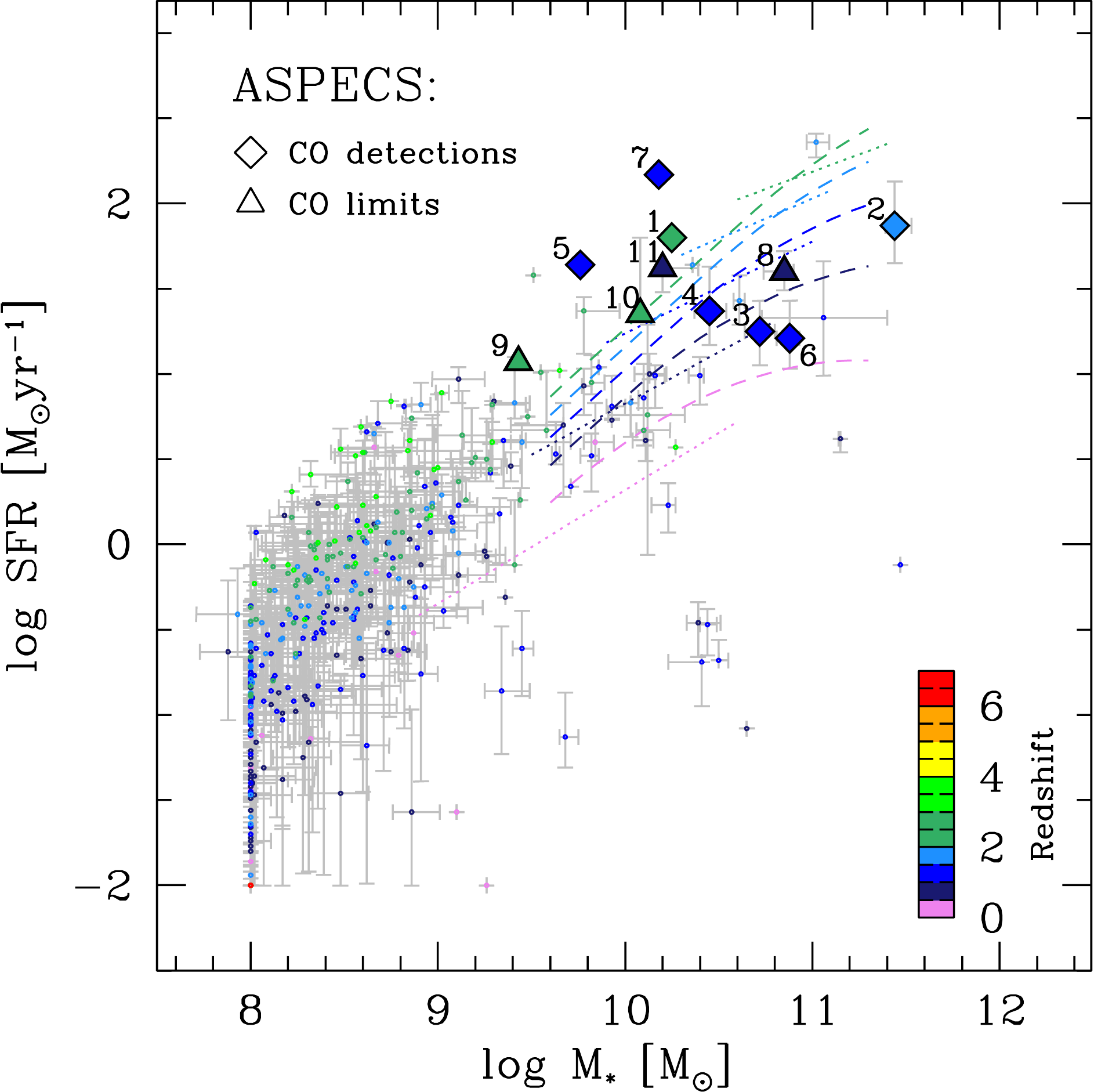}
\end{center}
\caption{Star formation rates and stellar masses of all the galaxies in our field, color-coded by redshift. Inferred parameters are derived using the high-$z$ extension of MAGPHYS \citep{dacunha08,dacunha15}. The sample discussed here is highlighted with big symbols: Diamonds refer to the CO detections, while galaxies in the present sample that are not detected in CO are marked with triangles. We stress that only galaxies with secure spectroscopic redshifts are considered in the present analysis. The loci of the main sequence in various redshift bins are shown as dotted lines \citep[from][]{whitaker12} and dashed lines \citep[from][]{schreiber15}. Half of the galaxies in our sample lie along the main sequence at their respective redshifts. ID.5, 7 and 11 occupy the `starburst' region above the main sequence, while ID.3 and 6 exhibit a SFR $\sim3 \times$ lower than what is typically observed in main sequence galaxies at the same redshifts and stellar masses.}\label{fig_ms}
\end{figure}

Within a radius of $34''$ from the ALMA 3mm pointing center (approximately the size of the primary beam of our 3mm observations), there are $1302$\,galaxies from the combination of all the available photometric catalogs. We use the high-$z$ extension of MAGPHYS \citep{dacunha08,dacunha15} to fit the Spectral Energy Distributions (SEDs) of all of them. The input photometry includes 26 broad and medium filters ranging from observed $U$ band to {\em Spitzer} IRAC 8$\mu$m. Additionally, we include the ASPECS 1mm continuum photometry for those sources where $>2$-$\sigma$ emission is reported in our 1mm continuum image. We do not include any {\em Spitzer} MIPS or {\em Herschel} PACS photometry because the angular resolution of those instruments is not sufficient to accurately pinpoint the emission\footnote{For instance, including MIPS and PACS in the SED fits yields to an overestimate (by a factor $\sim3$) of the SFR in the brightest source in our sample, but with a poor SED fit quality, because of the contamination of foreground sources; on the other hand, the second brightest galaxy, which appears isolated, shows consistent results if the fit is performed with or without MIPS and PACS photometry.}. Our MAGPHYS analysis provides us with a posterior probability distribution of the stellar mass ($M_{\rm *}$), the star formation rate, the specific SFR (sSFR=SFR/$M_{\rm *}$), the dust mass ($M_{\rm dust}$) and the IR luminosity for each galaxy in the field. We take the 14\% and 86\% quartiles of the posterior distributions as the uncertainties in the parameters, and we account for an additional fiducial 10\% uncertainty due to systematics (subtleties in the photometric analysis adopted in the input catalogs, such as aperture corrections and deblending assumptions; zero point uncertainties; etc). Fig.~\ref{fig_ms} shows the SFR as a function of $M_*$ for all the 1302 galaxies in our field.

\begin{figure*}\begin{center}
\includegraphics[width=0.89\textwidth]{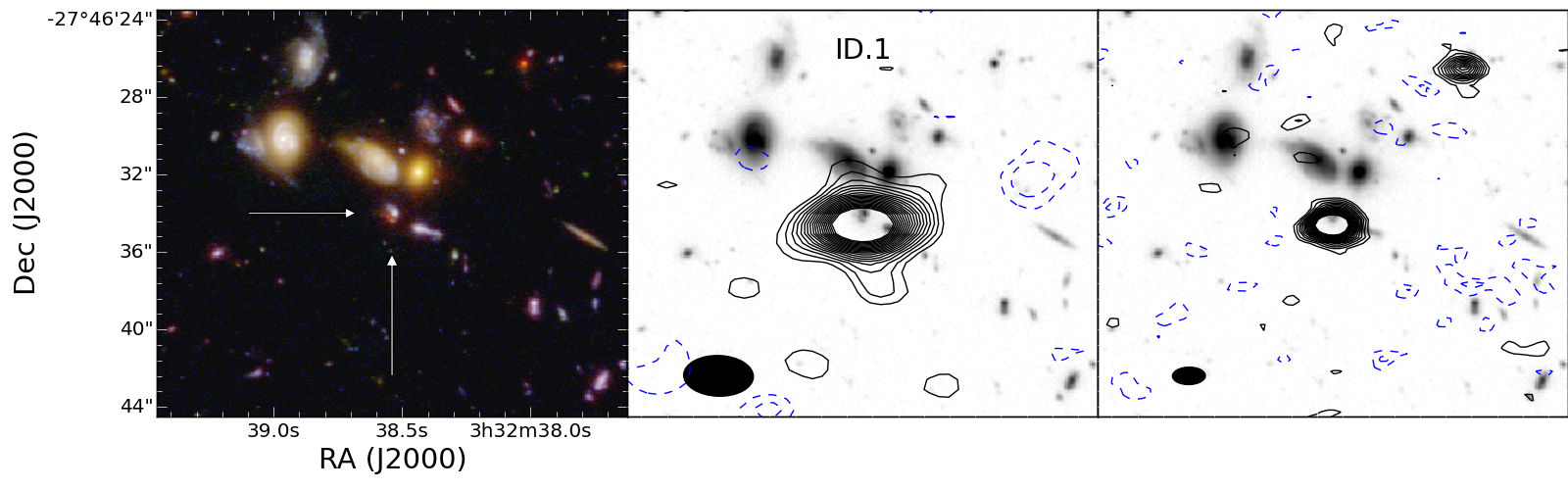}
\includegraphics[width=0.99\columnwidth]{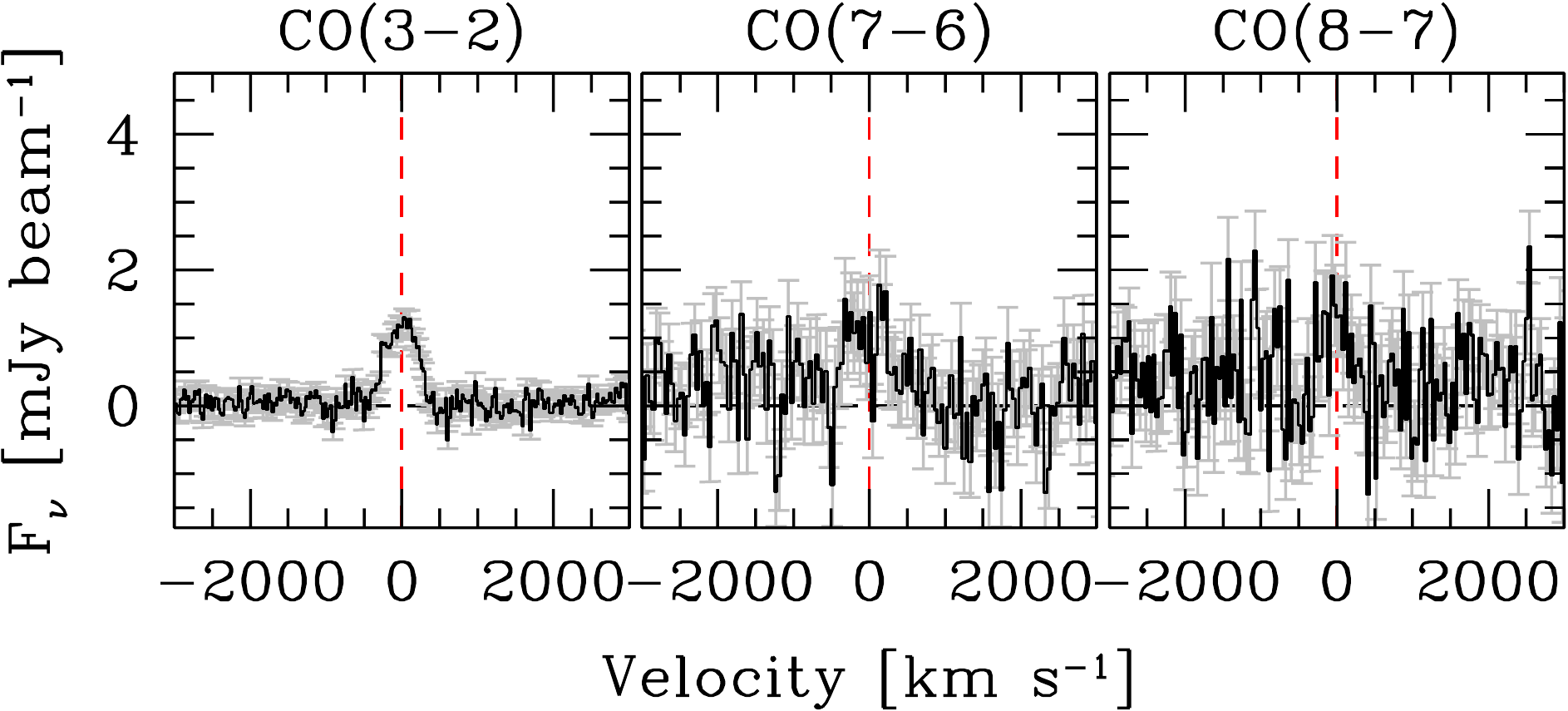}
\includegraphics[width=0.99\columnwidth]{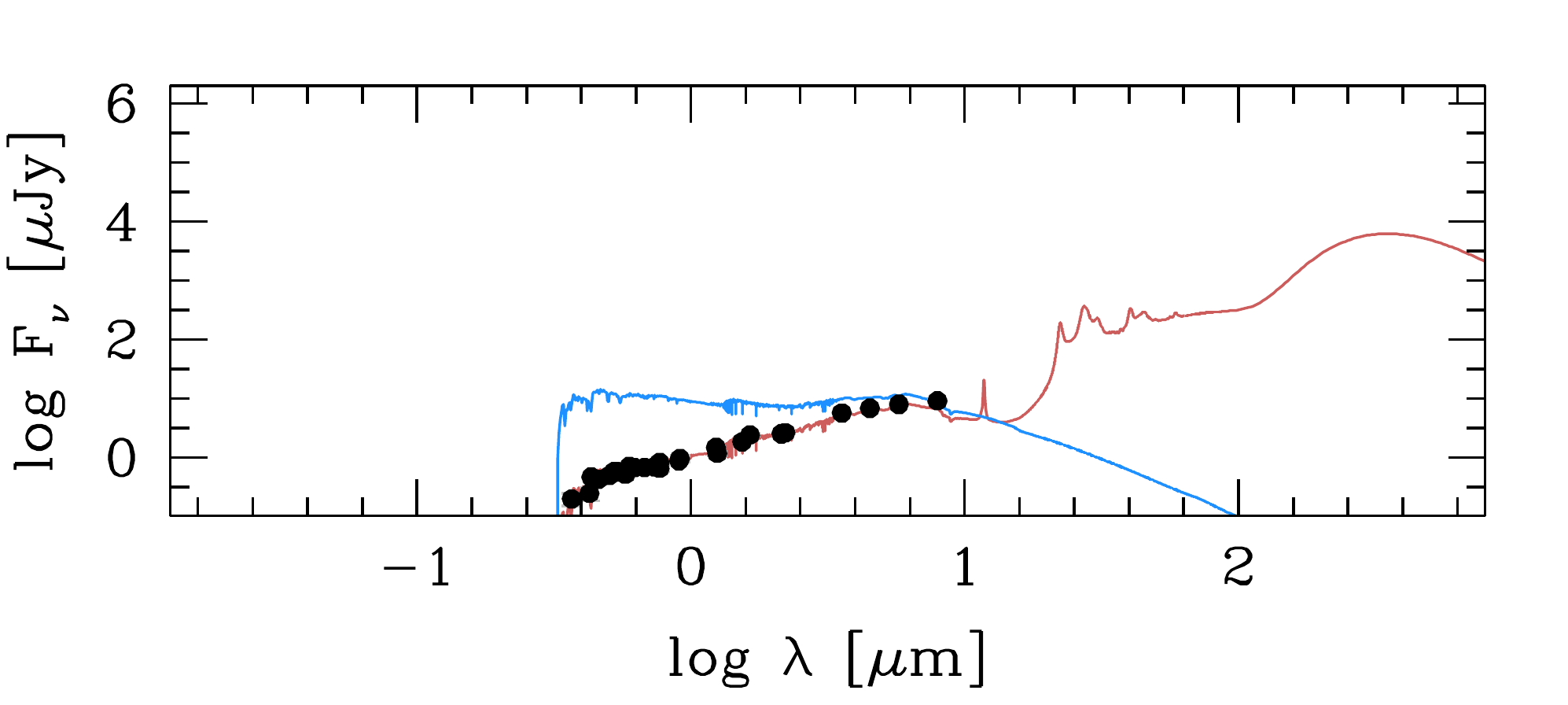}\\
\end{center}
\caption{{\em Top left:} {\em HST} F105W/F775W/F435W RGB image of ID.1. The postage stamp is $20''\times20''$. {\em Top center:} {\em HST} F125W image of the same field. The map of the lowest-J accessible CO transition (in this case, CO[3-2]) is shown as contours ($\pm 2,3,$\ldots,$20$-$\sigma$ [$\sigma$(ID.1)=0.78\,mJy\,beam$^{-1}$]; solid black lines for the positive isophotes, dashed blue lines for the negative). The synthesized beam is shown as a black ellipse. {\em Top right:} Same as in the center, showing the 1.2mm dust continuum. {\em Bottom left:} Spectra of the CO lines encompassed in our spectral scan. {\em Bottom right:} Spectral Energy Distribution. The red line shows the best MAGPHYS fit of the available photometry (black points), while the blue line shows the corresponding model for the unobscured stellar component. The main output parameters are quoted. Similar plots for all the sources in our sample are available in the appendix.
}\label{fig_id1}
\end{figure*}

\section{The sample}\label{sec_sample}

We focus our discussion on those galaxies in the field covered by ASPECS that we originally expected to detect in CO emission. Our expectations are based on the MAGPHYS predictions discussed in the previous section. Fig.~\ref{fig_ms} shows the stellar masses and star formation rates of all $1302$ galaxies in the field (color--coded by redshift). Out of these, 56 galaxies have secure spectroscopic redshifts within our CO redshift coverage, and are brighter than 27.5\,mag in the filters F850LW or F105W ($z$ and $Y$ band respectively)\footnote{This flux cut allows us to reject sources that are too faint for a reliable SED analysis.}. We further restrict our analysis to the redshift windows for which ASPECS covers at least one of the following low--J CO transitions: J=2$\rightarrow$1, J=3$\rightarrow$2, or J=4$\rightarrow$3.

From these galaxies, we select the 11 galaxies for which the MAGPHYS SED analysis yields an IR luminosity $L_{\rm IR}>10^{11}$\,\Lsun{} at $>1$-$\sigma$ significance. These sources are marked by symbols in Fig.~\ref{fig_ms}, and spectroscopic redshifts are available for all of  these sources. The IR luminosity of a galaxy (derived from the SED fitting) has been found to correlate with the CO luminosity (see also Sec.~\ref{sec_co_lum}). Following the best fit of the relation in \citet{carilli13}, the IR--luminosity cut above corresponds to $L'_{\rm CO(1-0)}>3\times 10^9$\,\Kkmspc{}, i.e. similar to the line luminosity limit of our survey (see Paper I, Walter et al.\ 2016). Consequently, we should be able to detect CO, or at least place meaningful limits, on these 11 galaxies.

Tab.~\ref{tab_sample} summarizes the main optical/near-IR properties of the galaxies considered in this paper. In Fig.~\ref{fig_id1}, we show, for one of the sources, the {\em HST} image compared with the CO and dust continuum maps; the CO spectra; and the SED data and modeling. Similar plots are presented for all sample galaxies in the Appendix.

Four of the galaxies in our sample match some of the CO lines identified in our blind search (see Paper I). The ASPECS name for these sources is also reported in Tab.~\ref{tab_sample}. Three additional galaxies show CO emission, although at lower significance. Finally, four sources remain undetected in CO.

\begin{figure*}
\includegraphics[width=0.99\textwidth]{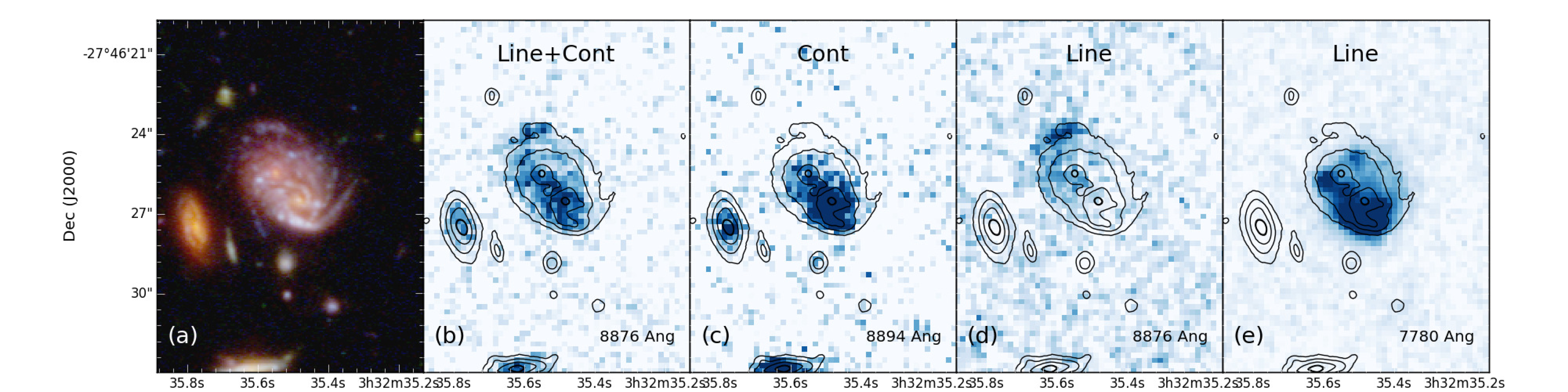}\\
\includegraphics[width=0.49\textwidth]{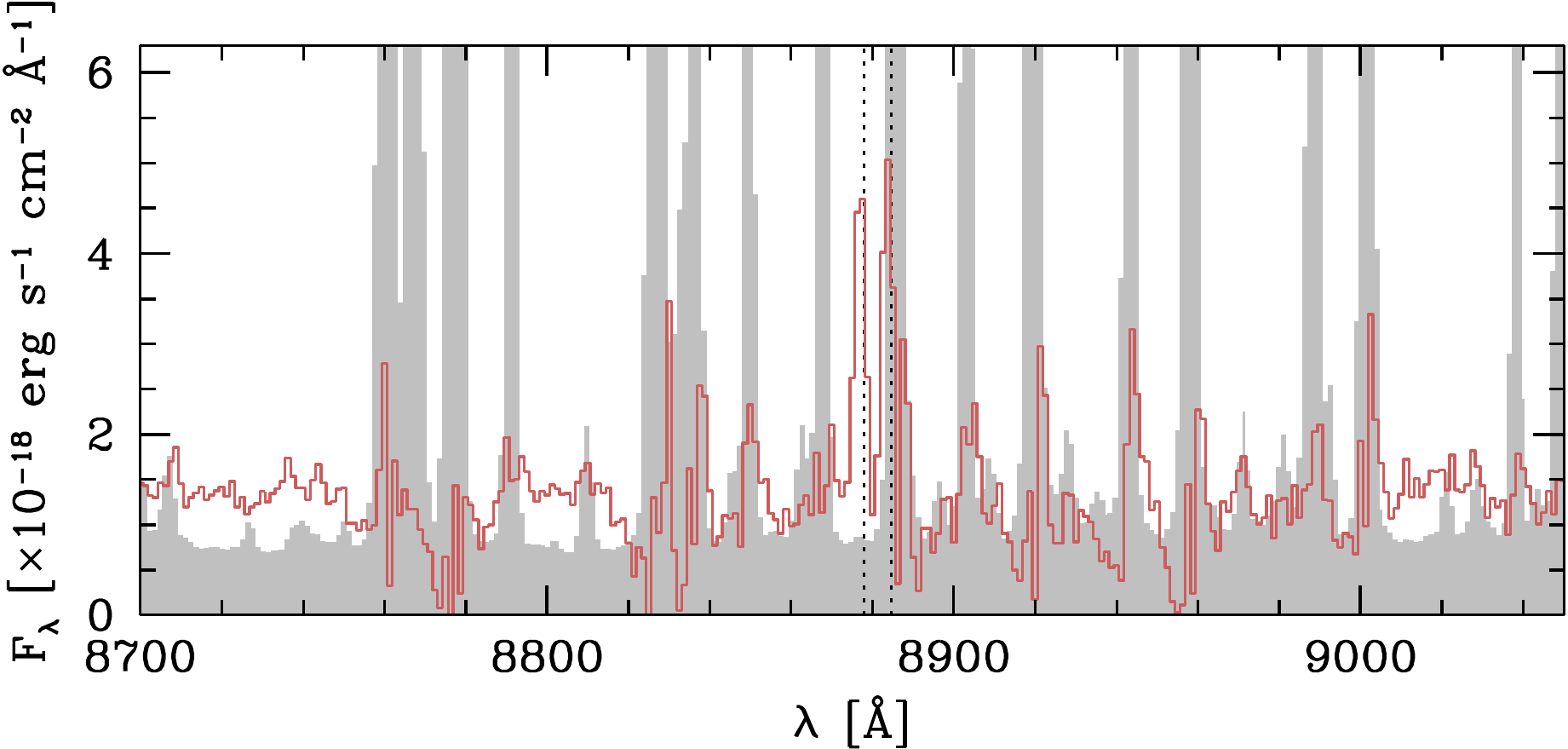}
\includegraphics[width=0.49\textwidth]{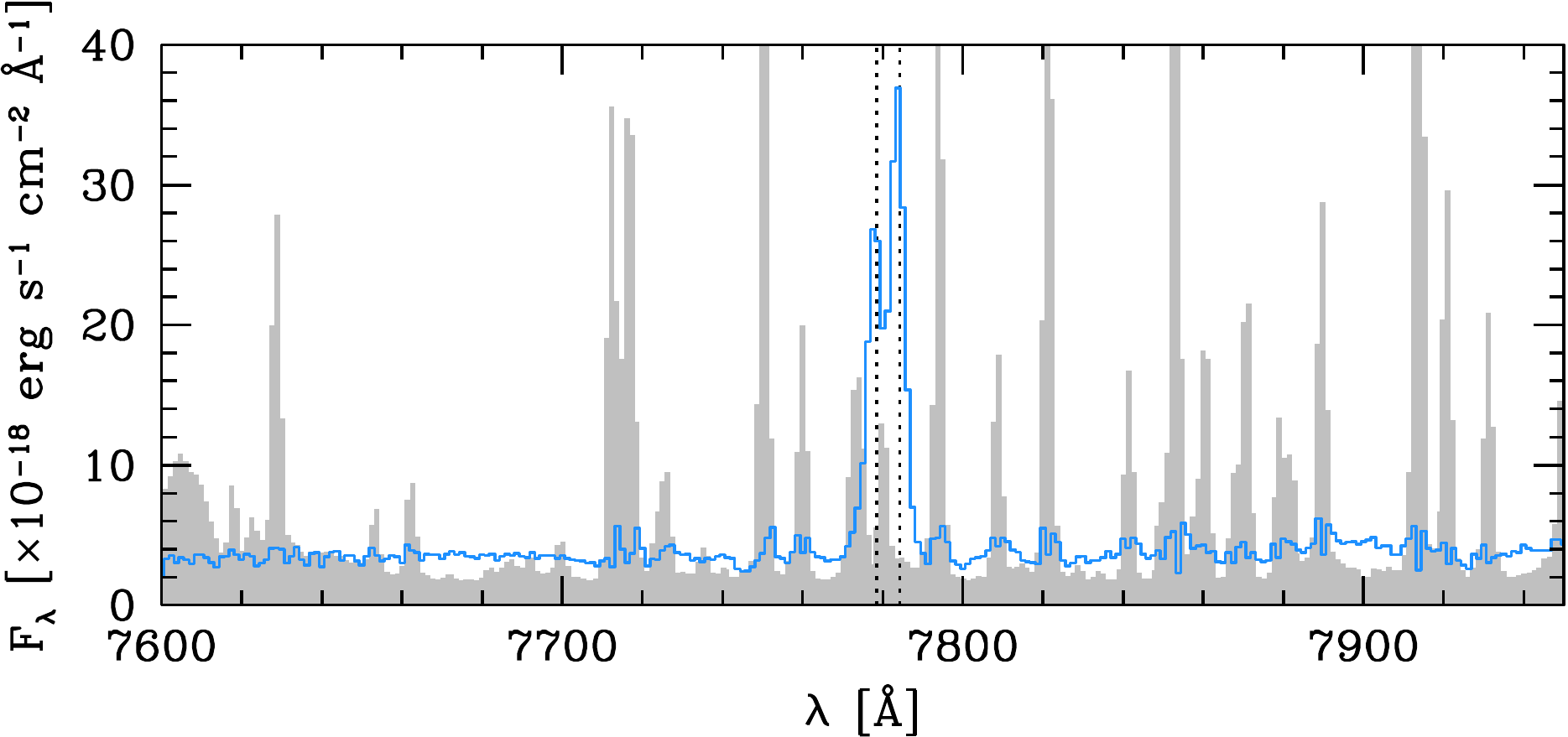}\\
\caption{MUSE optical spectroscopy of the counterparts of ID.3 and ID.4. {\em Top panels ---}(a) {\em HST} RGB image (see Fig.~\ref{fig_id1} for details). (b) MUSE channel map at 8876 Ang, integrated over $\sim 5$ \AA{}, showing the \Oii{} emission of the background component plus the starlight continuum from both the galaxies. The {\em HST}/F160W contours are overplotted in black to guide the eye on the position of the sources. (c) MUSE channel map at 8894 \AA{}, i.e., a few \AA{} off the \Oii{} line, showing the continuum emission only. The map is integrated over $\sim10$ \AA{}. (d) Difference between panels (b) and (d). The continuum emission is effectively subtracted, as confirmed by the disappearance of all the field sources. The residual emission is the \Oii{} line emission from the background object, which thus resides at $z=1.382$ (consistent with the CO redshift of ASPECS 3mm.3). (e) Same as panel (d), but this time centered at 7780 \AA{}, thus highlighting the \Oii{} emission of the foreground galaxy at $z=1.088$. {\em Bottom panels ---} MUSE optical spectra of the \Oii{} lines of the counterparts of ASPECS 3mm.3 (left) and 3mm.5 (right). The vertical, dotted lines mark the wavelengths corresponding to the \Oii{} doublet at $z$=$z_{\rm CO}$. The gray shading shows the noise in the spectra (which, at these wavelengths, is dominated by sky emission lines). The \Oii{} emission is clearly seen in both sources.}\label{fig_muse_dspiral}
\end{figure*}

\subsection{Notes on individual galaxies}

\begin{table*}
\caption{\rm The sample of galaxies examined in this work, and their optical/near-IR global properties. The sorting is based on the significance of the CO detection. (1) Source ID. (2) ASPECS name for blind CO detections (3mm.X, see Paper I) and for the blind 1.2mm continuum detections (CX, see Paper II). (3-4) Optical coordinates in \citet{skelton14}. (5) Redshift. (6) J$_{\rm up}$ of the CO transitions encompassed in our ASPECS data. (7-10) MAGPHYS-derived stellar mass ($M_*$), star formation rate (SFR), specific star formation rate (sSFR), IR luminosity ($L_{\rm IR}$). (11) Effective radius from the near-IR analysis by \citet{vanderwel12}.} \label{tab_sample}
\begin{center}
\begin{tabular}{ccccccccccc}
\hline
ID   & ASPECS& Optical RA  & Optical Dec & $z$   & Obs.CO    &     $M_*$	      &   SFR		      &    sSFR 	      & $L_{\rm IR}$    & $R_{\rm e}$ \\
     & name  &             &             &	 & trans.    &[$\times 10^{9}$\,\Msun]& [\Msun\,yr$^{-1}$]    &    [Gyr$^{-1}$]       &[$\times 10^{11}$\,\Lsun{}]& [kpc] \\
 (1) & (2)   & (3)         & (4)         & (5)   &  (6)      &     (7)  	      &   (8)		      &    (9)  	    & (10)		  &    (11)			     \\ 
\hline
  1  & 3mm.1$^*$,C1 & 03:32:38.54 & -27:46:34.0 & 2.543 & 3,7,8    & $17.8_{-1.7}^{+1.8}$  & $63_{-6}^{+ 6}$	   & $3.4_{-0.31}^{+0.34}$  &	 $12.3_{-1.1}^{+1.2}$	 & 1.7 \\
  2  & 3mm.2,C2     & 03:32:39.74 & -27:46:11.2 & 1.551 & 2,5,6    &  $275_{-40}^{+70}$    & $74_{-30}^{+60}$     & $0.27_{-0.14}^{+0.27}$ &	 $12.0_{-4.3}^{+8.6}$	 & 8.3 \\
  3  & 3mm.3        & 03:32:35.55 & -27:46:25.5 & 1.382 & 2,5      & $52_{-10}^{+12}$      & $18_{-7}^{+ 9}$	   & $0.42_{-0.25}^{+0.13}$ &	 $ 1.9_{-0.9}^{+1.3}$	 & 8.3 \\
  4  & 3mm.5,C6     & 03:32:35.48 & -27:46:26.5 & 1.088 & 2,4      & $28_{-5}^{+ 7}$	    & $23_{-9}^{+20}$	   & $0.9_{-0.4}^{+0.9}$    &	 $ 2.8_{-1.2}^{+2.4}$	 & 5.8 \\
  5  &              & 03:32:36.43 & -27:46:31.8 & 1.098 & 2,4      & $5.8_{-0.5}^{+0.6}$   & $44_{-4}^{+ 4}$	   & $7.41_{-0.7}^{+0.7}$   &	 $15.5_{-1.4}^{+1.5}$	 & 6.0 \\
  6  &       C7     & 03:32:35.78 & -27:46:27.5 & 1.094 & 2,4      & $75_{-13}^{+12}$      & $16_{-6}^{+11}$	   & $0.21_{-0.08}^{+0.17}$ &	 $ 3.1_{-1.1}^{+1.5}$	 & 3.8 \\
  7  &              & 03:32:39.08 & -27:46:01.8 & 1.221 & 2,5      & $15.1_{-1.4}^{+1.5}$  & $148_{-13}^{+15}$    & $9.3_{-0.8}^{+0.9}$    &	 $49.0_{-4.4}^{+4.9}$	 & 0.7 \\
  8  &              & 03:32:36.66 & -27:46:31.0 & 0.999 & 4	    & $70_{-17}^{+11}$      & $40_{-9}^{+14}$	   & $0.54_{-0.05}^{+0.40}$ &	 $ 7.1_{-2.5}^{+1.5}$	 & 6.6 \\
  9  &              & 03:32:39.41 & -27:46:22.4 & 2.447 & 3,7,8    & $2.6_{-0.2}^{+ 0.3}$  & $11.8_{-1.1}^{+1.2}$ & $4.2_{-0.4}^{+0.4}$    &	 $ 1.35_{-0.12}^{+0.13}$ & 5.8 \\
 10  &              & 03:32:37.07 & -27:46:17.2 & 2.224 & 3,6,7    & $12.0_{-1.2}^{+ 1.2}$ & $22_{-2}^{+41}$	   & $1.86_{-0.17}^{+3.53}$ &	 $ 1.95_{-0.18}^{+5.3}$  & 2.7 \\
 11  &              & 03:32:36.33 & -27:46:00.1 & 0.895 & 4	    & $15.9_{-1.4}^{+ 9.0}$ & $42_{-12}^{+4}$	   & $2.7_{-1.5}^{+0.27}$   &	 $ 5.8_{-1.6}^{+0.6}$	 & 1.2 \\
\hline
\end{tabular}
\begin{tablenotes}
      \small
      \item $^*$ Also 1mm.1 and 1mm.2, see Paper I.
\end{tablenotes}
\end{center}
\end{table*}

{\it ID.1} (Tab.~\ref{tab_sample}) is a compact galaxy at $z\approx 2.5$. \citet{momcheva16} report a grism redshift $z=2.561$, based on the detection of the \Oii{} line in the 3D-HST data. This redshift is improved by our blind detection of three CO transitions (ASPECS 3mm.1, 1mm.1, 1mm.2; see Paper I), clearly pininng down the redshift to $z=2.543$. The {\em HST} images show a blue component in the North and a red component in the South (or possibly a single, relatively blue component, partially reddened in the South by a thick dust lane). A group of bright galaxies is present a few arcsec North of this galaxy, but their spectroscopic redshifts show that the group is in the foreground, with only one other source lying at $z\sim2.5$ (the galaxy $\sim2''$ West of ID.1). The starlight emission coincident with the CO detection is compact, with a scale radius $R_e\approx 1.7$\,kpc \citep{vanderwel12}. Chandra reveals X-ray emission associated with this galaxy. The measured X-ray flux is $F_{\rm X}=5.7\times10^{-17}$\,erg\,s$^{-1}$\,cm$^{-2}$, yielding an X-ray luminosity $L_{\rm X}=3.0\times10^{42}$\,erg\,s$^{-1}$ \citep{xue11}.

{\it ID.2} has a {\em HST} morphology consistent with a large disk galaxy at $z$=1.552. Its slit redshift \citep[$z$=$1.552$,][]{kurk13} matches well our CO line detection (ASPECS 3mm.2), assuming CO(2-1). The disk has an inclination of $\sim 60^\circ$ (based on the aspect ratio, \citealt{vanderwel12}), with an effective radius of $8.3$\,kpc. The galaxy is detected with Chandra. \citet{xue11} report a flux of $F_{\rm X}=3.6\times10^{-15}$\,erg\,s$^{-1}$\,cm$^{-2}$ \citep[but $2.6\times10^{-15}$\,erg\,s$^{-1}$\,cm$^{-2}$ in][]{lehmer05}, yielding an X-ray luminosity $L_{\rm X}=5.5\times10^{43}$\,erg\,s$^{-1}$, suggesting that ID.2 hosts an AGN.

{\it ID.3} and {\it ID.4} are the two components of an apparent pair of overlapping spiral galaxies. The southern component exibits bright \Oii{} emission at $\sim7784$\,\AA{} (see Fig.~\ref{fig_muse_dspiral}), clearly placing it at $z$=$1.088$ (in agreement with the CO redshift of ASPECS 3mm.5); the northern component shows bright CO emission (ASPECS 3mm.3) which could be interpreted as CO(2-1) at $z$=$1.382$. Our careful analysis of the MUSE data around 8880\,\AA{} reveals faint \Oii{} emission (although contaminated by sky line emission), supporting the CO identification (see Fig.~\ref{fig_muse_dspiral}). The disk of ID.4 has a scale radius of $5.8$\,kpc based on HST imaging \citep{vanderwel12}; for ID.3, the estimated radius is $8.3$\,kpc (but the overlap with the southern component may partially affect this estimate). ID.4 appears as an upper limit in the X--ray catalog by \citet{xue11} ($F_{\rm X}<6.7\times10^{-17}$\,erg\,s$^{-1}$\,cm$^{-2}$, $L_{\rm X}<4.3\times10^{41}$\,erg\,s$^{-1}$). 

{\it ID.5} and {\it ID.8} lie in a crowded region of our field. \citet{skelton14} report a spectroscopic redshift $z$=$1.047$ for ID.5. However, the inspection of the MUSE data reveals two, clearly distinguished line sets of the \Oii{} doublet, at $z$=$1.038$ and $z$=$1.098$. The latter matches the redshift of two CO lines which are slightly too faint to be selected in our blind search (S/N$\approx$4.8, see Paper I). ID.8 on the other hand is found at another redshift ($z$=$0.999$). No CO emission is found at this position and frequency, although the lowest-J transition that we encompass is CO(4-3) at 1mm. ID.8 is detected in the X--rays \citep{xue11}. Its faintness ($F_{\rm X}=8.2\times 10^{-17}$\,erg\,s$^{-1}$\,cm$^{-2}$, $L_{\rm X}=4.3\times10^{41}$\,erg\,s$^{-1}$) seems consistent with a starburst rather than an AGN \citep{ranalli03}.

{\it ID.6} is located $\sim 4''$ East of ID.3, and probably belongs to a common physical structure (together with other galaxies with a spectroscopic $z\approx1.09$). It is detected in the 1mm continuum, and its CO spectrum shows a $\sim 3$-$\sigma$ excess at the frequency of the expected CO(2-1) line. The CO(4-3) transition is also detected with similar significance, although the best gaussian fit of the line suggests a velocity shift of $\sim200$\,\kms{} between the two transition. This is likely due to the poor S/N of the two lines.

{\it ID.7} appears as a very compact source ($R_{\rm e}=0.7$\,kpc) at $z$=$1.221$. Its {\em Chandra} image reveals the presence of a bright AGN ($F_{\rm X}=1.01\times10^{-14}$\,erg\,s$^{-1}$\,cm$^{-2}$ in \citealt{lehmer05}; $8.3\times10^{-15}$\,erg\,s$^{-1}$\,cm$^{-2}$ in \citealt{evans10}; $6.3\times10^{-15}$\,erg\,s$^{-1}$\,cm$^{-2}$ in \citealt{xue11}), yielding an X-ray luminosity of $L_{\rm X}=5.4\times10^{43}$\,erg\,s$^{-1}$). It is not detected in the 1mm dust continuum. A 3-$\sigma$ excess is measured at the expected frequency of the CO(2-1) transition.

{\it ID.9} and {\it ID.10} are both at $z\sim2.3$. They are among the faintest galaxies in our sample in terms of $L_{\rm IR}$, just above the $10^{11}$\,\Lsun{} cut. ID.9 appears as a compact bulge. ID.10 appears as a spiral galaxy with disturbed morphology. ASPECS data cover 3 CO transitions in these galaxies: CO(3-2), CO(7-6), and CO(8-7). None of these lines is detected.

{\it ID.11} is a compact ($R_{\rm e}$=$1.2$\,kpc) galaxy at $z=0.895$. As for ID.8, the lowest-J CO transition in the ASPECS coverage is the CO(4-3), which remains undetected.

\section{CO--based measurements}\label{sec_results}

\subsection{CO luminosities and associated H$_2$ masses}\label{sec_MH2}

We measure the line fluxes (or place limits) for all the CO transitions covered in both the 3mm and 1mm line scans. We extract the CO spectra at the position of the optical coordinates of the sources in our sample. We fit the lowest-J transitions accessible with ASPECS data with a gaussian profile; in the case of a detection, we fit the higher-J lines imposing the same line width. We consider detections cases where the flux obtained in the gaussian fit is $>$3$\times$ its uncertainty. If the line is not detected, we assume a fiducial line width of 300\,\kms{} and we use the upper boundary of the 3-$\sigma$ confidence range on the flux as upper limit. Tab.~\ref{tab_lines1} reports the CO line fluxes, shifts compared with the nominal redshift, and the line width. The detected sources in our sample have a median CO flux of 0.19\,Jy\,\kms{} (considering only the lowest-J transition observed in each object). For a comparison, the detected main-sequence galaxies in \citet{tacconi13} have a median CO flux of 0.57\,Jy\,\kms{}, i.e., $3\times$ higher than the median flux of our detections.

The luminosity of the lowest-J transitions observed in our molecular scans is transformed into the equivalent ground state luminosity $L'_{\rm CO(1-0)}$ using $L'_{\rm CO(J-[J-1])}/r_{J1}$, where we adopt the recent CO excitation ladder of main--sequence galaxies derived by \citet{daddi15}: $r_{21}$=$0.76\pm0.09$, $r_{31}=0.42\pm0.07$, and $r_{41}=0.31\pm0.06$\footnote{\citet{daddi15} do not measure CO(4-3) in the galaxies in their sample. The value of $r_{41}$ adopted here is extrapolated from their measurements of $r_{31}$ and $r_{51}$, in the case of a CO ladder that peaks around J$\approx$5 (see their Fig.~10, left). As uncertainty, we adopt conservative 20\% error.}. The uncertainty in $L'_{\rm CO(1-0)}$ accounts for both the measured flux uncertainty and the standard deviation in the $r_{J1}$ values in the sample studied by \citet{daddi15}. The molecular gas masses are then derived as:

\begin{equation}\label{eq_MH2}
\frac{M_{\rm H2}}{\rm M_\odot} = \frac{\alpha_{\rm CO}}{r_{J1}}\,\frac{L'_{\rm (J-[J-1])}}{\rm K\,km\,s^{-1}\,pc^2}
\end{equation}

We adopt $\alpha_{\rm CO}$=3.6\,\Msun\,(K \kms{} pc$^{2}$)$^{-1}$ \citep{daddi10b}. This conversion factor has been demonstrated to be appropriate for main--sequence galaxies, through comparisons with dynamical masses \citep{daddi10b}, CO line SED--fitting \citep{daddi15} and detailed dust-SED modeling \citep{genzel15}. In Sec.~\ref{sec_co_vs_dust} we further discuss the implications of our $\alpha_{\rm CO}$ assumption. Tab.~\ref{tab_lines2} lists the values of molecular gas masses that we derive for each source. We then combine these measurements or limits on the molecular gas mass with properties of the galaxies inferred from the SED fitting (in particular, the stellar mass $M_*$ and the SFR), to compute the molecular--to--stellar mass ratio $M_{\rm H2}/M_*$ and the depletion time scale $t_{\rm dep}=M_{\rm H2}/{\rm SFR}$ (see Tab.~\ref{tab_lines2}).

\begin{table}
\caption{\rm CO lines in the galaxies of our sample. (1) Source ID. (2) Upper J of the CO transition. (3) Velocity shift, compared with the redshift quoted in Tab.~\ref{tab_sample}. (4) Line flux. (5) Line width, expressed as full width at half maximum (FWHM) from the gaussian fit.} \label{tab_lines1}
\begin{center}
\begin{tabular}{ccccc}
\hline
ID   & J$_{\rm up}$ & $\Delta v$ & $F_{\rm line}$ & FWHM    \\
     &              & [\kms]     & [Jy\,\kms]     & [\kms{}] \\
 (1) & (2)          & (3)        & (4)            & (5)      \\ 
\hline
1  &3&$ -45\pm  8$  &  $0.723_{-0.003}^{+0.003}$ &  $504\pm12$ \\ 
1  &7&$ -150\pm120$ &  $0.786_{-0.006}^{+0.006}$ &  $504^*   $ \\ 
1  &8&$ -45\pm70$   &  $1.098_{-0.005}^{+0.005}$ &  $504^*   $ \\ 
\hline
2  &2&$ 135\pm 9$   &  $0.443_{-0.007}^{+0.007}$ &  $538\pm13$ \\ 
2  &5&$ 135\pm45$   &  $0.502_{-0.090}^{+0.090}$ &  $538^*   $ \\ 
2  &6&$ -45\pm45$   &  $0.820_{-0.100}^{+0.100}$ &  $538^*   $ \\ 
\hline
3  &2&$ -37\pm 8$   &  $0.135_{-0.003}^{+0.003}$ &  $ 57\pm12$ \\ 
3  &5&   ---        & $<0.021$                   &  $ 57^*   $ \\ 
\hline
4  &2&$  52\pm  7$  &  $0.180_{-0.006}^{+0.006}$ &  $ 82\pm11$ \\ 
4  &4&   ---        & $<0.121$                   &  $ 82^*   $ \\ 
\hline
5  &2&$ 220\pm 35$  &  $0.190_{-0.040}^{+0.040}$ &  $352\pm11$ \\ 
5  &4&$ -28\pm40$   &  $0.390_{-0.065}^{+0.065}$ &  $352^*   $ \\ 
\hline
6  &2&$ -160\pm 70$ &  $0.340_{-0.070}^{+0.060}$ &  $530\pm11$ \\ 
6  &4&$  230\pm 70$ &  $0.370_{-0.090}^{+0.090}$ &  $182^*   $ \\ 
\hline
7  &2&$ 150\pm17$   &  $0.104_{-0.029}^{+0.019}$ &  $150\pm11$ \\ 
7  &5&   ---        &  $<0.106$                  &  $150^*   $ \\ 
\hline
8  &4&       ---       & $<0.059$                   &     ---     \\ 
\hline
9  &3&       ---       & $<0.076$                   &     ---     \\ 
9  &7&       ---       & $<0.012$                   &     ---     \\ 
9  &8&       ---       & $<0.230$                   &     ---     \\ 
\hline
10 &3&       ---       & $<0.048$                   &     ---     \\ 
10 &6&       ---       & $<0.144$                   &     ---     \\ 
10 &7&       ---       & $<0.465$                   &     ---     \\ 
\hline
11 &4&       ---       & $<0.015$                   &     ---     \\ 
\hline
\end{tabular}
\begin{tablenotes}
      \small
      \item $^*$ Fixed from the fit of a lower J line.
\end{tablenotes}
\end{center}
\end{table}

\begin{table*}
\caption{\rm CO luminosities and CO-based galaxy parameters. (1) Source ID. (2) Redshift. (3) Observed transition. (4) Line luminosity. (5) Equivalent CO(1-0) luminosity, assuming the $r_{J1}$ ratios in \citet{daddi15}. (6) Molecular gas mass, assuming $\alpha_{\rm CO}$=$3.6$ \Msun{} (\Kkmspc)$^{-1}$. (7) Molecular--to--stellar mass ratio, $M_{\rm H2}/M_*$. (8) Depletion time, $t_{\rm dep}=M_{\rm H2}/{\rm SFR}$.} \label{tab_lines2}
\begin{center}
\begin{tabular}{cccccccc}
\hline
ID   & $z$ & J$_{\rm up}$ & $L'$     &  $L'_{\rm CO(1-0)}$ & $M_{\rm H2}$  &  $M_{\rm H2}/M_{*}$  & $t_{\rm depl}$ \\
     &     &              & [$\times10^9$\,\Kkmspc] & [$\times10^9$\,\Kkmspc] & [$\times 10^9$ \Msun] &   &  [Gyr] \\
 (1) & (2) &  (3)         & (4)      & (5)                 & (6)           & (7)                  & (8)     \\ 
\hline
 1&2.543&3& $24.03_{-0.10}^{+0.10}$ &	$57_{-10}^{+10}$    & $206_{-34}^{+34}$    & $12_{-2}^{+2}$         &   $3.3_{-0.6}^{+0.7}$	  \\
 2&1.551&2& $13.71_{-0.27}^{+0.21}$ &	$18_{-2}^{+2}$      &  $65_{-8}^{+8}$      & $0.24_{-0.05}^{+0.05}$ &   $0.9_{-0.4}^{+0.6}$	  \\
 3&1.382&2& $3.364_{-0.08}^{+0.07}$ &	$4.4_{-0.5}^{+0.5}$ & $15.9_{-1.9}^{+1.9}$ & $0.30_{-0.07}^{+0.08}$ &   $0.9_{-0.3}^{+0.6}$	  \\
 4&1.088&2& $2.831_{-0.09}^{+0.09}$ &	$3.7_{-0.5}^{+0.5}$ & $13.4_{-1.7}^{+1.7}$ & $0.48_{-0.11}^{+0.13}$ &   $0.6_{-0.3}^{+0.4}$	  \\
 5&1.098&2& $3.089_{-0.66}^{+0.70}$ &	$4.1_{-1.0}^{+1.0}$ & $15_{-4}^{+4}$       & $2.5_{-0.6}^{+0.7}$    &   $0.33_{-0.09}^{+0.09}$	  \\
 6&1.094&2& $5.388_{-1.16}^{+0.91}$ &	$7.1_{-1.7}^{+1.5}$ & $25_{-6}^{+5}$       & $0.34_{-0.09}^{+0.10}$ &   $1.6_{-0.7}^{+0.9}$	  \\
 7&1.221&2& $2.047_{-0.57}^{+0.37}$ &	$2.7_{-0.8}^{+0.6}$ & $10_{-3}^{+2}$       & $0.6_{-0.2}^{+0.16}$   &   $0.066_{-0.020}^{+0.016}$ \\
 8&0.999&4& $<0.20$                 &	$<0.63$ 	    & $<2.3$               & $<0.03$                &   $<0.06$                   \\
 9&2.447&3& $<2.4$                  &	$<5.6$  	    & $<21$                & $<8$                   &   $<1.8$                    \\
10&2.224&3& $<2.2$                  &	$<5.3$  	    & $<19$                & $<1.6$                 &   $<0.9$                    \\
11&0.895&4& $<0.53$                 &	$<1.7$  	    & $<6.2$               & $<0.4$                 &   $<0.15$                   \\
\hline
\end{tabular}
\end{center}
\end{table*}

\begin{figure*}
\includegraphics[width=0.49\textwidth]{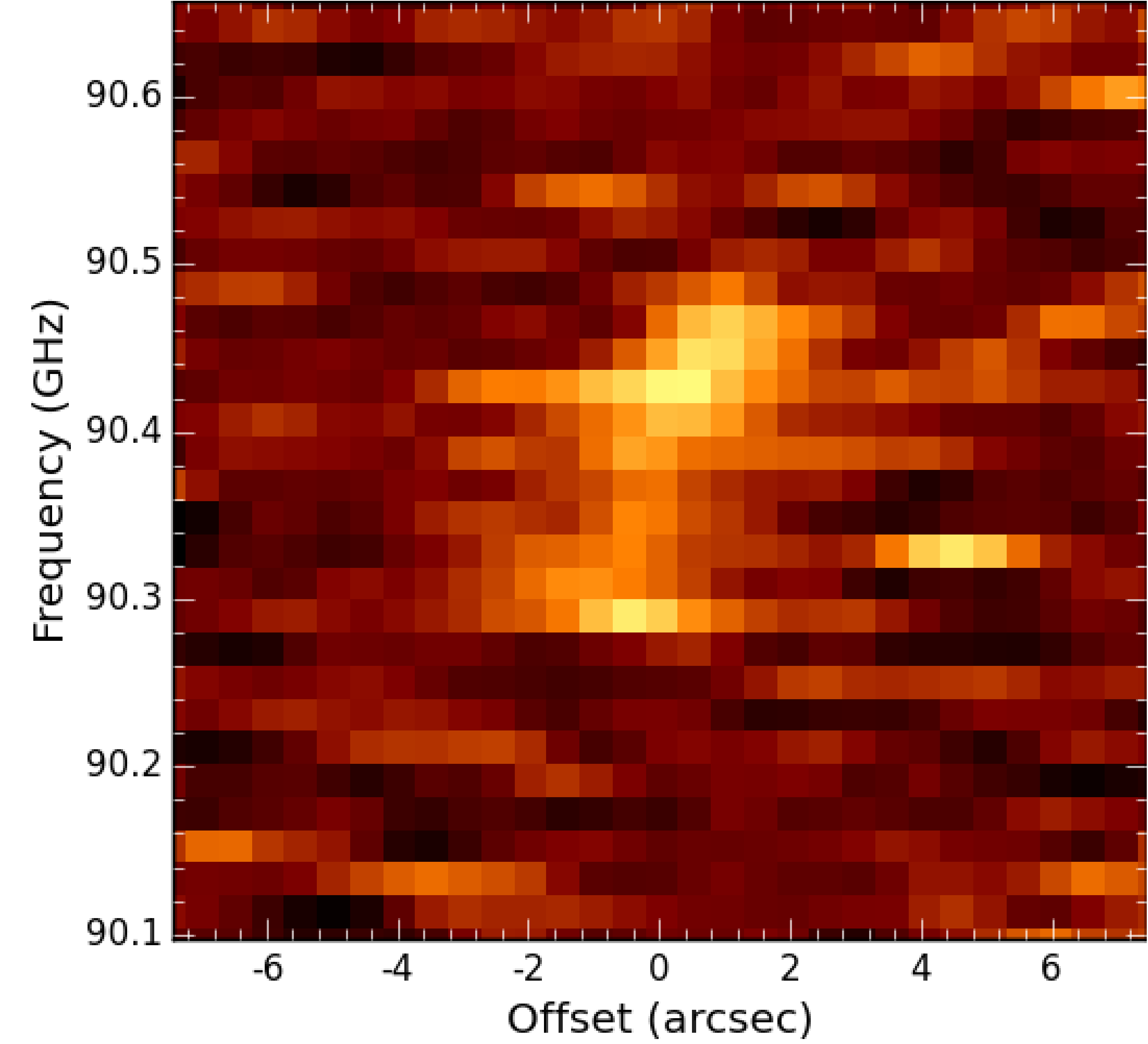}
\includegraphics[width=0.49\textwidth]{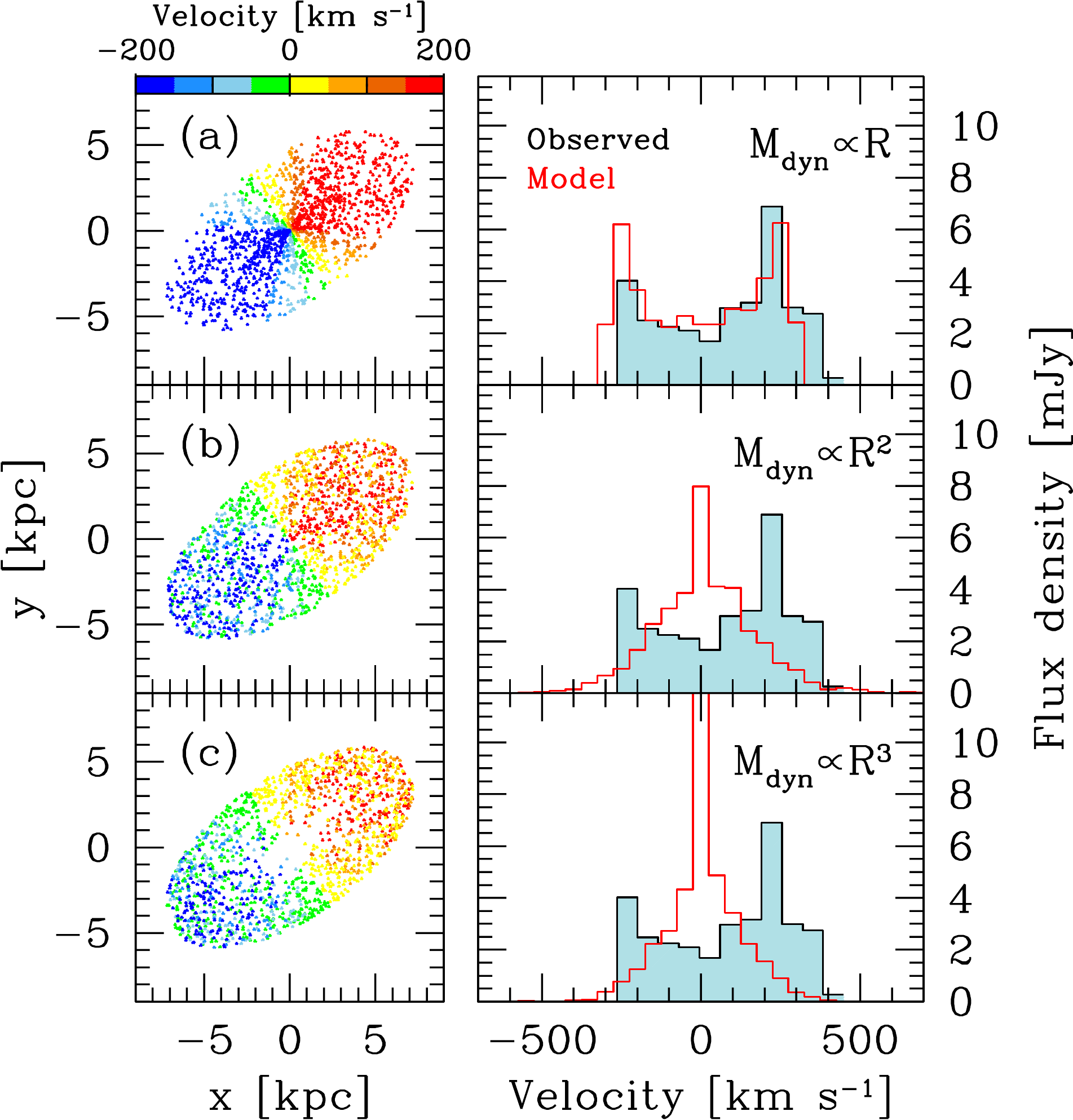}
\caption{{\em Left ---} Position-velocity diagram of the CO(2-1) emission in ID.2, extracted along the major axis of the galaxy. A velocity gradient is apparent. {\em Right ---} Simulated velocity maps of ID.2, assuming that the gas is emitted in a disk geometry and that the CO(2-1) emission traces the mass distribution. The three models refer to different radial scaling of the dynamical mass: (a) $M_{\rm dyn}\propto R$ (thus $v_{\rm rot}$={\it const}); (b) $M_{\rm dyn}\propto R^2$ (i.e., constant surface mass density in the disk; $v_{\rm rot}\propto \sqrt{R}$); (c) $M_{\rm dyn}\propto R^3$ (i.e., constant volume mass density; $v_{\rm rot}\propto R$, i.e., solid rotator). All models assume a dynamical mass $M_{\rm dyn}=2\times10^{11}$\,\Msun{} at $R$=8.3\,kpc. The expected line profiles (red histograms) are compared with the observed one (black dots). The flat rotation curve model seems to best reproduce the observed line profile.}\label{fig_mdyn}
\end{figure*}

\subsection{Size of the CO-emitting region in ID.2}\label{sec_mdyn}

In the case of ID.2, our ALMA observations spatially resolve the CO(2-1) emission over $>15$\,kpc, despite the relatively coarse spatial resolution of the 3mm data. A clear velocity gradient is observed in the line emission, as shown in Fig.~\ref{fig_mdyn}. While the resolution and the signal-to-noise are too poor for an accurate modeling of the gas dynamics, we obtain an estimate of the dynamical mass assuming that the gas is rotating in a disk with the inclination derived from the {\em HST} near-IR imaging (PA=-55$^\circ$, inclination=60$^\circ$ with respect to the line of sight). We then assume a radial distribution of the mass that scales as $M_{\rm dyn}\propto R^\gamma$, where $\gamma$=1 yields the flat rotation curves typically observed in galaxies; $\gamma$=2 implies a constant surface density of mass in the disk, and yields $v_{\rm rot}\propto \sqrt{R}$; and $\gamma$=3 corresponds to a solid rotator ($v_{\rm rot}\propto R$). We then generated mock velocity maps for these three cases, assuming that the CO light traces the mass distribution; and we inferred expected line profiles (see Fig.~\ref{fig_mdyn}). The $\gamma$=1 case shows the typical ``double-horned'' profile observed in local spiral galaxies. This seems to provide a better description of the observed CO(2-1) line than the other two models, which fail to reproduce the extension of the blue wing of the line. The implied dynamical mass is $M_{\rm dyn}\approx 2\times 10^{11}$\,\Msun{} at $R$=8.3\,kpc (= the effective radius). We stress however that this estimate is highly dependent on the model assumptions. A firmer estimate of the dynamical mass in this galaxy requires deeper data at higher spatial resolution.

ID.2 also appears in the SINFONI Integral field spectroscopy survey in the Near-IR \citep[SINS;][]{foerster09} as GMASS-1084 \citep[see also][]{kurk13}. SINS investigated the morphology and kinematics of ionized gas (as traced by the H$\alpha$ Hydrogen line) in a sample of galaxies at $z=1-3.5$. The H$\alpha$ line in ID.2 is emitted on a smaller region (half-light radius $R_{1/2}=3.1\pm1.0$\,kpc) than the CO. The observed H$\alpha$ circular velocity is $67\pm9$\,\kms{}, which is corrected into 230\,\kms{} by assuming a low inclination angle ($\sim 20^\circ$). This yields a dynamical mass of $1.2\times 10^{11}$\,\Msun{}, roughly consistent with our estimate, especially if one considers that the high level of dust reddening ($A_V=2.4$\,mag from our global MAGPHYS fit) in this source may be responsible of suppressing H$\alpha$ in parts of this galaxy. We note however that the SED fit of this source in \citet{foerster09} yields a stellar mass of only $M_*=3.61_{-0.60}^{+0.34}\times10^{10}$\,\Msun{} and a large SFR=$490_{-31}^{+190}$\,\Msun{}\,yr$^{-1}$ (i.e., $L_{\rm IR}\approx 5.7\times10^{12}$\,\Lsun). This last estimate disagrees with our dust continuum measurements: e.g., assuming a modified black body template with $\beta$=1.6 and $T_{\rm dust}$=25\,K, such a high SFR would imply a dust continuum flux density of 11\,mJy at 1.2mm (observed: $0.22\pm0.02$\,mJy) and of 32\,mJy at 160\,$\mu$m (observed: $6.9\pm0.3$\,mJy). 


As seen from Fig.~\ref{fig_mdyn}, the molecular gas, as traced through CO emission, is extended on scales of $>15$\,kpc ($> 2''$ at $z$=1.552), i.e., comparable to that of the stellar disk. On the other hand, the 1\,mm dust continuum is unresolved at $1.5''\times1.0''$ resolution, i.e., it is significantly smaller than that of the CO (see Fig.~\ref{fig_id1_a}). This is not an effect of interferometric filtering or sensitivity of the 1mm data. If we convolve the 1mm continuum data to the synthesized beam of the 3mm data, we do not recover the size seen in CO emission. This serves as a cautionary note that CO and dust sizes may not be the same. As a consequence, the masses deduced from these measurements may trace different regions or components in the galaxy \citep[for other examples of mismatch between CO and dust morphology in high-redshift galaxies, see][]{riechers11,hodge15,spilker15}. 
This may explain some of the differences between gas mass estimates derived from CO and dust imaging, with the gas masses derived from dust emission being typically smaller than those derived from CO (see Sec.~\ref{sec_co_vs_dust}): At the observed wavelength (1.2 mm), dust is optically thin (with the only exception of ID.1, all the sources in our sample globally have $\Sigma_{\rm gas}\ll 10^4$\,\Msun{}\,pc$^{-2}$, i.e., $N_{\rm H2}\ll 10^{24}$\,cm$^{-2}$; this yields $\tau$[242\,GHz]$\ll$0.1 for solar metallicities, adopting the \citealt{draine84} formalism). The CO low-J emission, on the other hand, is optically thick practically everywhere in galaxies. 


\subsection{CO excitation}

\begin{figure}
\includegraphics[width=0.99\columnwidth]{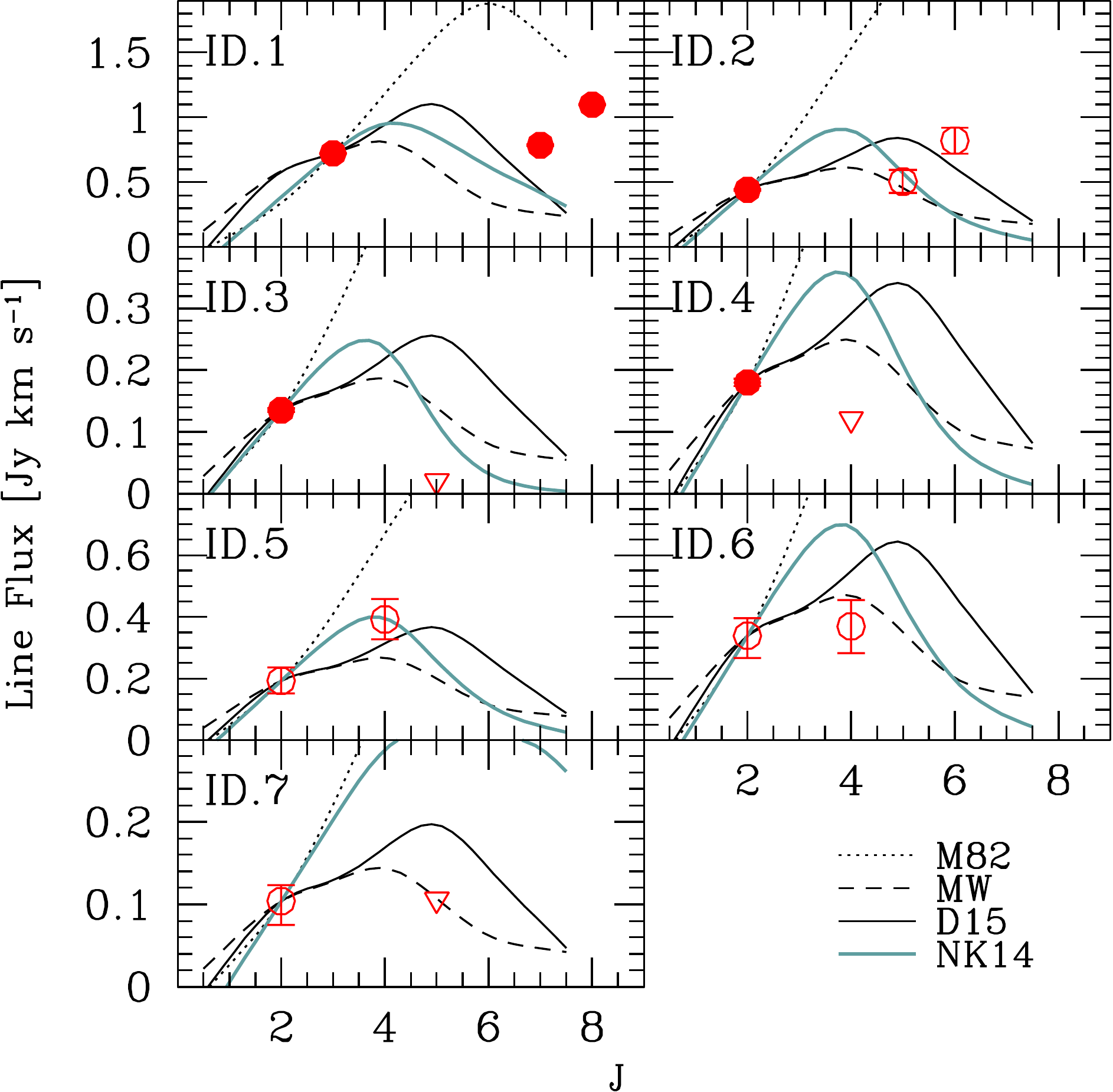}
\caption{CO ladder for the galaxies of our sample detected in CO. Filled symbols mark the transitions detected in our blind search (see Paper I), while empty symbols mark lines that do not match the blind detection requirements. Upper limits, marked with triangles, correspond to 3-$\sigma$ limits. The excitation templates of the Milky Way and M82 are taken from \citet{weiss07}, while the main sequence galaxy template is from \citet{daddi15} (D15). Finally, the theoretical predictions based on the SFR surface density are based on \citet{narayanan14} (NK14). All templates are scaled to match the observed CO flux of the lowest J transition detected in ASPECS. The galaxies in our sample typically show a modest to very low CO excitation. ID.1 (=ASPECS 3mm.1, 1mm.1,2) and ID.5 show slightly higher CO excitation than the template by \citet{daddi15}, although still well below the high-excitation case of the M82 starburst template.}\label{fig_co_excit}
\end{figure}

As shown in Tab.~\ref{tab_sample}, ASPECS cover 2--3 different CO transitions in 9 out of 11 galaxies in our sample. Fig.~\ref{fig_co_excit} shows the inferred constraints on the CO excitation ladder. In ID.1, all three observed transitions [CO(3-2), CO(7-6), and CO(8-7)] were detected in our blind search for line emission (ASPECS 3mm.1, 1mm.1, and 1mm.2, respectively). In ID.2, the CO(2-1) line appears in the results of our blind search (ASPECS 3mm.2). The CO(5-4) and CO(6-5) lines are also observed, but because of their lower significance, they were not detected in our blind search. In particular, the CO(6-5) line is very noisy as it is found at the high-frequency end of the 1mm spectral scan, and it is spatially located at the edge of our mosaic. The CO(2-1) transitions in ID.3 and ID.4 are also identified in our blind search (ASPECS 3mm.3 and 3mm.5, respectively). However, the CO(5-4) line in ID.3 and the CO(4-3) line in ID.4 are not detected. In particular for ID.3, this places very strong limits on the CO excitation of this galaxy, significantly below the average CO ladder of the Milky Way disk (see Fig.~\ref{fig_co_excit}). In ID.5 and ID.6, we detect both CO(2-1) and CO(4-3). Finally, in ID.7 we only have a tentative detection of CO(2-1), while the CO(5-4) transition remains undetected. No other line is detected in the remainder of our sample.

In Fig.~\ref{fig_co_excit}, we compare our measurements and limits with the CO excitation templates of the Milky Way disk, and of the starburst in M82 \citep{weiss07}. Additionally, we compare with the average template for high-$z$ main sequence galaxies by \citet{daddi15} and with the theoretical predictions based on the SFR surface density by \citet{narayanan14} (see Tab.~\ref{tab_sample} and the discussion in sec.~\ref{sec_ks}). In no case do we find starburst-like CO excitation, comparable with the center of M82 \citep{weiss07} or with what is typically observed in high-$z$ SMGs \citep[e.g.,][]{bothwell13,spilker14}. ID.1 shows a CO(7-6)/CO(3-2) ratio $r_{73}$=$0.2$, consistent with a high-density photon-dominated region \citep{meijerink07}. On the other hand, the CO(8-7) transition appears brighter, implying that a high-excitation component of the ISM might be in place. Interestingly, ID.2 shows a lower excitation (in particular in the CO[5-4]/CO[2-1] ratio, which is consistent with Milky Way excitation). This difference in CO excitation is remarkable if one considers that ID.1 is not detected with Chandra ($L_{\rm X}<6\times10^{42}$\,erg\,s$^{-1}$), whereas ID.2 shows a bright X--ray detection, indicative of the presence of a central AGN. The X-ray emission from the AGN can boost the emission of high-J CO transitions. The CO(7-6)/CO(3-2) ratio $r_{73}$ is typically $0.16-0.63$ in high-density photon-dominated regions powered by star formation (as in ID.2), but it can reach values as high as $r_{73}$=30 in presence of intense X-ray illumination \citep{meijerink07}. This might explain the higher CO excitation observed in high-$z$ QSOs with respect to sub-mm galaxies \citep{carilli13}. The lack of such high excitation feature suggests that the central AGN activity in ID.2 has no major impact on its global CO properties. We attribute the higher excitation in ID.1 to the much more compact emission in this galaxy. As shown in Tab.~\ref{tab_sample}, ID.1 has a radius that is only $\sim$1/5 of that of ID.2, which translates into a difference in surface area of $\sim$24. Our MAGPHYS-based SFR estimates are comparable ($\sim 70$\,\Msun{}\,yr$^{-1}$), thus the surface density of star formation ($\Sigma_{\rm SFR}$) is much higher in ID.1. This is also discussed in Sec.~\ref{sec_sfr_law} below. The increased radiation field intensity caused by the high star formation rate surface density and/or the higher gas density are likely the reason for the increased CO excitation (see \citealt{narayanan14}). 

\section{Discussion}\label{sec_discussion}

In the following we discuss the sources of our sample in the broad context of gas properties in high--redshift galaxies. 

\subsection{Location in the galaxy `main sequence' plot}

The stellar masses of the galaxies in our sample range between $(2.8-275)\times 10^{9}$\,\Msun{} (two orders of magnitude). The $L_{\rm IR}>10^{11}$\,\Lsun{} cut in our sample definition selects sources with SFR$>$10\,\Msun\,yr$^{-1}$. The measured SFRs range between 12--150\,\Msun{}\,yr$^{-1}$ \footnote{We note that the FAST analysis by \citet{skelton14} yields consistent SFRs for ID.1, ID.3, and ID.10 but different values (by a factor $2\times$ or more) for ID.2 ($6$\,\Msun\,yr$^{-1}$), ID.4 (50\,\Msun\,yr$^{-1}$), ID.5 (21\,\Msun\,yr$^{-1}$), ID.6 ($3.7$\,\Msun{}\,yr$^{-1}$), ID.7 (230\,\Msun\,yr$^{-1}$), ID.8 ($0.01$\,\Msun\,yr$^{-1}$), ID.11 ($2.6$\,\Msun\,yr$^{-1}$). No FAST-based SFR estimate is available for ID.9. These differences are likely due to 1) different assumptions on the source redshifts; 2) different coverage of the SED photometry, in particular thanks to the addition of the 1mm continuum constraint in our MAGPHYS analysis; 3) different working assumptions in the two codes. In particular, FAST relies on relatively limited prescriptions for the dust attenuation and star formation history, and does not model the dust emission.}.

Fig.~\ref{fig_ms} shows the location of our galaxies in the $M_{\rm *}$--SFR (`main sequence') plane. We plot all the galaxies in the field with a F850LP or F160W magnitude brighter than $27.5$\,mag (this cut allows us to remove sources with highly uncertain SED fits). The galaxies in the present sample are highlighted with large symbols.
The different redshifts of the sources are indicated by different colors. As expected from the known evolution of the `main sequence' of star-forming galaxies \citep[e.g.,][]{whitaker12,schreiber15}, sources at higher redshifts tend to have higher SFR per unit stellar mass. Comparing with the {\em Herschel}--based results by \citet{schreiber15}, we find that half of the galaxies in our sample (ID.1, 2, 4, 8, 9, 10) lie on the main sequence (within a factor $3\times$) at their redshift. Three galaxies (ID.5, 7, 11) are above the main sequence (in the `starburst' region), and the remaining two galaxies (ID.3 and ID.6) show a SFR $\sim 3\times$ lower than main sequence galaxies at those redshifts and stellar masses. Similar conclusions are reached if we compare our results with the main sequence fits by \citet{whitaker12} (see Fig.~\ref{fig_ms}).

\subsection{Star formation law}\label{sec_sfr_law}

The relationship between the total infrared luminosity ($L_{\rm IR}$, a proxy for the star formation rate) and the total CO luminosity ($L'_{\rm CO}$, a proxy for the available gas mass) of galaxy samples is typically referred to as the `integrated Schmidt--Kennicutt' law \citep{schmidt59,kennicutt98,kennicutt12}, or, more generally, the `star formation' law. Sometimes average surface density values are derived from these quantities, resulting in average surface star formation rate densities ($\Sigma_{\rm SFR}$) and gas densities ($\Sigma_{\rm gas}$). We here explore both relations.

\subsubsection{Global star formation law: IR vs.\ CO luminosities}\label{sec_co_lum}

\begin{figure}
\includegraphics[width=0.99\columnwidth]{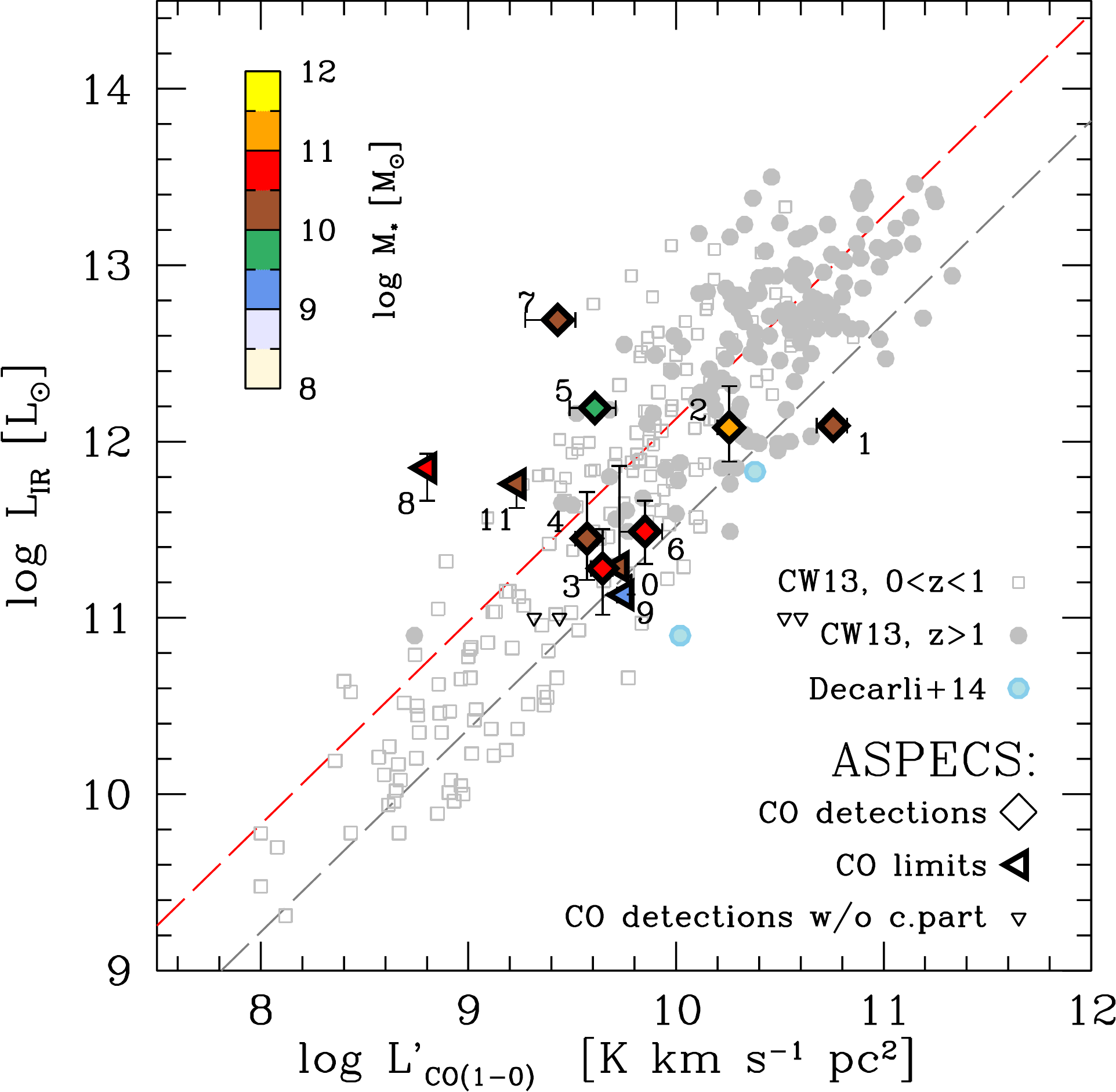}
\caption{IR luminosity as a function of the CO(1-0) luminosity for both local galaxies (grey open symbols) and high-redshift sources ($z>1$, grey filled symbols) from the compilation in \citet{carilli13}. The sources in our sample are shown with big symbols, using the same coding as in Fig.~\ref{fig_ms}. In addition, we also plot the x-axis position of the remaining CO lines found in our 3mm blind search (down-ward triangles; see Paper I). The two parallel sequences of `normal' and `starburst' galaxies \citep{daddi10b,genzel10} are shown as dashed lines (in grey and red, respectively). Our sources cover a wide range of luminosities, both in the CO line and in the IR continuum. Most of the sources in our sample lie along the sequence of `main sequence' galaxies. Four sources lie above the relation: ID.5, which still falls close to the high-$z$ starburst region; ID.7, in which the AGN contamination may lead to an excess of IR luminosity; and ID.8 and ID.11, which are undetected in CO, and that could be shifted towards the relation if one assumes very low CO excitation (as observed in other galaxies of our sample). Conversely, most of the sources detected in CO in our blind search (see Paper I) which lack of an optical/IR counterpart lie significantly below the observed relation.}\label{fig_co_lum}
\end{figure}

In Fig.~\ref{fig_co_lum} we compare the IR and CO(1-0) luminosities of our sources with respect to a compilation of galaxies both at low and high redshift from the review by \citet{carilli13}, and with the secure blind detections in \citet{decarli14}. For galaxies in our sample that are undetected in CO, we plot the corresponding 3-$\sigma$ limit on the line luminosities. The IR--CO luminosity empirical relation motivates the $L_{\rm IR}$ cut in our sample selection, as galaxies with $L_{\rm IR}>10^{11}$\,\Lsun{} should have CO emission brighter than $L'\approx 3\times10^9$\,\Kkmspc{} (i.e., our typical sensitivity limit in ASPECS; see Paper I). {\em All} the galaxies in our sample should therefore be detected in CO.
We find that most of the CO--detected galaxies in our sample lie along the 1-to-1 relation followed by local spiral galaxies as well as color-selected main sequence galaxies at $1<z<3$ \citep{daddi10b,genzel10,genzel15,tacconi13}. Only two galaxies significantly deviate: ID.5, which appears on the upper envelope of the IR--CO relation, close to high-redshift starburst galaxies; and ID.7, which is largely underluminous in CO for its bright IR emission. As discussed in the previous section, these two galaxies appear as starbursts in Fig.~\ref{fig_ms}. Moreover, ID.7 hosts a bright AGN. If the AGN contamination at optical wavelengths is significant, our MAGPHYS-based SFR estimate is likely in excess (since MAGPHYS would associate some of the AGN light at rest-frame optical and UV wavelengths to a young stellar population), thus explaining the big vertical offset of this galaxy with respect to the `star formation law' shown in Fig.~\ref{fig_co_lum}. Notably, out of the 4 CO non-detections in our sample, ID.9 and ID.10 are still consistent with the relation, while ID.8 and ID.11 are not. These two galaxies are located at $z=0.999$ and $z=0.895$ respectively. The lowest-J transition sampled in our study is CO(4-3). Their non-detections might be explained if the excitation in these two sources was much lower than what we assumed to infer $L'_{\rm CO(1-0)}$ ($r_{41}=0.31$; see Sec.~\ref{sec_MH2}).

The sources that are also detected in the blind search for CO (ID.1, 2, 3, 4) tend to lie on the lower `envelope' of the plot. This is expected, as these galaxies have been selected based on their CO luminosity (x--axis). 

Fig.~\ref{fig_co_lum} also shows the x-axis position of the remaining CO blind detections from the 3mm search in Paper I. The CO luminosities of these lines are uncertain (the line identification is ambiguous in many cases, and a fraction of these lines is expected to be a false-positive; see Paper I); however, it is interesting to note that these sources typically populate ranges of line luminosities that were previously unexplored at $z>1$ \citep[see similar examples in][]{chapman08,chapman15b,casey11}, and comparable with or even lower than the typical dust luminosities of local spiral galaxies. We emphasize that a significant fraction of these lines is expected to be real (see Paper I). Deeper data are required to better characterize these candidates.

\subsubsection{Average surface densities: SFR vs.\ gas mass}\label{sec_ks}

We infer average estimates of $\Sigma_{\rm SFR}$ and $\Sigma_{\rm gas}$ by dividing the global SFR and $M_{\rm H2}$ of the galaxy by a fiducial area set by the size of the stellar component, as CO and optical radii are typically comparable \citep{schruba11,tacconi13}. We thus use the information from the stellar morphology derived by \citet{vanderwel12} and reported in Tab.~\ref{tab_sample} to infer $\Sigma_{\rm SFR}$=SFR/($2\,\pi \, R_e^2$) and $\Sigma_{\rm H2}$=$M_{\rm H2}$/($2\,\pi \, R_e^2$), where $M_{\rm H2}$ is our CO-based measurement of the molecular gas mass, and the factor 2 is due to the fact that the $R_e$ includes only half of the light of the galaxy \citep[see a similar approach in][]{tacconi13}. 

\begin{figure}
\includegraphics[width=0.99\columnwidth]{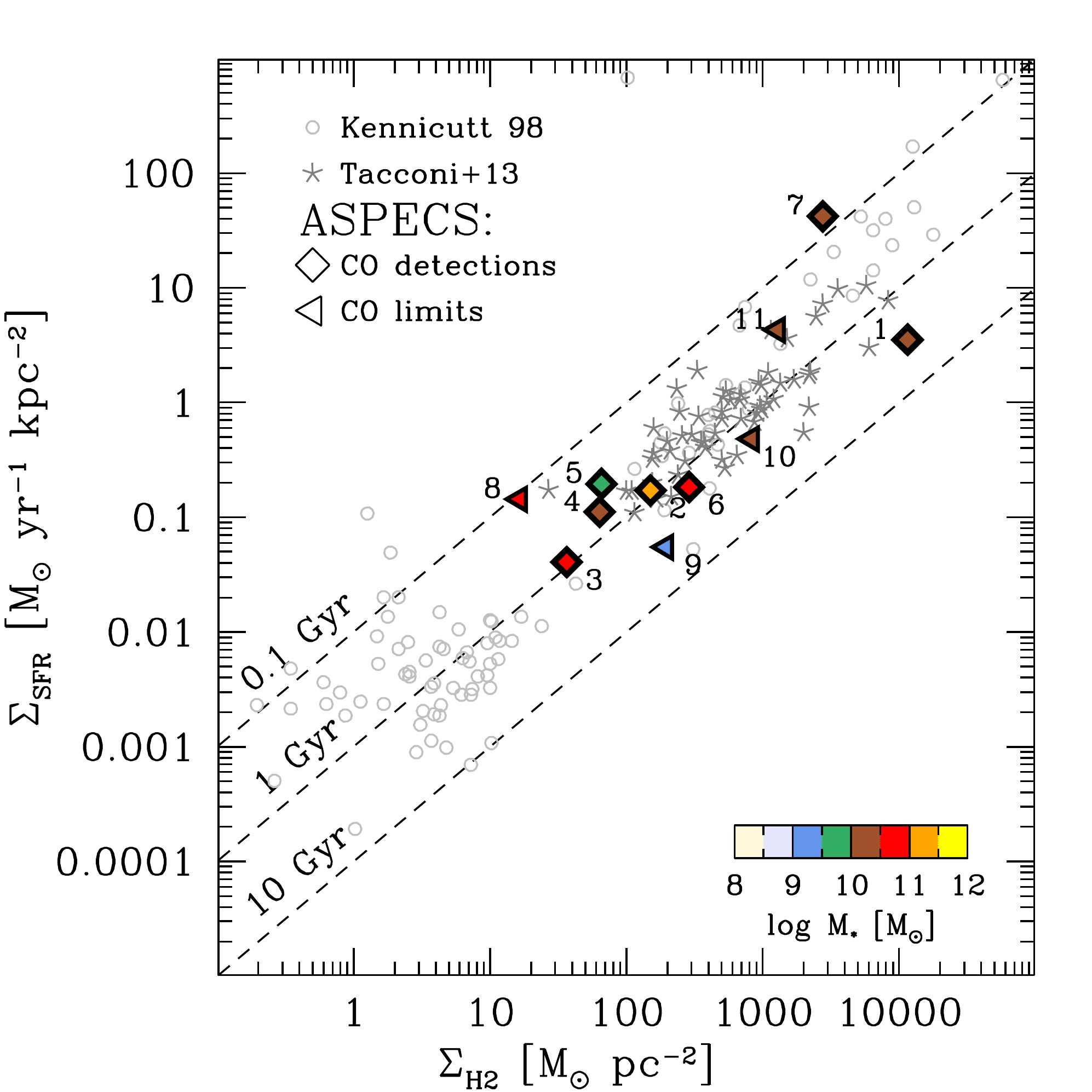}
\caption{The `global' star-formation law relates the average star formation rate surface density ($\Sigma_{\rm SFR}$) with the average gas density in galaxies. Here we consider only the molecular gas phase ($\Sigma_{\rm H2}$). Each point in the plot refers to a different galaxy. We plot the reference samples from \citet{kennicutt98} \citep[corrected for the updated SFR calibration in][]{kennicutt12}, as well as the PHIBSS galaxies from \citet{tacconi13}. Data from the literature have been corrected to match the same $\alpha_{\rm CO}=3.6$\,\Msun{}(\Kkmspc)$^{-1}$ assumed in this work. The symbol code is the same as in Fig.~\ref{fig_co_lum}. The galaxies in our sample align along the $t_{\rm depl}\approx 1$\,Gyr, with the only exception of ID.7 and ID.8 which show a short depletion time. It is interesting to note that the two CO-brightest galaxies in our sample, ID.1 and ID.2, populate opposite extremes of the high-$z$ galaxy distribution, with the former being very compact (thus displaying higher SFR and gas densities), and the latter being very extended (thus showing lower SFR and gas densities).}\label{fig_ks}
\end{figure}

In Fig.~\ref{fig_ks} we show the star-formation law for average surface densities. Global measurements of local spiral galaxies and starbursts are taken from \citet{kennicutt98}, and corrected for the updated SFR calibration following \citet{kennicutt12} and to the $\alpha_{\rm CO}$ value adopted in this paper. We also plot the galaxies in the IRAM Plateau de Bure HIgh-z Blue Sequence Survey \citep[PHIBSS;][]{tacconi13}, again corrected to match the same $\alpha_{\rm CO}$ assumption used in this work, and the secure detections in \citet{decarli14}. Interestingly, the two CO-brightest galaxies in our sample, ID.1 and ID.2 appear to populate opposite extremes of the density ranges observed in high-$z$ galaxies: ID.1 appears very compact, thus reaching the top-right corner of the plot ($\Sigma_{\rm gas}\approx 10000$\,\Msun{}\,pc$^{-2}$). On the other hand, in ID.2 the vast gas reservoir is spread over a large area (as apparent in Fig.~\ref{fig_mdyn}), thus yielding a globally low $\Sigma_{\rm gas}$. We also find that most of the sources in our sample lie along the $t_{\rm depl}\approx 1$\,Gyr line, in agreement with local spiral galaxies and the PHIBSS main sequence galaxies. Only ID.7 and ID.8 lie closer to the $t_{\rm depl}\approx0.1$\,Gyr line. In particular, the offset of ID.7 with respect to the bulk of the sample in the context of the global star-formation law (Fig.~\ref{fig_co_lum}) is combined here with the very compact size of the emitting region, thus isolating the source in the top-left corner of the plot (see Fig.~\ref{fig_ks}). Once again, a significant AGN contamination in the estimates of both the rest-frame optical/UV luminosity and in the size of the emitting region could explain such outlier. We also caution that, in some of these galaxies, optical and CO radii might differ.

\subsection{Depletion times}

\begin{figure}
\includegraphics[width=0.99\columnwidth]{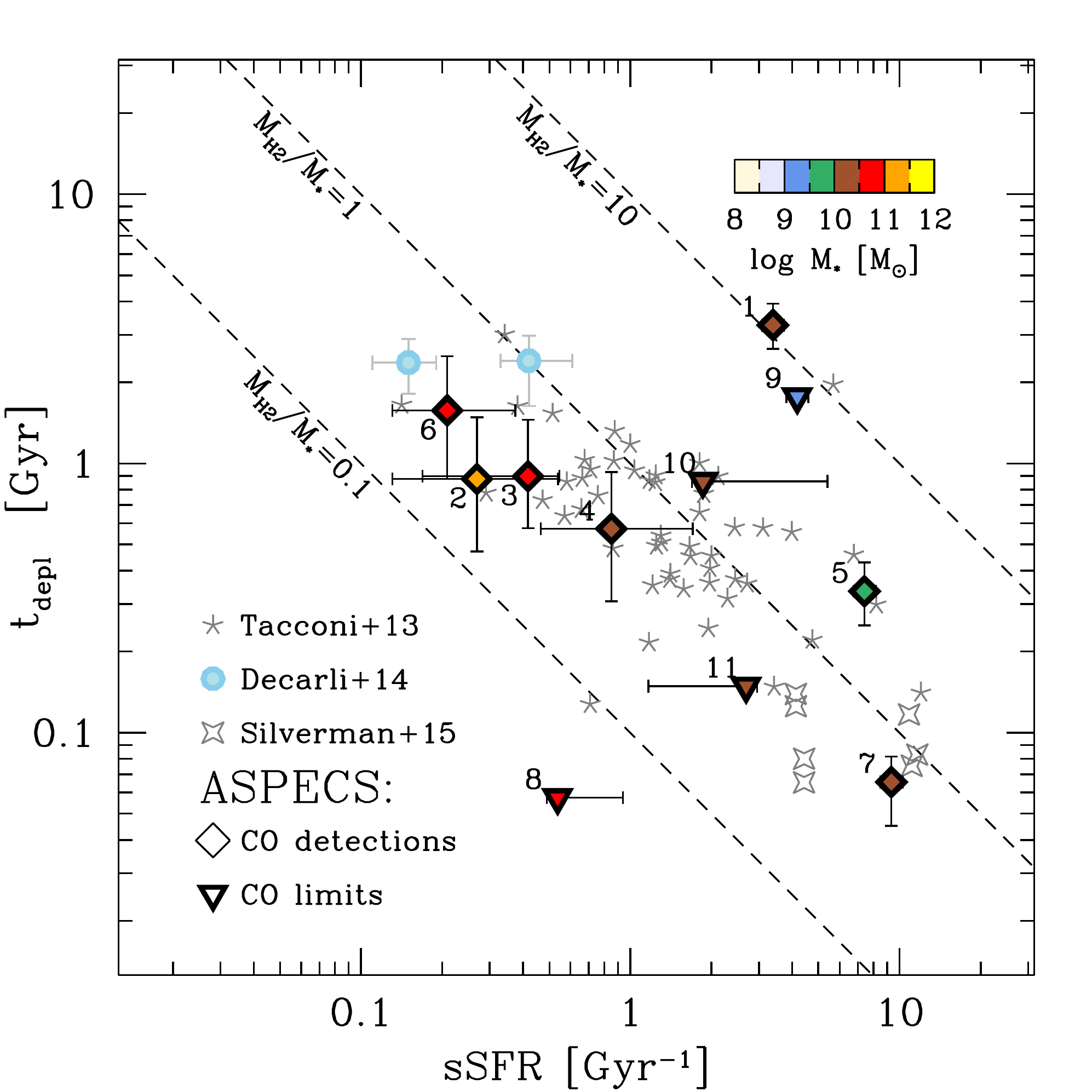}
\caption{The depletion time $t_{\rm depl}=M_{\rm H2}/{\rm SFR}$ as a function of the specific star formation rate sSFR=SFR/$M_{\rm *}$ for the galaxies in our sample, the secure blind detections in \citet{decarli14}, the PHIBSS sample by \citet{tacconi13}, and the starburst sample in \citet{silverman15}. The symbol code is the same as in Fig.~\ref{fig_co_lum}. Starburst galaxies typically reside in the bottom-right corner of the plot. Our ASPECS sources cover a wide range in parameter space, highlighting the diverse properties of these galaxies. }\label{fig_tdepl_ssfr}
\end{figure}

Fig.~\ref{fig_tdepl_ssfr} shows the depletion time, $t_{\rm depl}=M_{\rm H2}/{\rm SFR}$, as a function of the specific star formation rate. This timescale sets how quickly the gas is depleted in a galaxy given the currently observed SFR (ignoring any gas repleneshing). Our data are compared again with the secure blind detections in \citet{decarli14}, with the PHIBSS sample, and with the sample of starburst galaxies studied by \citet{silverman15} (in the latter case, we do not change the adopted value of $\alpha_{\rm CO}=1.1$\,\Msun(\Kkmspc)$^{-1}$, as these are not main sequence galaxies). Starburst galaxies tend to reside in the bottom-right corner of the plot (they are highly star-forming given their stellar mass, and they are using up their gaseous reservoir fast). Galaxies with large gas reservoirs and mild star-formation populate the top-left corner of the plot. Since the IR luminosity is proportional to the SFR, and the CO luminosity is used to infer $M_{\rm H2}$, the y-axis of this plot conceptually corresponds to a diagonal line (top-left to bottom-right) in Fig.~\ref{fig_co_lum}. Also, diagonal lines in Fig.~\ref{fig_tdepl_ssfr} mark the loci of constant molecular--to--stellar mass ratio $M_{\rm H2}/M_*$. 

The sources in our sample range over almost 2 dex in sSFR and $t_{\rm depl}$. Noticeably, ID.1 is highly star-forming (it resides slightly above the main sequence of star forming galaxies at $z\sim 2.5$, see Fig.~\ref{fig_ms}), so we would expect it to reside in the bottom-right corner of Fig.~\ref{fig_tdepl_ssfr}; however, its gaseous reservoir is very large for its IR luminosity (see also Fig.~\ref{fig_co_lum}), thus placing ID.1 in the top-right corner of the plot ($M_{\rm H2}/M_*=12$). On the other hand, ID.2 hosts an enormous reservoir of molecular gas, but because of its even larger stellar mass (yielding low sSFR), it resides on the left side of the plot ($M_{\rm H2}/M_*=0.24$). Their depletion time scales however are comparable (1--3 Gyr). We stress that these results are based on very high-S/N CO line detections, and on very solid descriptions of the galaxy SEDs (see Fig.~\ref{fig_id1_a}). The sources that populate the starburst region in Fig.~\ref{fig_ms} and reside in the top-left part of Fig.~\ref{fig_co_lum} (in particular, ID.5 and ID.7) consistently appear in the bottom-right corner of Fig.~\ref{fig_tdepl_ssfr}, among starbursts.



\subsection{Gas to stellar mass ratios}

\begin{figure}
\includegraphics[width=0.99\columnwidth]{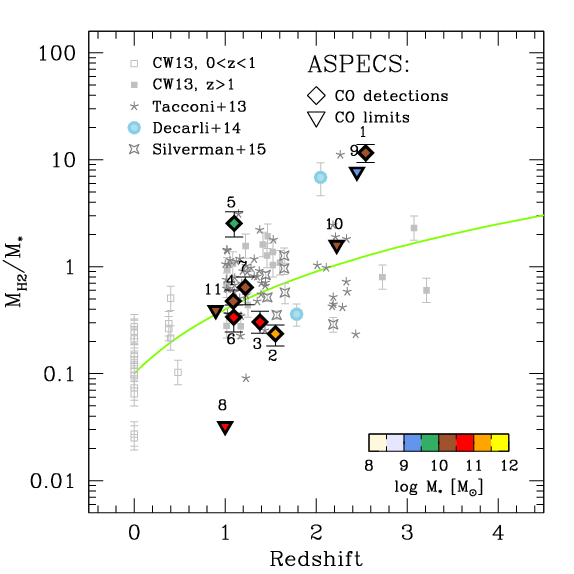}
\caption{Gas mass fraction (defined as $M_{\rm H2}/M_*$) as a function of redshift from various samples of galaxies in the literature (grey, from the compilation in \citealt{carilli13}), compared with the secure CO detections in \citet{decarli14}, the PHIBSS sample \citep{tacconi13}, the starburst sample in \citet{silverman15}, and our results from this work. The symbol coding is the same as in Fig.~\ref{fig_ms}. Our data seem to support the picture of a generally increasing $M_{\rm H2}/M_*$ ratio in main sequence galaxies as a function of redshift, as highlighted by the $f_{\rm gas}=0.1 \times (1+z)^2$ green line \citep{geach11,magdis12}. In particular, ID.2 appears as a starburst with respect to its position above the `main sequence' in Fig.~\ref{fig_ms}, and shows a large $M_{\rm H2}/M_*$ ratio. On the other hand, we also point out that significant upper limits are present (triangles).}\label{fig_fgas_z}
\end{figure}

A useful parameter to investigate the molecular gas content in high-$z$ galaxies is the molecular gas to stellar mass ratio, $M_{\rm H2}/M_{\rm *}$. We prefer this parameter rather than the molecular gas fraction, $f_{\rm gas}=M_{\rm H2}/(M_*+M_{\rm H2})$, as the two involved quantities ($M_{\rm H2}$ and $M_*$) appear independently at the numerator and denominator of the fraction, so that the parameter is well defined even if we only have upper limits on $M_{\rm H2}$. Fig.~\ref{fig_fgas_z} shows the dependence of $M_{\rm H2}/M_*$ on redshift in the galaxies of our samples, and in galaxies from the literature. This plot informs us on the typical gas content as a function of cosmic time, and can help us shed light on the origin of the cosmic star-formation history (see, e.g., \citealt{geach11}, \citealt{magdis12}, and Paper III of this series). Color-selected star-forming galaxies close to the epoch of galaxy assembly are claimed to show large $M_{\rm H2}/M_*$, with reservoirs of gas as big as (or even larger than) the stellar mass (i.e., $M_{\rm H2}/M_*\sim1$; see, e.g., \citealt{daddi10a}, \citealt{tacconi10,tacconi13}). Indeed, we find examples of very high gas fractions:  ID.1 ($M_{\rm H2}/M_*=12$) and the starburst galaxy ID.5 ($M_{\rm H2}/M_*=2.5$) are the most extreme cases. However, it is interesting to note that we also find galaxies with very modest gas fractions, such as ID.2 ($M_{\rm H2}/M_*=0.24$). The CO-detected galaxies at $1.0<z<1.7$ in our sample show an average $M_{\rm H2}/M_*$ ratio that is $\sim 2\times$ lower than the average value for the PHIBSS sample at the same redshift, and closer to the global trend established in \citet{geach11} and \citet{magdis12}. The non-detection of CO in ID.8 places particularly strict limits ($M_{\rm H2}/M_*<0.03$). If the lack of detection is attributed to the very low CO excitation in this galaxy, it would take a 10$\times$ lower $r_{41}$ (i.e., $r_{41}\approx0.03$) to shift ID.8 on the average trend reported by \citet{geach11}.

\subsection{CO vs. Dust-based ISM masses}\label{sec_co_vs_dust}

In addition to the CO line measurements, six of the 11 galaxies in our sample also have detections in the 1\,mm dust continuum. We can thus estimate the mass of the molecular gas independently of the CO data. The Rayleigh-Jeans part of the dust emission is only weakly dependent on the dust temperature, thus it can be used to trace the mass of dust. Using the dust-to-gas scaling \citep[see, e.g.,][]{sandstrom13}, it is possible to infer the gas mass via the dust mass.

\citet{groves15} compare CO-based gas masses with the monochromatic luminosity of the dust continuum in the Rayleigh-Jeans tail. Their analysis relies on a detailed study of 37 local spiral galaxies in the KINGFISH sample \citep{kennicutt11}. The galaxy luminosity in the {\em Herschel}/SPIRE 500$\mu$m band is found to scale almost linearly with the gas mass, yielding:
\begin{equation}\label{eq_groves}
\frac{M_{\rm gas}}{10^{10}\,{\rm M_\odot}}=28.5 \, \frac{\nu L_\nu (500\mu{\rm m})}{10^{10}\,\rm L_\odot}
\end{equation}
We compute the rest-frame luminosity $\nu L_\nu$(500$\mu$m) from the observed 1mm continuum of the galaxies in our sample. For the $k$-correction, we adopt a modified black body with $T_{\rm dust}$=25\,K and $\beta$=$1.6$ \citep[see, e.g.,][]{beelen06}, shifted at the redshift of each source. Since the observing frequency (242\,GHz) falls close to the rest-frame 500$\mu$m (as most of our sources reside at $z\sim1.2$), and we are sampling the Rayleigh-Jeans tail (which is almost insensitive to the dust temperature), the differences in the corrections due the adopted templates are negligible for the purposes of this analysis. The adopted values for the $k$ correction, as well as the resulting gas masses, are listed in Tab.~\ref{tab_Mism}.

A similar approach was presented by \citet{scoville14,scoville15}. This calibration is tuned on a set of relatively massive [$(0.2-4)\times10^{11}$\,\Msun] star-forming galaxies (30 local star-forming galaxies, 12 low-redshift ULIRGs, and 30 SMGs at $z$=1.4--3.0), all having literature observations of the CO(1-0) transition. The tight relation observed between CO(1-0) luminosity and the rest-frame 850$\mu$m monochromatic luminosity \citep[see Fig.~1 in][]{scoville15} suggests that a simple conversion factor can be used to derive gas masses from monochromatic dust continuum observations. Setting the dust temperature to $T_{\rm dust}=25$\,K \citep[following][]{scoville14}, from eq.~12 in their paper we derive $M_{\rm ISM}$ from our 1\,mm flux densities as follows:
\begin{equation}\label{eq_scoville}
\frac{M_{\rm ISM}}{10^{10}\,{\rm M_\odot}}=\frac{1.20}{(1+z)^{4.8}}\,\,\frac{F_\nu}{\rm mJy} \, \left(\frac{\nu}{\rm 350\,GHz}\right)^{-3.8} \, \frac{\Gamma_0}{\Gamma_{\rm RJ}} \, \left(\frac{D_{\rm L}}{\rm Gpc}\right)^2
\end{equation}
where $F_\nu$ is the observed dust continuum flux density at the observing frequency $\nu$ ($242$\,GHz in our case), $D_{\rm L}$ is the luminosity distance, and $\Gamma_{\rm RJ}$ is a unitless correction factor that accounts for the deviation from the $\nu^2$ scaling of the Rayleigh-Jeans tail. In the reference sample of local galaxies, low-redshift ULIRGs and high-$z$ SMGs that \citet{scoville14} used to calibrate eq.~\ref{eq_scoville}, $\Gamma_{\rm RJ}=\Gamma_0=0.71$. The resulting ISM masses are listed in Tab.~\ref{tab_Mism}.

\begin{table*}
\caption{\rm Gas mass estimates based on the dust continuum. Only sources detected at 1mm in ASPECS are considered. (1) Source ID. (2) Redshift. (3) Observed 242\,GHz = 1.2\,mm continuum flux density (see Paper II). (4) $k$ correction, expressed as the ratio between the flux density computed at $\lambda_{\rm rest\,frame}=500$\,$\mu$m and the one at $\lambda_{\rm obs}=1.2$\,mm, assuming a modified black body template for the dust emission with $\beta$=1.6 and $T_{\rm dust}=25$\,K. (5) Gas mass based on the 1mm flux density, derived following eq.~\ref{eq_groves} \citep{groves15}. (6) Gas mass based on the 1mm flux density, derived following eq.~\ref{eq_scoville} \citep{scoville14,scoville15}. (7) Gas mass derived from the dust mass estimate resulting from MAGPHYS SED fitting, assuming a dust-to-gas ratio DGR=1/100.} \label{tab_Mism}
\begin{center}
\begin{tabular}{ccccccc}
\hline
ID   & $z$ & $F_{\nu}$(1.2mm) & $k$-corr & log\,$M_{\rm gas,\,Groves}$ & log\,$M_{\rm ISM,\,Scoville}$  & log\,$M_{\rm gas,\,MAGPHYS}$ \\
     &     & [$\mu$Jy]           &                              & [\Msun]		      & [\Msun] 		       &  [\Msun]		   \\
 (1) & (2) &  (3)                & (4)                          & (5)                     & (6)                         & (7)                              \\ 
\hline
  1 & 2.543 & $552.7\pm13.8$ & 0.374 & $11.02_{-0.011}^{+0.011}$ & $10.69_{-0.011}^{+0.011}$ & $10.53_{-0.17}^{+0.17}$   \\  
  2 & 1.551 & $223.1\pm21.6$ & 0.919 &  $10.63_{-0.04}^{+0.04}$  &  $10.33_{-0.04}^{+0.04}$  & $10.09_{-0.14}^{+0.13}$   \\  
  4 & 1.088 & $ 96.5\pm24.7$ & 1.665 &  $10.24_{-0.10}^{+0.13}$  &   $9.95_{-0.13}^{+0.10}$  & $ 9.78_{-0.20}^{+0.18}$   \\  
  5 & 1.098 & $ 46.4\pm14.9$ & 1.641 &  $ 9.93_{-0.12}^{+0.17}$  &   $9.63_{-0.17}^{+0.12}$  & $ 9.25_{-0.19}^{+0.14}$   \\  
  6 & 1.094 & $ 69.6\pm18.9$ & 1.650 &  $10.10_{-0.10}^{+0.14}$  &   $9.81_{-0.14}^{+0.10}$  & $ 9.58_{-0.18}^{+0.17}$   \\  
 10 & 2.224 & $ 36.7\pm13.8$ & 0.478 &  $ 9.86_{-0.14}^{+0.20}$  &   $9.53_{-0.20}^{+0.14}$  & $ 9.25_{-0.20}^{+0.17}$   \\  
\hline
\end{tabular}
\end{center}
\end{table*}

Finally, we can infer an estimate of $M_{\rm gas}$ from the estimate of the dust mass, $M_{\rm dust}$, that we obtain via our MAGPHYS fit of the available SED, simply scaled by a fixed dust-to-gas mass ratio (DGR). \citet{sandstrom13} investigate the dust and gas content in a sample of local spiral galaxies, and find DGR$\approx$1/70. \citet{genzel15} and \citet{berta16} perform a detailed analysis of both gas and dust mass estimates in galaxies at $0.9<z<3.2$ observed with {\em Herschel}, and find a lower value of DGR$\approx$1/100, which is the value we adopt here. We stress that there is a factor $>2\times$ scatter in the estimates of DGR due to its dependence on $M_*$ and metallicity \citep{sandstrom13,berta16}. Following the fundamental metallicity relation in \citet{mannucci10}, we estimate that galaxies in our sample typically have solar metallicities (the lowest metallicity estimates are for ID.9: $Z$=$0.6$\,Z$_\odot$; and ID.5: $Z$=$0.7$\,Z$_\odot$), so we do not foresee large intra-sample variations of DGR. For simplicity, in our analysis we thus assume a fixed DGR=1/100. While SED fits are available for all the galaxies in our sample, we consider here only those with a 1mm detection, in order to best anchor the Rayleigh-Jeans tail of the dust emission. The resulting masses are listed in Tab.~\ref{tab_Mism}.

\begin{figure}
\includegraphics[width=0.99\columnwidth]{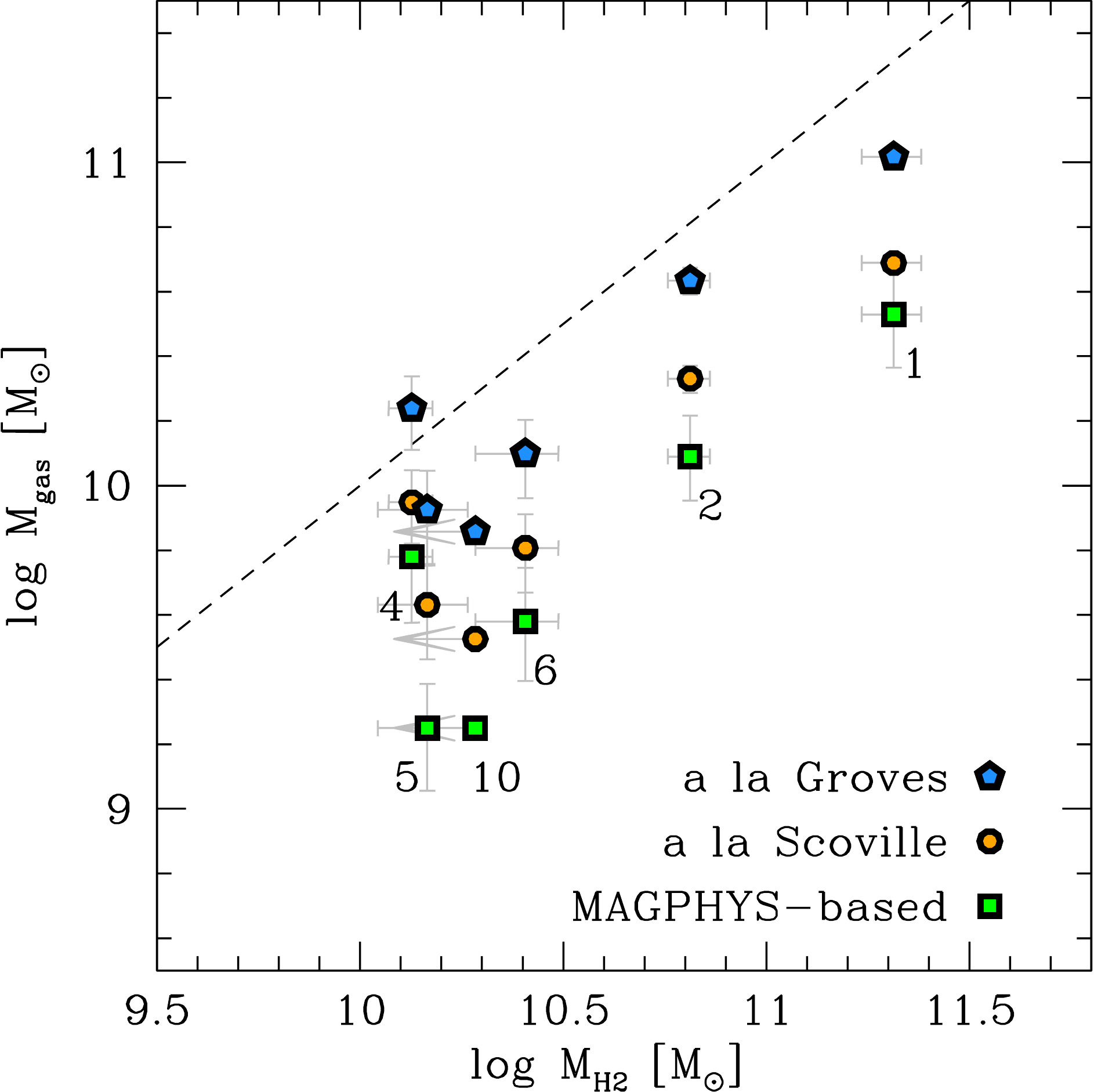}
\caption{Comparison between the H$_2$ masses that we derive from CO for the sources in our sample (x--axis), and the gas masses inferred from the 1mm continuum, following eq.~\ref{eq_groves} \citep{groves15}, eq.~\ref{eq_scoville} \citep{scoville14,scoville15}, and based on the MAGPHYS-based estimates of $M_{\rm dust}$, assuming a dust-to-gas ratio of 1/100 \citep{genzel15} (y--axis). The dashed line shows the 1-to-1 case. Only sources with a 1mm continuum detection are shown. The dust-based estimates are correlated with each other, due to the strong dependence on the 1mm continuum emission. The various mass estimates are also correlated with the CO-based ones over 1.5 dex. There are however systematic offsets among the various gas mass recipes, with dust-based masses that appear lower than the ones inferred from CO.}\label{fig_mass_comparison}
\end{figure}

Fig.~\ref{fig_mass_comparison} compares the gas estimates based on Eq.~\ref{eq_groves}, following \citet{groves15}; the ones obtained via Eq.~\ref{eq_scoville}, following \citet{scoville14}; and the estimates based on dust from the MAGPHYS SED fits, with our CO-based estimate (assuming $\alpha_{\rm CO}=3.6$\,\Msun{}[\Kkmspc]$^{-1}$). The dust--based gas estimates obtained with different approaches are strongly correlated with each other, as expected because they all scale (almost linearly) with $F_\nu$(1mm). They also correlate well with the CO-based H$_2$ mass estimates over one and a half dex of dynamic range. However, systematic offsets are observed. The \citet{groves15} estimates are on average $1.5\times$ lower than those based on CO. The estimates based on \citet{scoville15} are another $2\times$ lower, and the masses based on MAGPHYS are a factor $1.7\times$ lower than those obtained following \citet{scoville15}.


What causes the discrepancies between these mass estimates? The CO masses might be overestimated because of our assumptions in terms of CO excitation and $\alpha_{\rm CO}$. A higher CO excitation would imply higher $r_{J1}$, thus lower CO(1-0) luminosity (see eq.~\ref{eq_MH2}). If we assume the M82 excitation template by \citet{weiss07} (see Fig.~\ref{fig_co_excit}), the inferred $M_{\rm H2}$ masses would be $1.3\times$ lower for ID.2--6, and $2.2\times$ lower for ID.1 and ID.10. This would solve the discrepancy with respect to the estimates based on the Groves recipe, and it would mitigate, but not solve, the discrepancy with the other gas mass estimates. However, such a high CO excitation scenario is ruled out by our 1mm line observations (see Fig.~\ref{fig_co_excit}). A lower value of $\alpha_{\rm CO}$ could also help. If we adopt the classical value for ULIRGs, $\alpha_{\rm CO}=0.8$\,\Msun{}(\Kkmspc)$^{-1}$ \citep{bolatto13}, the CO-based gas masses would be a factor 4.5 smaller, thus in good agreement with the ones from the dust. Fig.~\ref{fig_co_lum} shows that the majority of our sources lie along the relation of main sequence galaxies / local spiral galaxies in the $L_{\rm IR}$--$L'_{\rm CO}$ plot. This is irrespective of the choice of $\alpha_{\rm CO}$. Thorough studies of galaxies along this sequence support our choice for a larger value of $\alpha_{\rm CO}$ \citep[e.g.][]{daddi10a,genzel10,genzel15,sargent14}. Further support to our choice comes from the position of our sources along the `main sequence' of galaxies (Fig.~\ref{fig_ms}). Among the sources listed in Tab.~\ref{tab_Mism} and appearing in Fig.~\ref{fig_mass_comparison}, only ID.5 could be considered a starburst in this respect. Adopting a lower $\alpha_{\rm CO}$ for only this source would lower its molecular gas mass by a factor $\sim 4.5$, thus bringing it close to the bulk of the `main sequence' galaxies in terms of gas fraction (Fig.~\ref{fig_fgas_z}), but pushing it away from the sequence in the star formation law plot (Fig.~\ref{fig_ks}). It would also reduce its depletion time scale (Fig.~\ref{fig_tdepl_ssfr}) and bring the CO-based gas mass closer to the dust-based estimates (Fig.~\ref{fig_mass_comparison}). Similar considerations could also apply for ID.1, the CO--brightest galaxy in our sample. The compact morphology and the small separation from a companion galaxy, the rising CO emission at high J, the high values of $\Sigma_{\rm SFR}$ and $\Sigma_{\rm H2}$ and the very large $M_{\rm H2}$/$M_*$ all point toward a starburst scenario for this source; however, it is located along the main sequence of galaxies at $z\sim2.5$ in Fig.~\ref{fig_ms}, and the $L_{\rm IR}$--$L'_{\rm CO(1-0)}$ plot (Fig.~\ref{fig_co_lum}) shows that this source is located along the sequence of local spirals and main sequence galaxies (not along the sequence of starbursts), irrespective of the choice of $\alpha_{\rm CO}$, even if we assume the extreme case of thermalized CO(3-2) emission in order to derive $L'_{\rm CO(1-0)}$. Because of this, and because of the lack of any starburst signature (justifying a low $\alpha_{\rm CO}$) in all the other sources, the discrepancies between different gas mass estimates shown in Fig.~\ref{fig_mass_comparison} cannot be mended only by tuning our assumptions on the CO-based mass estimates.

The dust-based gas mass estimates could also be affected by systematic uncertainties. The offset between the estimates based on Eq.~\ref{eq_groves} and Eq.~\ref{eq_scoville} suggests a systematic in the calibration of the two recipes. E.g., the luminosity range used in \citet{groves15} to derive Eq.~\ref{eq_groves} does not cover the $>10^{10}$\,\Lsun{} range, where our galaxies are found. Eq.~\ref{eq_scoville}, based on \citet{scoville14}, is pinned down to a longer wavelength than what observed in ASPECS (850\,$\mu$m in the rest-frame, instead of $\sim 500$\,$\mu$m), thus the $k$ correction is significant and dependent on the adopted dust template. In particular, Eq.~\ref{eq_scoville} explicitly assumes $\beta$=$1.8$, which might not be universally valid (see discussion in Paper II). Our dust SED is only poorly sampled. Most remarkably, the comparison between maps of the CO and dust emission in ID.2 suggests that the gas is optically-thick over a large area, while the dust is not. We might be missing part of the dust emission due to surface brightness limits, thus affecting our estimates of the total ISM mass. In ID.3, the dust continuum emission is not detected at all, despite the bright CO emission. Since we do not detect any significant 1mm continuum associated with the extended disk of ID.2, and no dust emission at all in ID.3, it is hard to assess how big a correction one should consider. It is possible that a similar issue is present in other sources, in particular in galaxies that we see as CO emitters but for which we do not recover any 1mm continuum emission (see, e.g., Fig.~\ref{fig_co_lum}). Finally, the underlying assumption in the dust--based gas estimates is the dust-to-gas ratio. This can change significantly as a function of metallicity and other parameters in the galaxy \citep[see][for a detailed discussion]{sandstrom13}. A lower value of DGR (e.g., DGR$\sim$1/200) would halve the discrepancy between the MAGPHYS-based estimates and the CO-based ones. While this is a possibility at the low-mass end of Fig.~\ref{fig_mass_comparison}, we point out that the relatively large stellar mass of the galaxies at the bright end support metallicity values close to solar, thus disfavoring the large DGR values needed to reconcile the two gas mass estimates.

\section{Summary}\label{sec_conclusions}

We present a study of the molecular gas properties as derived from CO observations of high--redshift galaxies in ASPECS, the ALMA Spectroscopic Survey in the {\em Hubble} Ultra Deep Field (UDF). This dataset consists of a blind survey of molecular gas in the ALMA 3mm and 1mm bands targeting a region with the deepest {\em HST} imaging available, the so--called XDF. Our observations cover hundreds of high--redshift galaxies with well--characterized SEDs, i.e. we can test our expectations in terms of molecular gas content in galaxies without any prior selection through their optical or near-IR properties. This allows us to analyse our CO measurements and limits in the context of the global properties of the associated galaxies, thanks to an exquisite wealth of ancillary multi-wavelength information.

We focus on the galaxies for which a secure redshift is available, either via our ASPECS CO observations or from optical/near-IR spectroscopy reported in the literature. In particular, we consider those sources for which our sophisticated fit of the optical--to--mid-IR SED implies high IR luminosity ($L_{\rm IR}>10^{11}$\,\Lsun{}). These galaxies are expected to be detected in CO based on the empirical relation between CO and dust luminosity. We also restrict our analysis to those galaxies with a redshift such that a CO transition with J$_{\rm up}<5$ is covered in our ASPECS observations.

{\em Success of CO detection ---} Out of 11 sources selected in this way, 4 are also identified in our CO blind search (see Paper I of this series). Three additional galaxies are detected in CO, although with lower significance. The faintest galaxy detected in CO (at $\sim 3$-$\sigma$ level) harbors an AGN. This likely leads to an overestimate of the IR luminosity in our analysis (if the AGN component contributes significantly to the rest-frame optical/UV emission). Finally, four sources remain undetected in CO. In two of them, the lack of CO detection might be attributed to CO excitation, as the lowest J transition that we targeted in these sources is the CO(4-3) line. This however would point toward a very low-excitation scenario for these two sources. The other two undetected galaxies are just above our IR luminosity cut. They reside at relatively high redshift ($z=2.0-2.5$). In these cases, the lack of a detection might be attributed to insufficient depth of our ASPECS observations, and/or modest CO excitation. 

{\em CO excitation ---} As we cover CO emission in two separate ALMA bands, we constrain the CO excitation in all of our CO--detected galaxies.  In no case do we find evidence of high excitation, as observed in the center of M82 or in IR--luminous SMGs or QSOs at high redshift. The galaxy that has the highest excitation is a bright, compact galaxy, showing high star formation rate surface density (ID.1). We attribute the high CO excitation to either the increased radiation density (due to the locally intense star formation) or to the high density of the gas.  A second source (ID.5) shows CO excitation slightly higher than the average `main sequence' galaxy; this is consistent with this galaxy being a starburst, as suggested by other indicators (sSFR with respect to the `main sequence' at that redshift; IR--to--CO luminosity ratio; depletion time; etc). On the other hand, CO excitation is typically very low, often consistent or even lower than Milky Way excitation at least up to J=5. In one case, a $r_{52}$ value as low as $<0.025$ was measured (for a comparison, in the Milky Way we have $r_{52}=0.16$). An X-ray bright AGN with an extended gas reservoir (ID.2) also show modest CO excitation; in this case, any effect that the AGN may have in the center is diminished by the extended molecular gas emission in the disk. Interestingly, also the CO--brightest galaxy in our sample, ID.1, is detected in the X--rays. Its X--ray luminosity is modest however, and roughly consistent with the extrapolation of the SFR--$L_{\rm X}$ relation observed in local starbursts \citep{ranalli03}.

{\em Location with respect to the `main sequence' ---} We discuss our findings in context of previous molecular gas observations at high redshift (star formation law, gas depletion times, gas mass fractions), based on sophisticated SED modeling of their multi--wavelength properties using the high--redshift extension of the MAGPHYS code.  Half of the galaxies in our sample reside on the `main sequence' of star-forming galaxies at their redshift. Three galaxies are found in the starburst region (although in one of them the SFR might be over-estimated due to the contamination from an optically-bright AGN). Finally, two sources are found below the main sequence, suggesting that they are more quiescent systems.

{\em The `star formation law' ---} To first order, the CO--detected galaxies in the UDF cover the same parameter space as previous galaxy samples in the $L_{\rm IR}$--$L'_{\rm CO}$ diagram, although they preferentially reside towards the low-IR luminosity envelope of the relation, along the same sequence of local spiral galaxies and close to color-selected galaxies at $z>1$. Only two CO--detected sources lie on the opposite side, closer to the locus of high-$z$ starbursts. Two of the CO non--detections are found to be inconsistent with the $L_{\rm IR}$--$L'_{\rm CO}$ relation, suggesting that CO excitation in these sources must be low. Using {\em HST} imaging to derive the scale radii of the galaxies in our sample, we discuss their location in the `star formation law' diagram: on average, the sources agree with a depletion time of $\sim$1\,Gyr, as found in previous studies, but outliers (up to 1 dex) exist.

{\em Gas fractions ---}  With only two remarkable exceptions, the gas fractions observed in our study are slightly lower than what found in targeted observations of main sequence galaxies at similar redshift, but still significantly higher than what typically observed in the local universe.

{\em CO- vs. Dust-based estimates of gas mass ---} In a few cases, we have gas mass estimates derived from CO as well as the dust continuum, via different recipes involving the dust--to--gas ratio. The dust--based estimates are a factor of $\sim$2--5 smaller than those based on CO. This is consistent with recent reports in the literature that dust--based estimates of ISM masses of main sequence galaxies give significantly lower values than using the CO emission. All these methods use a number of assumptions: {\em CO:} extrapolation to a CO(1-0) flux from a higher--J transition + choice of CO--to--H$_2$ conversion factor, {\em dust:} assumption of temperature, dust SED template, optical depth, and dust--to--gas ratio. A larger sample of galaxies with well--defined dust SEDs is required to ultimately decide which gas--mass estimator is preferred.

In summary, accounting for detections as well as non--detections, we find large variations in the molecular gas properties of high--redshift galaxies. This might reflect the large variations in gas content of disk galaxies seen in semi-analytical models (see, e.g., \citealt{lagos11} and Fig.~9 in \citealt{popping14}). Perhaps not unexpectedly, global scaling relations cannot account for the large variations in gas content in individual high redshift galaxies. Our approach through blind frequency scans of well--characterized cosmological deep fields adds additional constraints to the studies of the molecular gas content in distant galaxies, and are thus complementary to dedicated studies of single galaxies that are pre--selected by their optical properties (e.g. SFR and stellar mass). Our study demonstrates that such studies are now feasible, even with early--cycle ALMA observations. Now that ALMA has reached completion, similar studies of larger fields will result in large, statistical samples, which are required to fully understand and beat down systematics and small number statistics. This will provide us with an entirely new approach to characterize the molecular gas content in distant galaxies.

\section*{Acknowledgments}

We thank the anonymous referee for her/his positive feedback and useful comments. RD thanks Laura Zschaechner for insightful discussions. FW, IRS, and RJI acknowledge support through ERC grants COSMIC--DAWN, DUSTYGAL, and COSMICISM, respectively. M.A. acknowledges partial support from FONDECYT through grant 1140099. DR acknowledges support from the National Science Foundation under grant number AST-1614213 to Cornell University. FEB and LI acknowledge Conicyt grants Basal-CATA PFB--06/2007 and Anilo ACT1417. FEB also acknowledge support from FONDECYT Regular 1141218 (FEB), and the Ministry of Economy, Development, and Tourism's Millennium Science Initiative through grant IC120009, awarded to The Millennium Institute of Astrophysics, MAS. IRS also acknowledges support from STFC (ST/L00075X/1) and a Royal Society / Wolfson Merit award. Support for RD and BM was provided by the DFG priority program 1573 `The physics of the interstellar medium'.  AK and FB acknowledge support by the Collaborative Research Council 956, sub-project A1, funded by the Deutsche Forschungsgemeinschaft (DFG). PI acknowledges Conict grants Basal-CATA PFB--06/2007 and Anilo ACT1417. RJA was supported by FONDECYT grant number 1151408. 
This paper makes use of the following ALMA data: \dataset[ADS/JAO.ALMA\# 2013.1.00146.S and 2013.1.00718.S.]{https://almascience.nrao.edu/aq}. ALMA is a partnership of ESO (representing its member states), NSF (USA) and NINS (Japan), together with NRC (Canada), NSC and ASIAA (Taiwan), and KASI (Republic of Korea), in cooperation with the Republic of Chile. The Joint ALMA Observatory is operated by ESO, AUI/NRAO and NAOJ. The 3mm-part of ALMA project had been supported by the German ARC.

\appendix

\begin{figure*}\figurenum{A.1}
\begin{center}
\includegraphics[width=0.89\textwidth]{brightCO_i1_ps.png}\\
\includegraphics[width=0.49\columnwidth]{fig_spc_co_bright_1.pdf}
\includegraphics[width=0.49\columnwidth]{fig_sed_co_bright_1.pdf}\\
\includegraphics[width=0.89\textwidth]{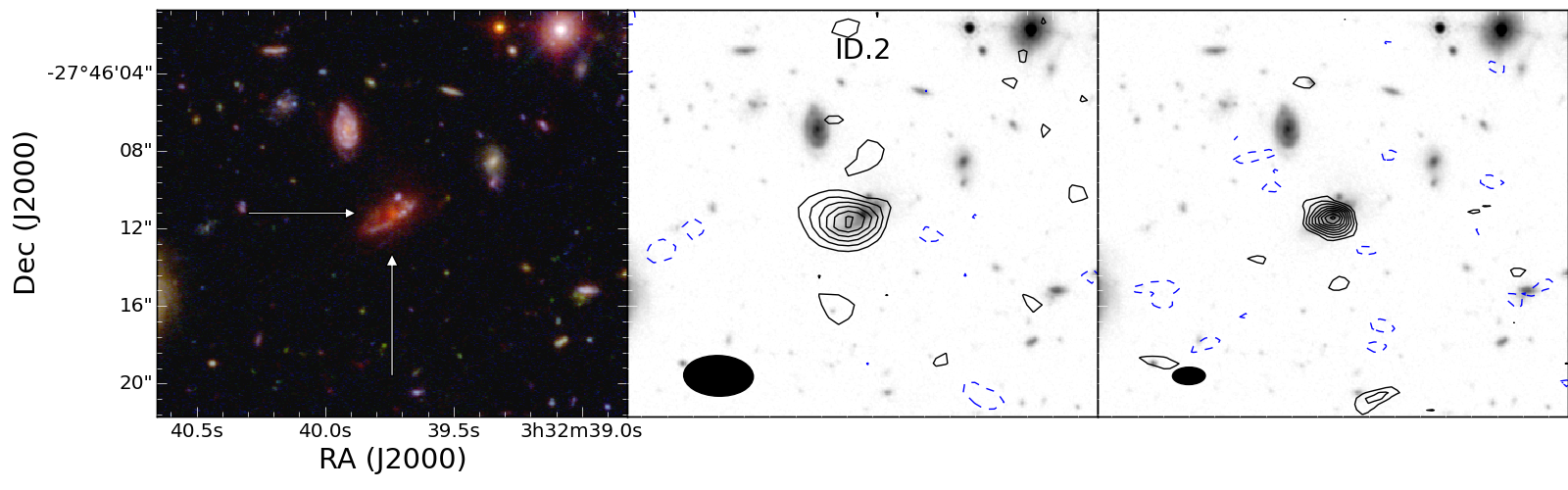}\\
\includegraphics[width=0.49\columnwidth]{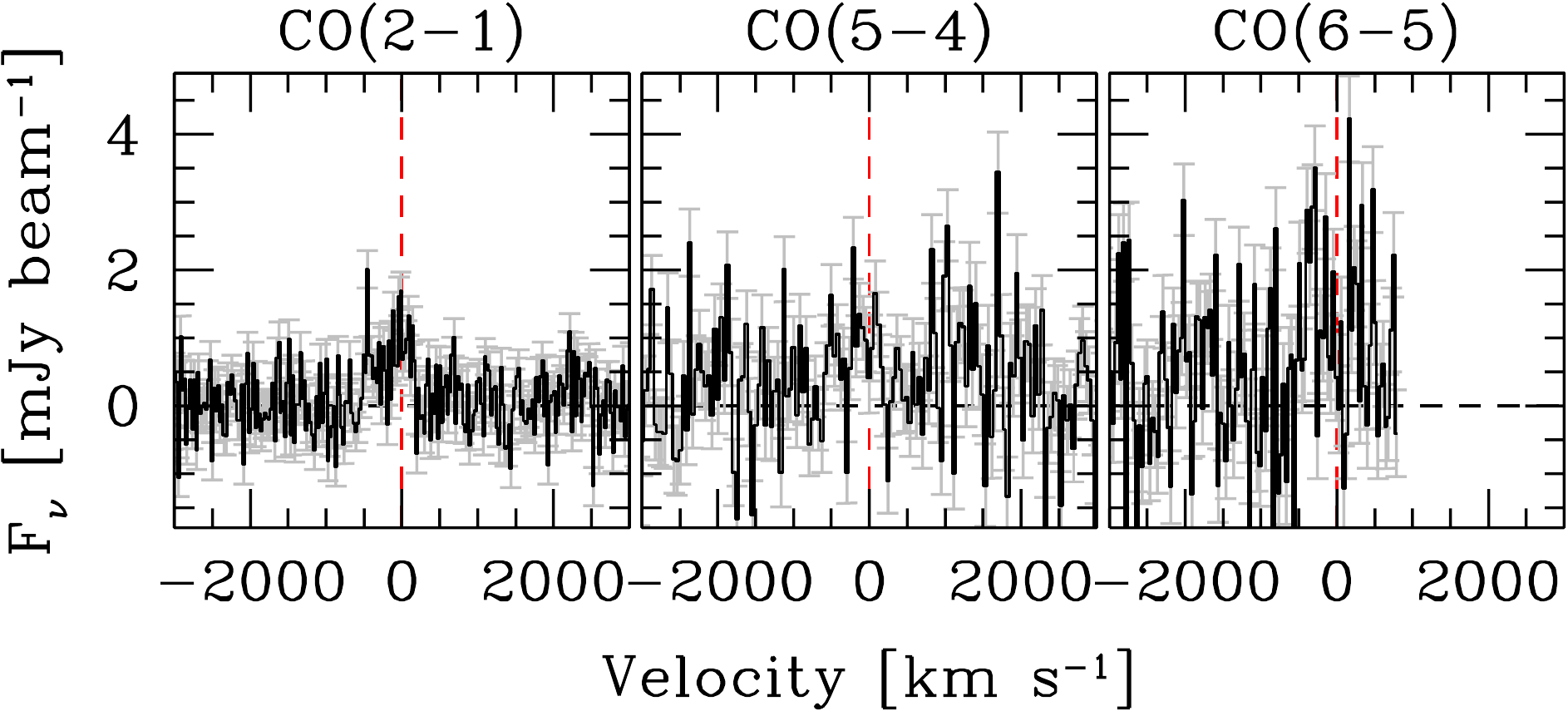}
\includegraphics[width=0.49\columnwidth]{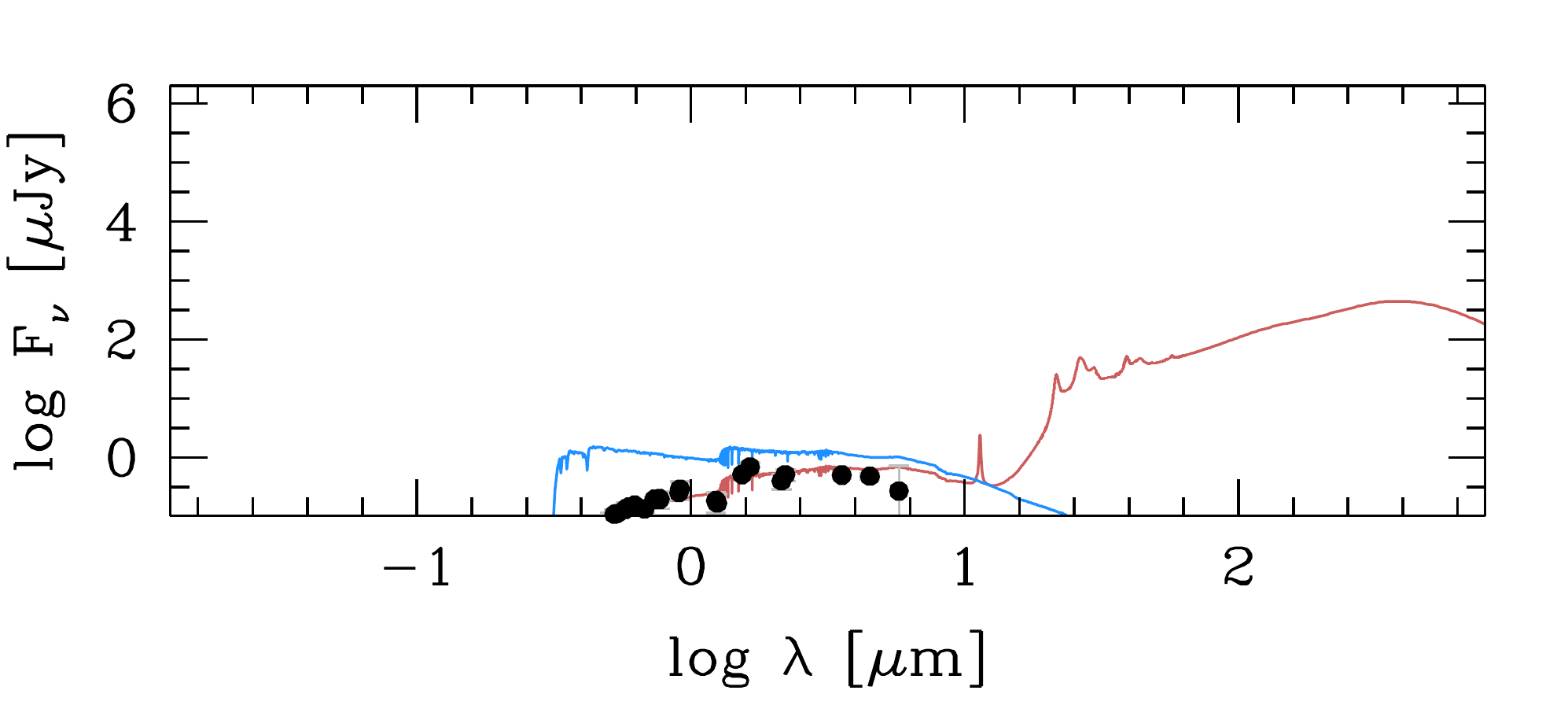}\\
\end{center}
\caption{{\em Top left:} {\em HST} F105W/F775W/F435W RGB image of ID.1 and 2. The postage stamp is $20''\times20''$. {\em Top center:} {\em HST} F125W image of the same field. The map of the lowest-J accessible CO transition (in this case, CO[3-2]) is shown as contours ($\pm 2,3,$\ldots,$20$-$\sigma$ [$\sigma$(ID.1)=0.78\,mJy\,beam$^{-1}$; $\sigma$(ID.2)=1.36\,mJy\,beam$^{-1}$]; solid black lines for the positive isophotes, dashed blue lines for the negative). The synthesized beam is shown as a black ellipse. {\em Top right:} Same as in the center, showing the 1.2mm dust continuum. {\em Bottom left:} Spectra of the CO lines encompassed in our spectral scan. {\em Bottom right:} Spectral Energy Distribution. The red line shows the best MAGPHYS fit of the available photometry (black points), while the blue line shows the corresponding model for the unobscured stellar component. The main output parameters are quoted. 
}\label{fig_id1_a}
\end{figure*}

\begin{figure*}\figurenum{A.2}\begin{center}
\includegraphics[width=0.89\textwidth]{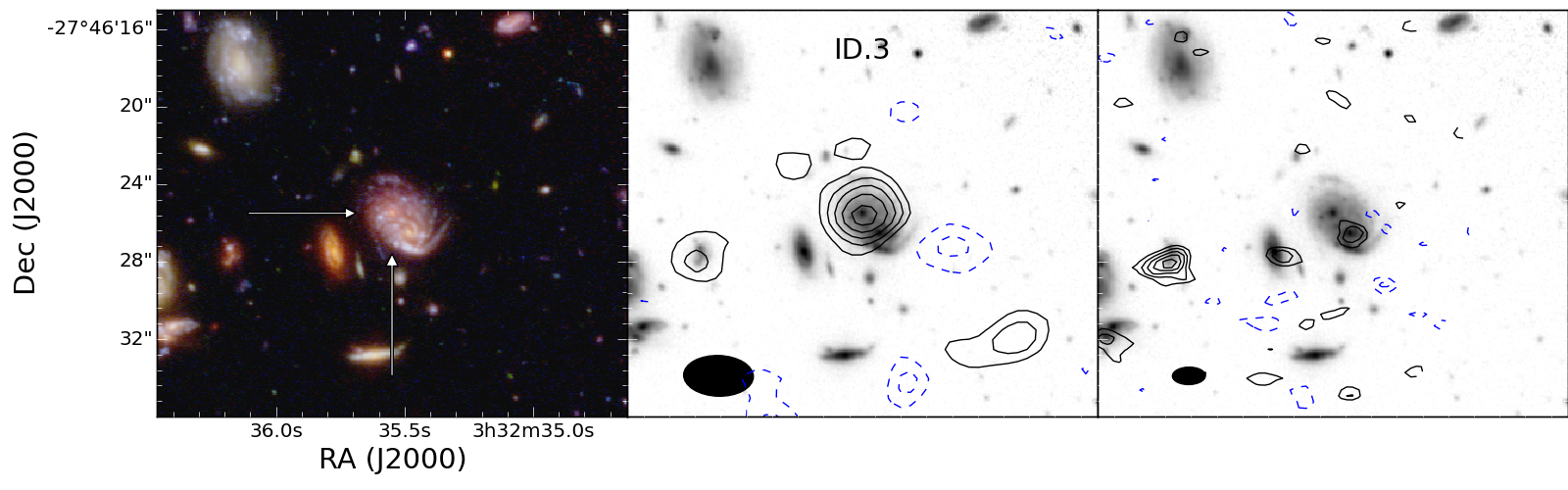}\\
\includegraphics[width=0.49\columnwidth]{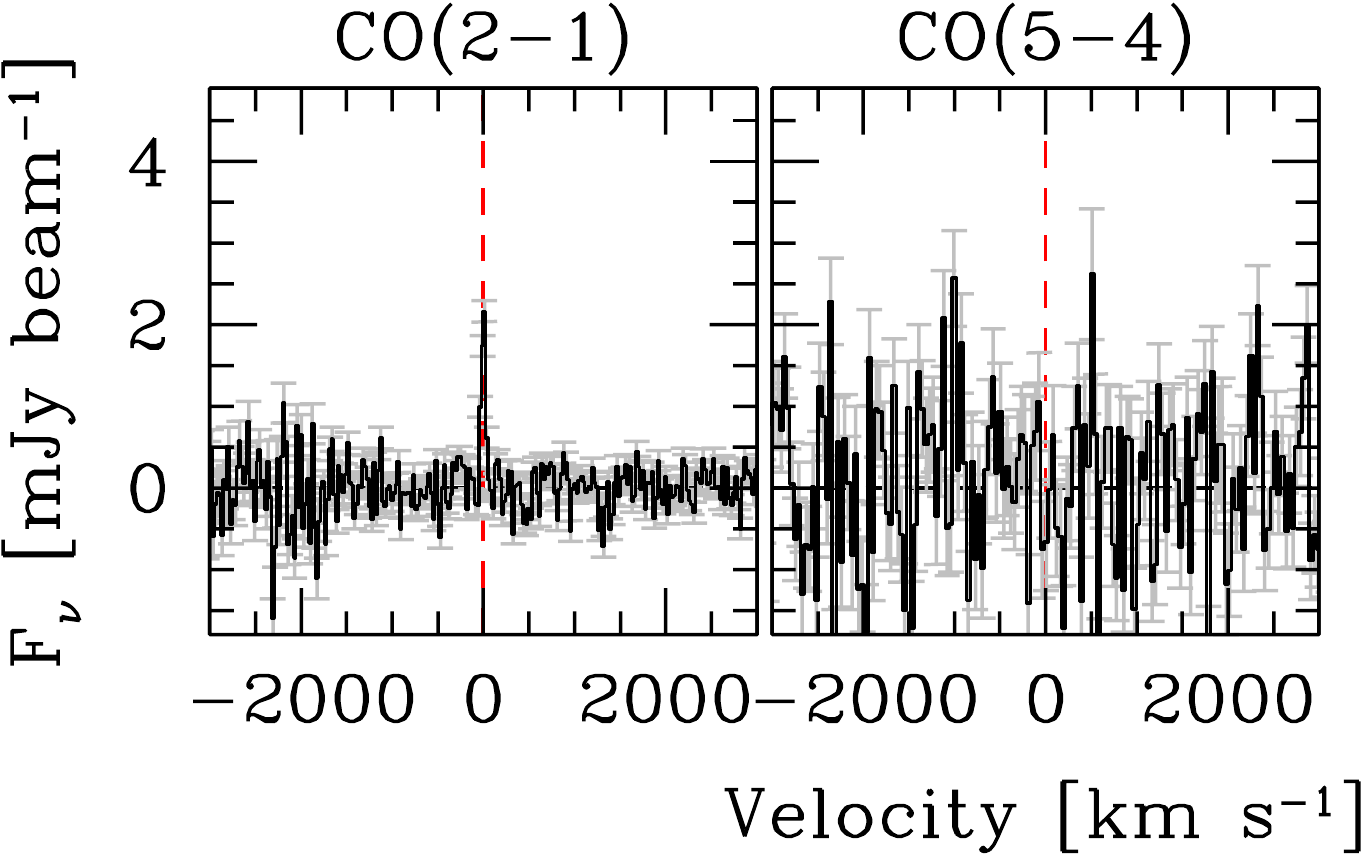}
\includegraphics[width=0.49\columnwidth]{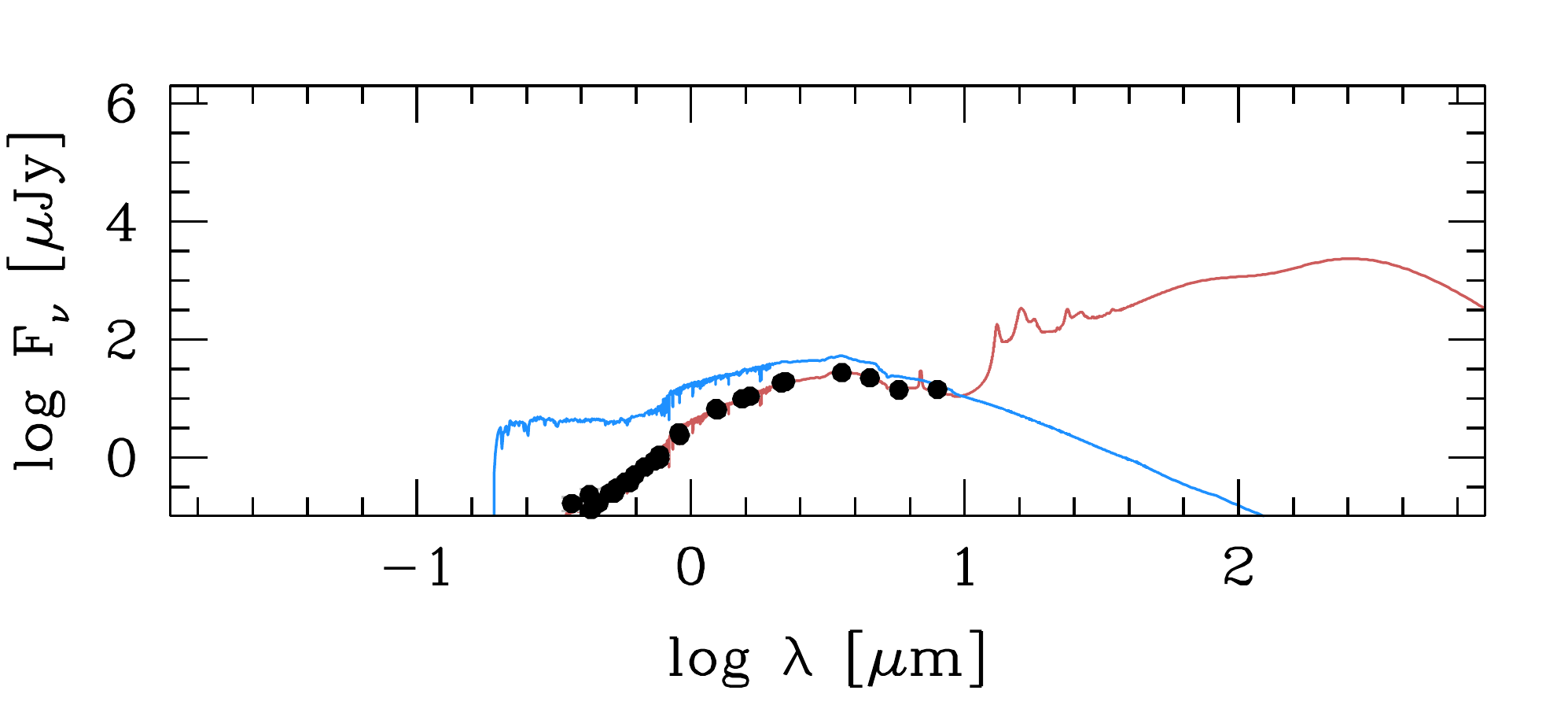}\\
\includegraphics[width=0.89\textwidth]{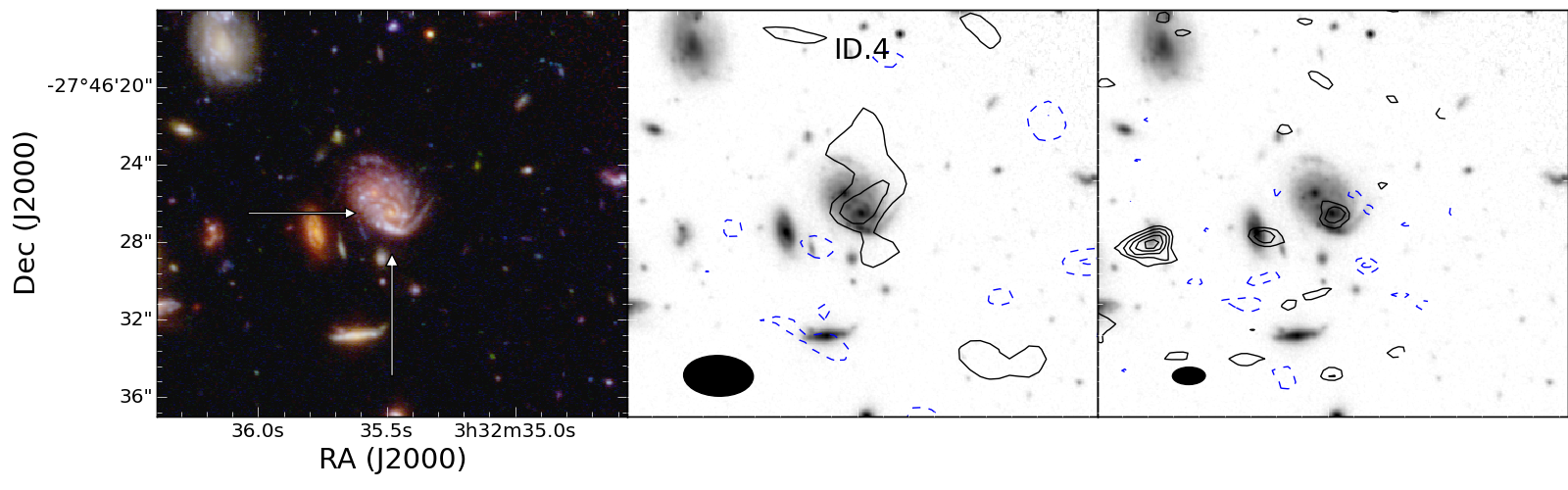}\\
\includegraphics[width=0.49\columnwidth]{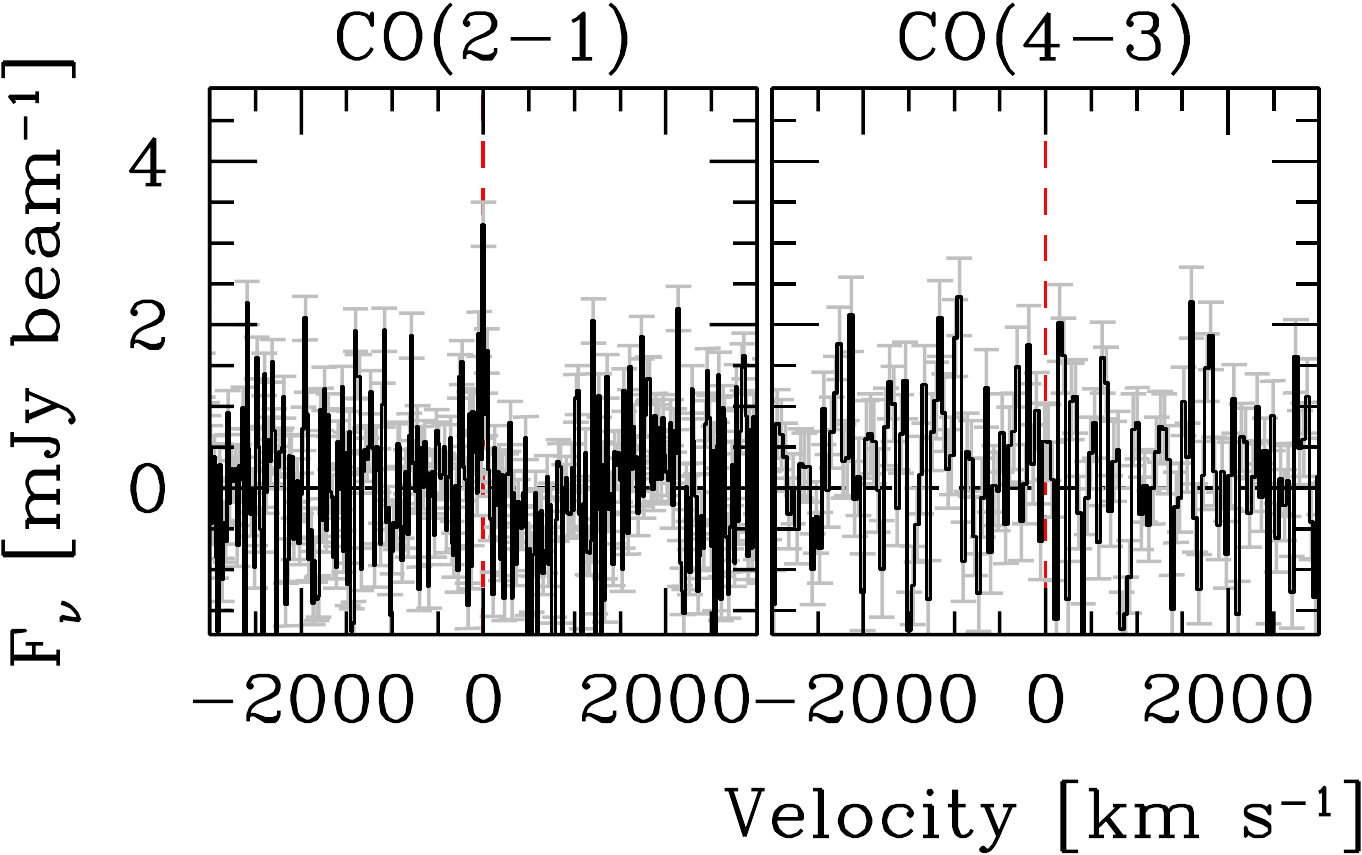}
\includegraphics[width=0.49\columnwidth]{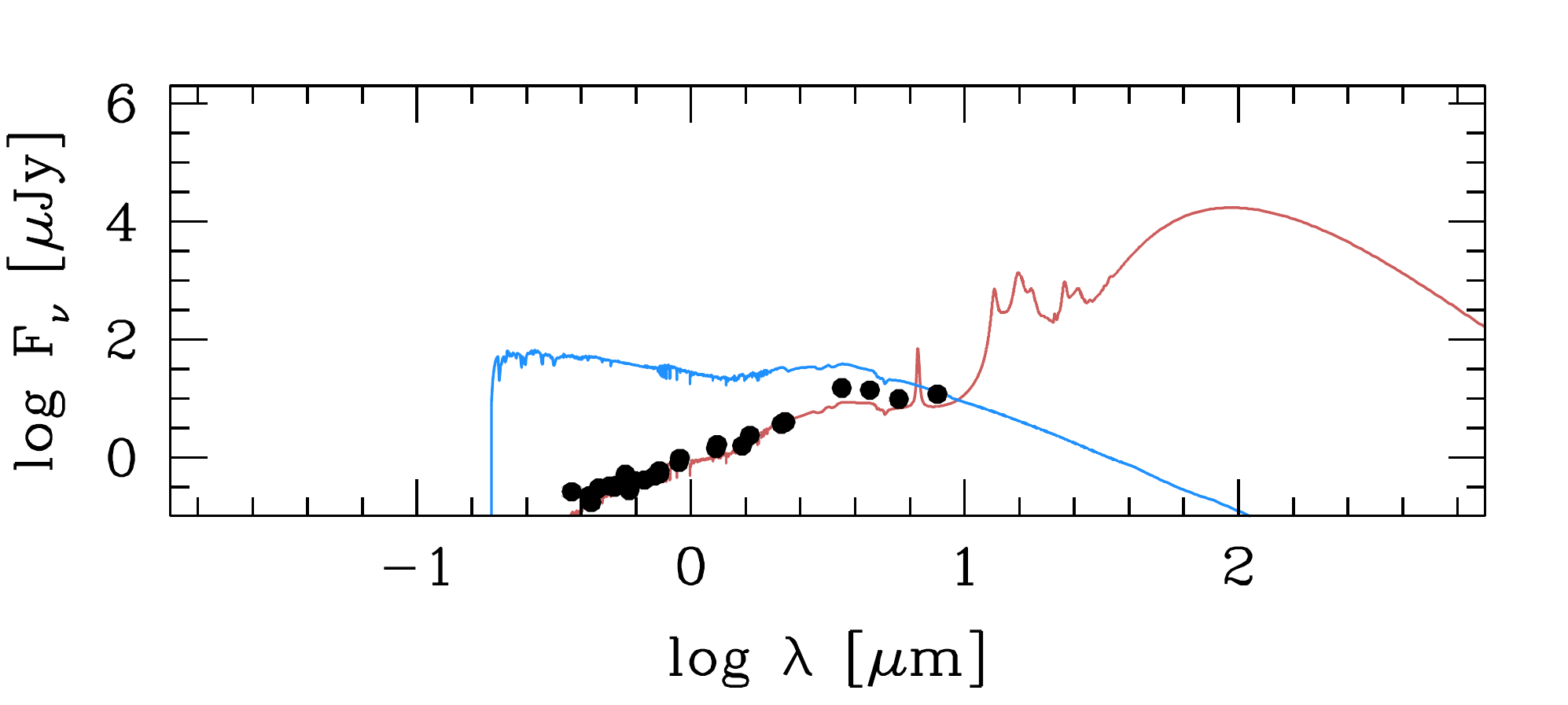}\\
\end{center}
\caption{{\em Top left:} {\em HST} F105W/F775W/F435W RGB image of ID.3 and 4. The postage stamp is $20''\times20''$. {\em Top center:} {\em HST} F125W image of the same field. The map of the lowest-J accessible CO transition (in this case, CO[3-2]) is shown as contours ($\pm 2,3,$\ldots,$20$-$\sigma$ [$\sigma$(ID.3)=0.27\,mJy\,beam$^{-1}$; $\sigma$(ID.4)=0.60\,mJy\,beam$^{-1}$]; solid black lines for the positive isophotes, dashed blue lines for the negative). The synthesized beam is shown as a black ellipse. {\em Top right:} Same as in the center, showing the 1.2mm dust continuum. {\em Bottom left:} Spectra of the CO lines encompassed in our spectral scan. {\em Bottom right:} Spectral Energy Distribution. The red line shows the best MAGPHYS fit of the available photometry (black points), while the blue line shows the corresponding model for the unobscured stellar component. The main output parameters are quoted. 
}\label{fig_id1_b}
\end{figure*}

\begin{figure*}\figurenum{A.3}\begin{center}
\includegraphics[width=0.89\textwidth]{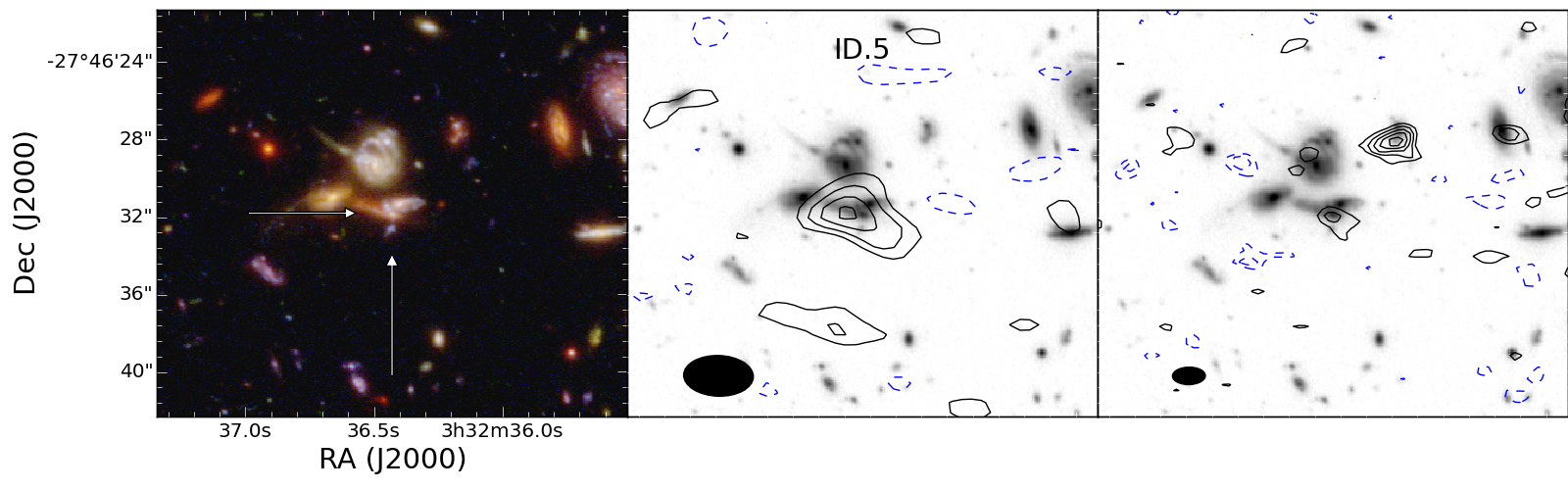}\\
\includegraphics[width=0.49\columnwidth]{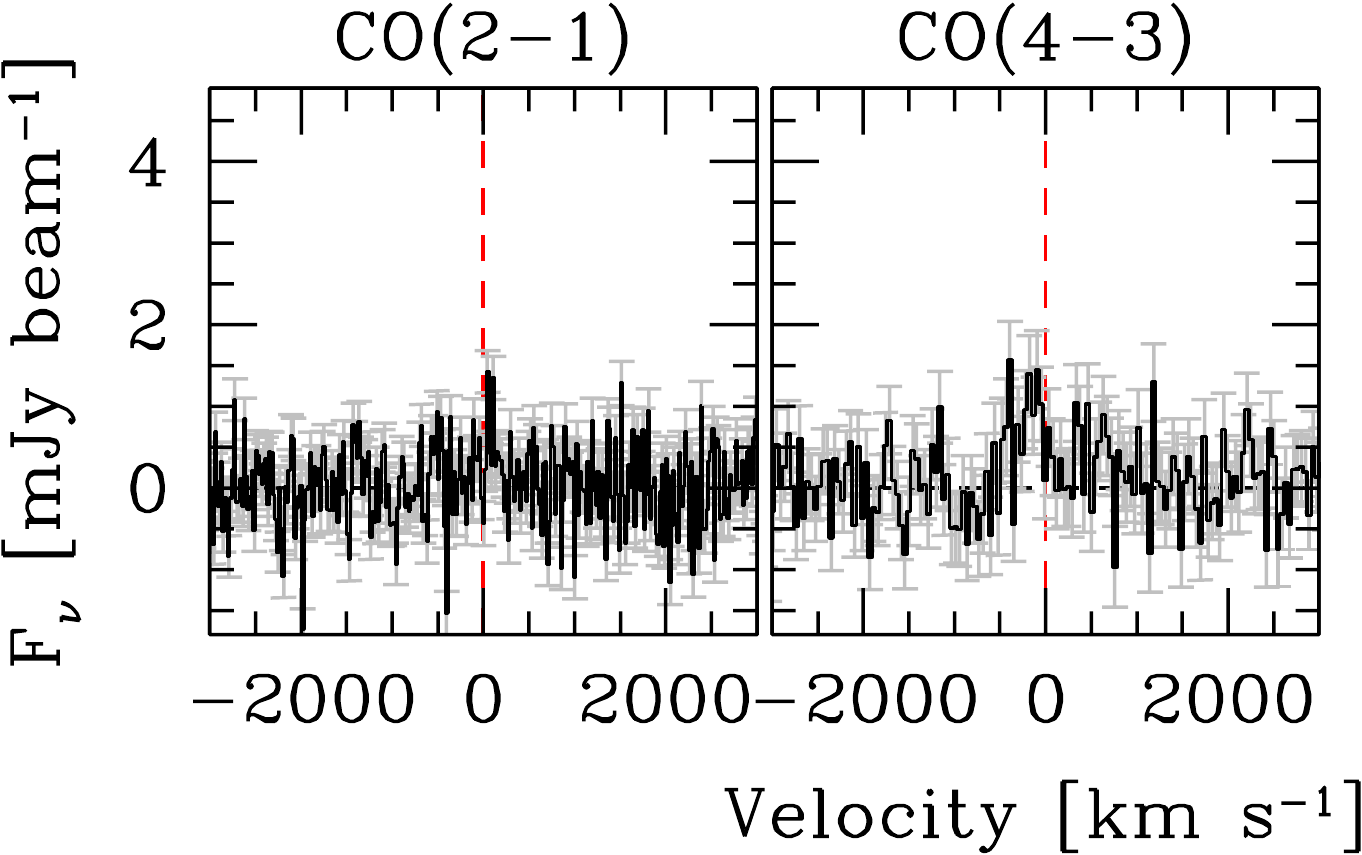}
\includegraphics[width=0.49\columnwidth]{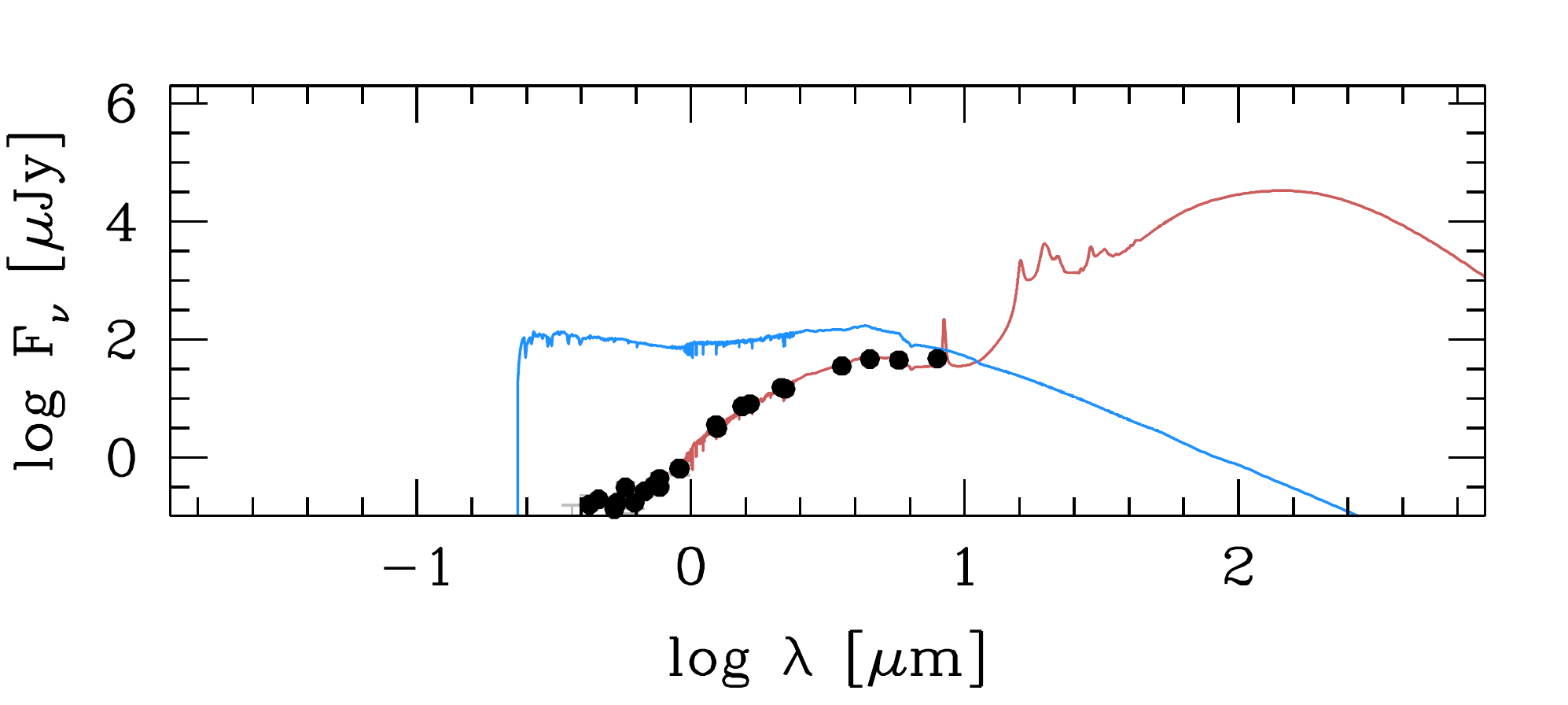}\\
\includegraphics[width=0.89\textwidth]{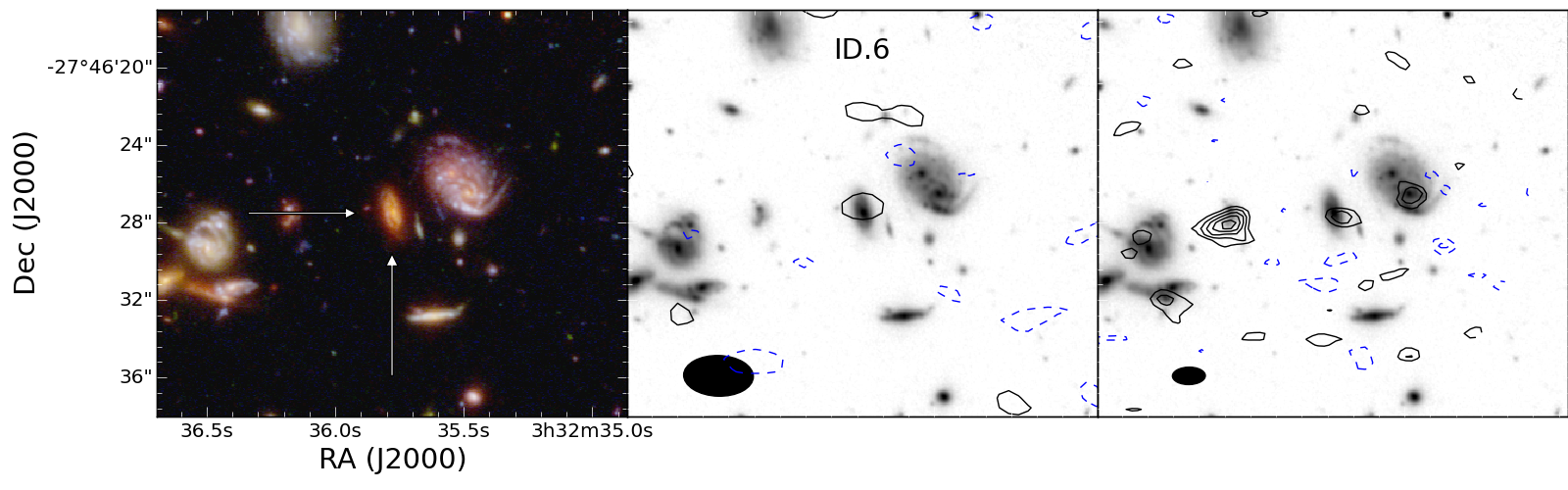}\\
\includegraphics[width=0.49\columnwidth]{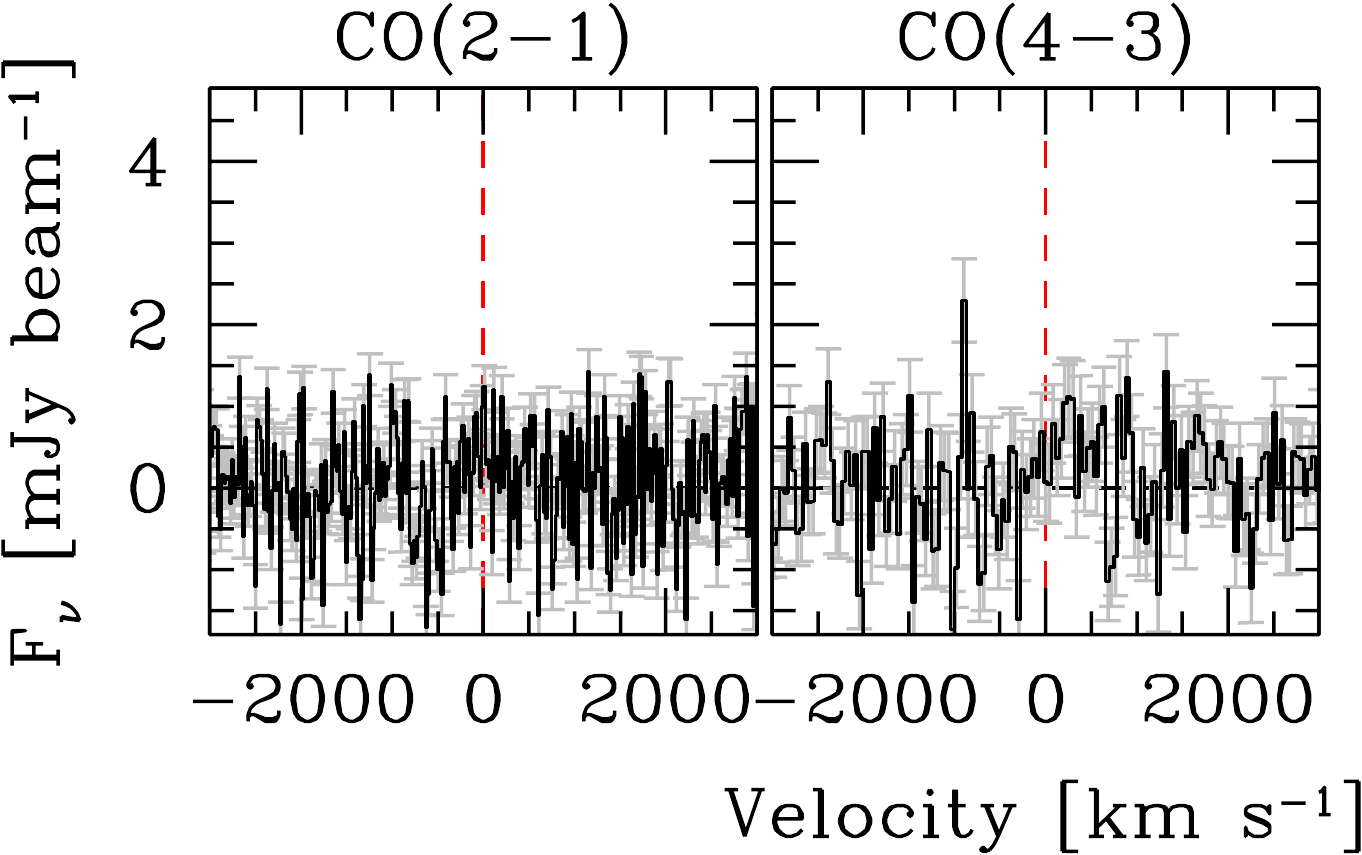}
\includegraphics[width=0.49\columnwidth]{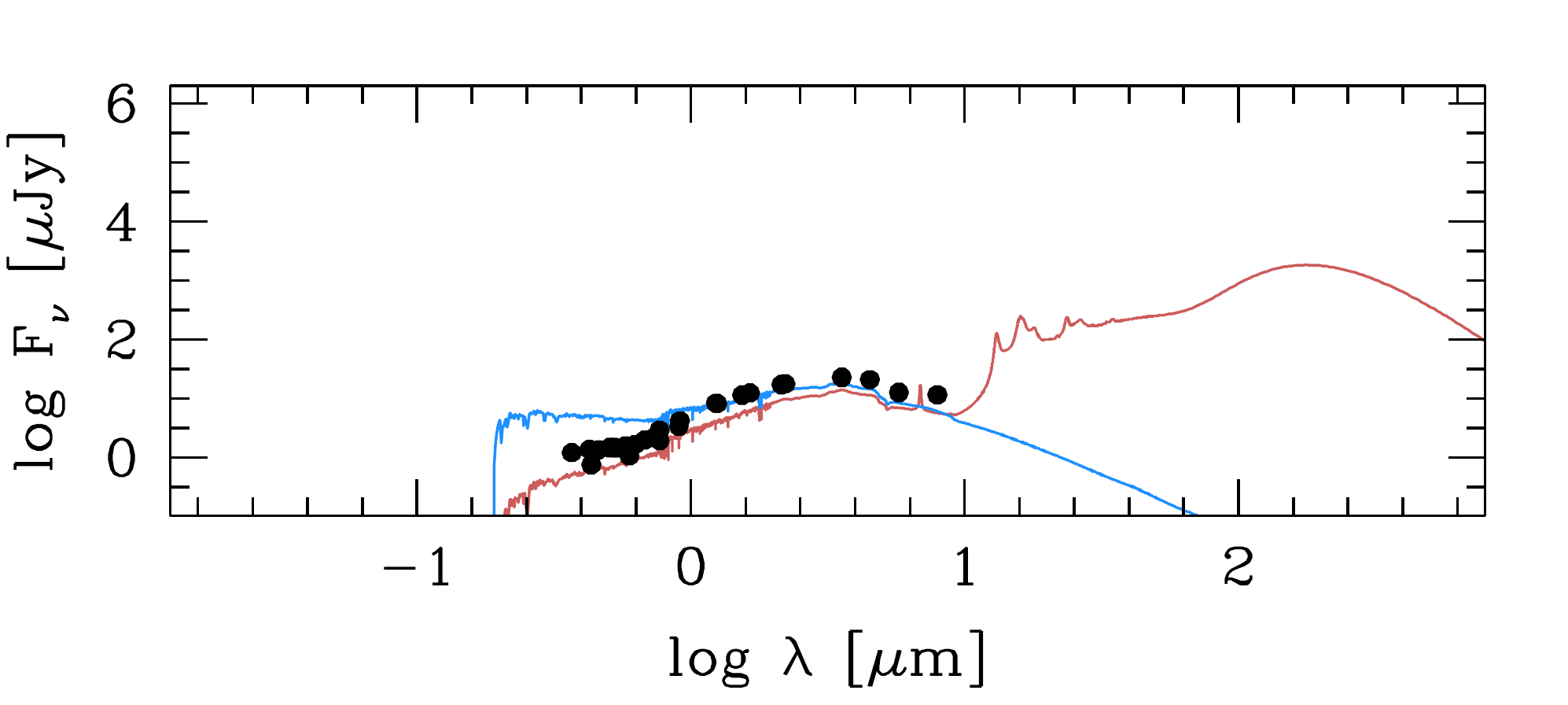}\\
\end{center}
\caption{{\em Top left:} {\em HST} F105W/F775W/F435W RGB image of ID.5 and 6. The postage stamp is $20''\times20''$. {\em Top center:} {\em HST} F125W image of the same field. The map of the lowest-J accessible CO transition (in this case, CO[3-2]) is shown as contours ($\pm 2,3,$\ldots,$20$-$\sigma$ [$\sigma$(ID.5)=1.13\,mJy\,beam$^{-1}$; $\sigma$(ID.6)=0.79\,mJy\,beam$^{-1}$]; solid black lines for the positive isophotes, dashed blue lines for the negative). The synthesized beam is shown as a black ellipse. {\em Top right:} Same as in the center, showing the 1.2mm dust continuum. {\em Bottom left:} Spectra of the CO lines encompassed in our spectral scan. {\em Bottom right:} Spectral Energy Distribution. The red line shows the best MAGPHYS fit of the available photometry (black points), while the blue line shows the corresponding model for the unobscured stellar component. The main output parameters are quoted. 
}\label{fig_id1_c}
\end{figure*}

\begin{figure*}\figurenum{A.4}\begin{center}
\includegraphics[width=0.89\textwidth]{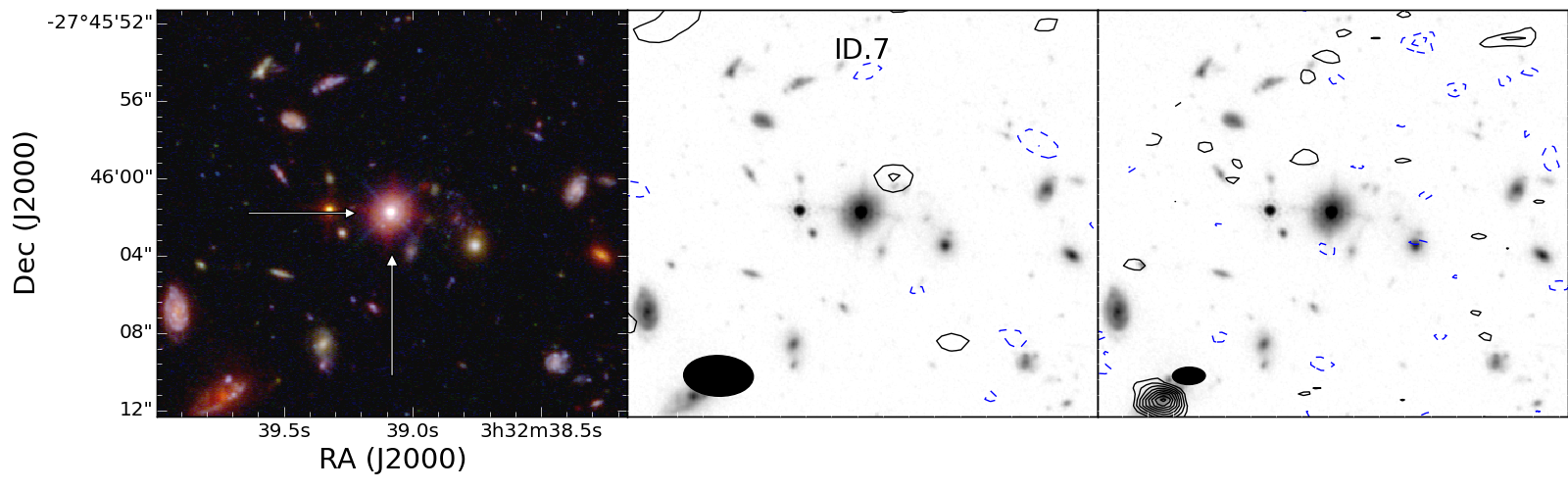}\\
\includegraphics[width=0.49\columnwidth]{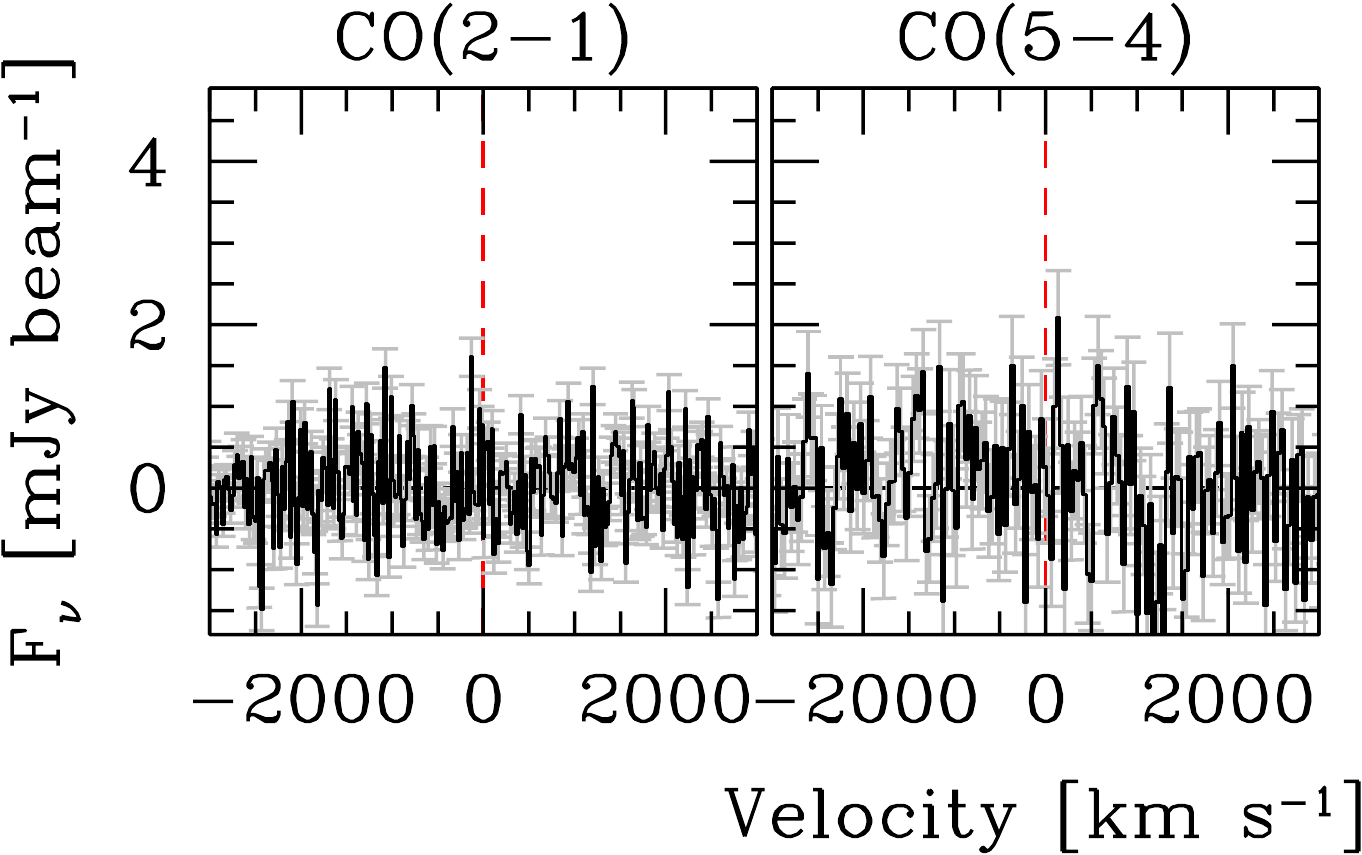}
\includegraphics[width=0.49\columnwidth]{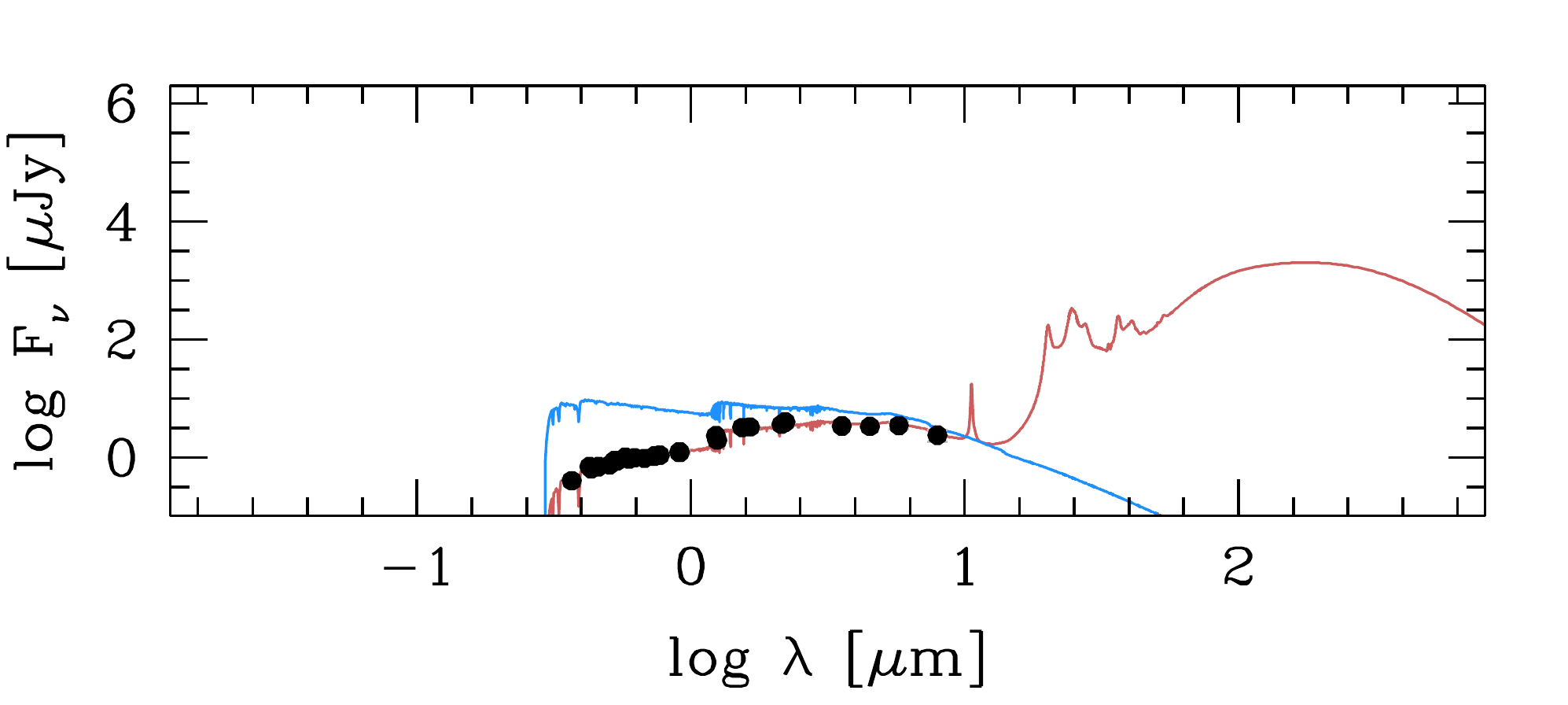}\\
\includegraphics[width=0.89\textwidth]{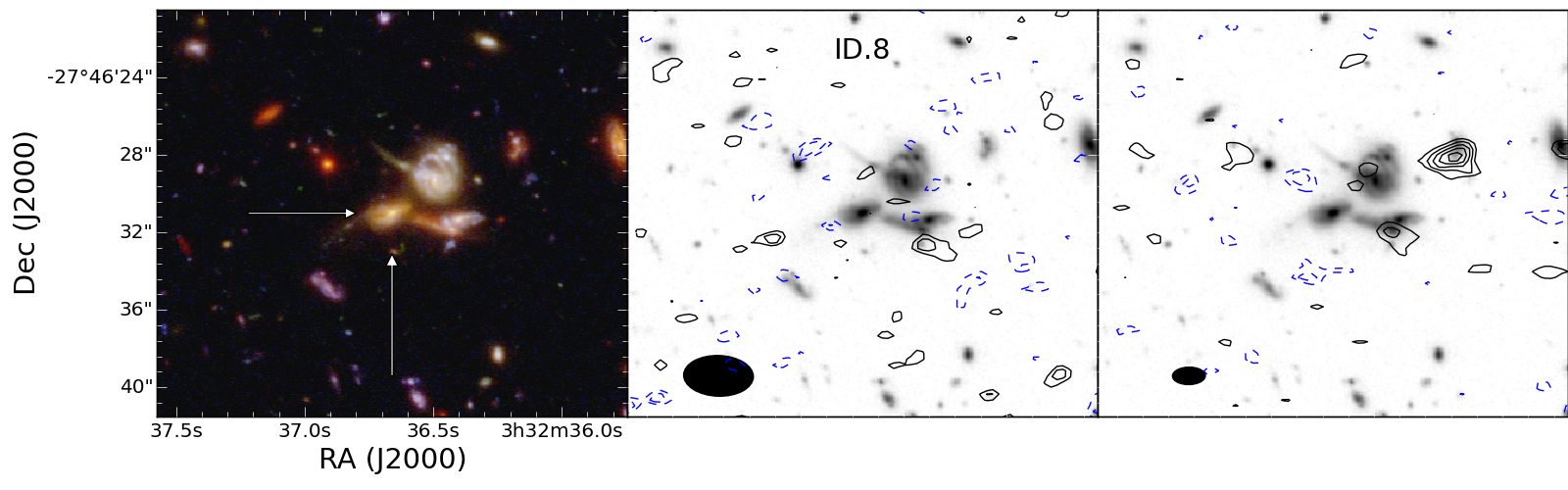}\\
\includegraphics[width=0.49\columnwidth]{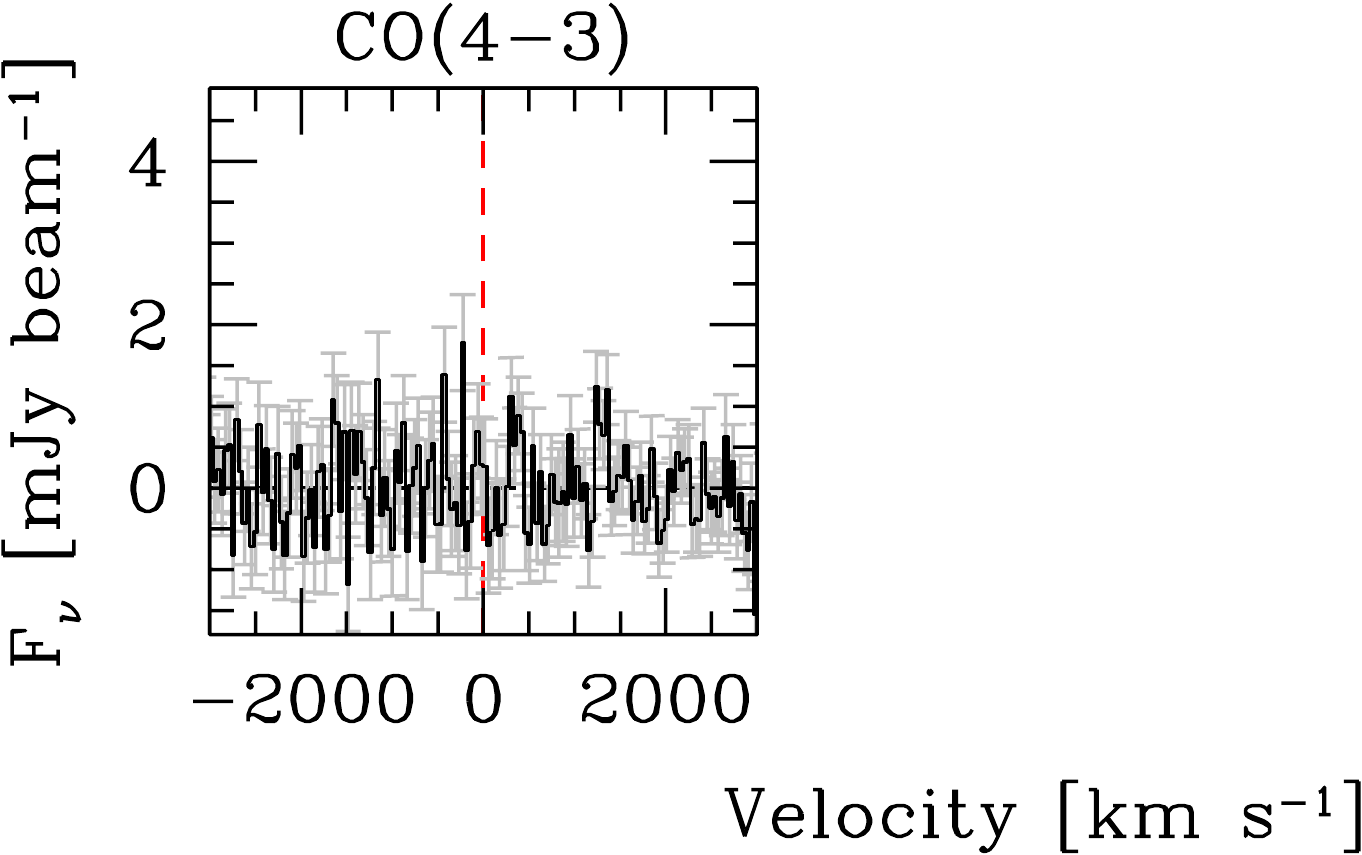}
\includegraphics[width=0.49\columnwidth]{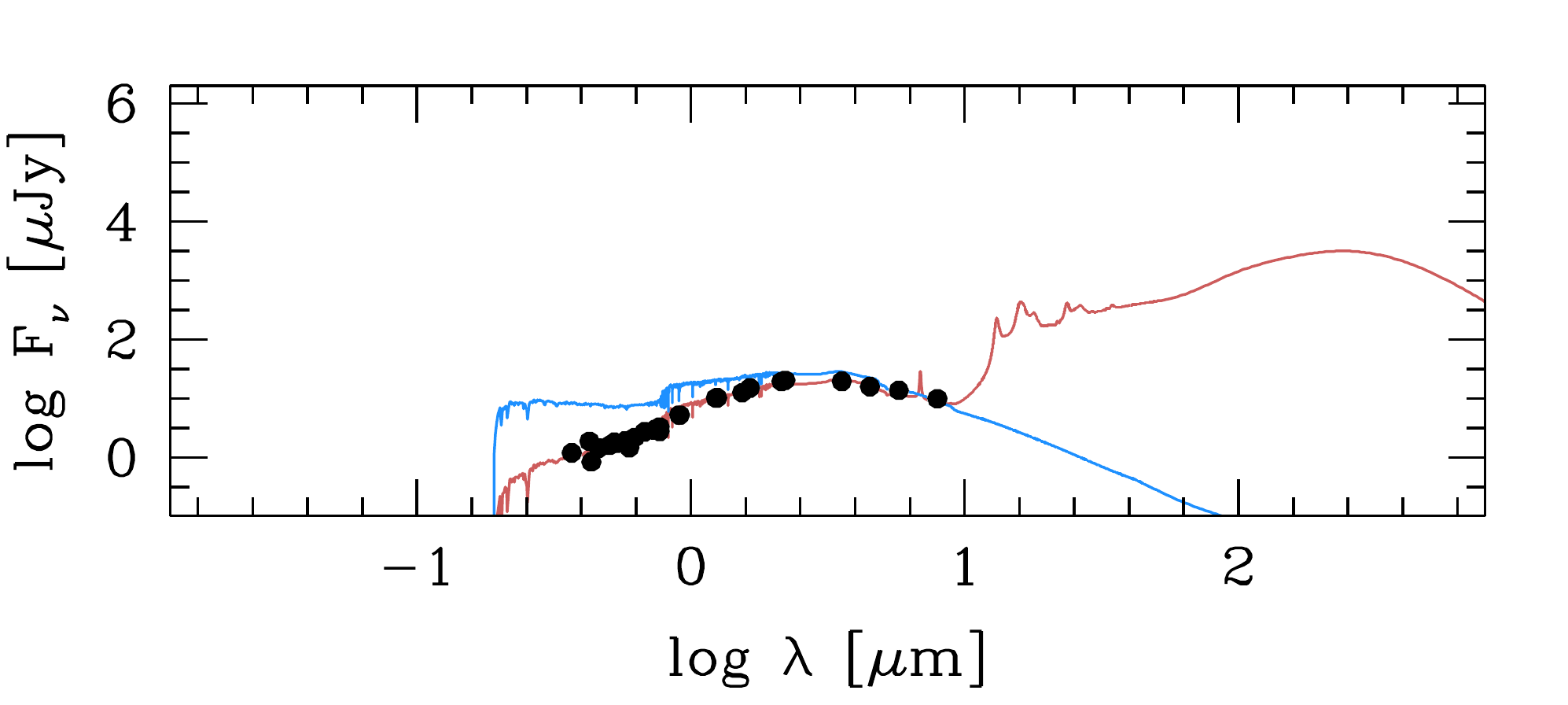}\\
\end{center}
\caption{{\em Top left:} {\em HST} F105W/F775W/F435W RGB image of ID.7 and 8. The postage stamp is $20''\times20''$. {\em Top center:} {\em HST} F125W image of the same field. The map of the lowest-J accessible CO transition (in this case, CO[3-2]) is shown as contours ($\pm 2,3,$\ldots,$20$-$\sigma$ [$\sigma$(ID.7)=0.67\,mJy\,beam$^{-1}$; $\sigma$(ID.8)=1.18\,mJy\,beam$^{-1}$]; solid black lines for the positive isophotes, dashed blue lines for the negative). The synthesized beam is shown as a black ellipse. {\em Top right:} Same as in the center, showing the 1.2mm dust continuum. {\em Bottom left:} Spectra of the CO lines encompassed in our spectral scan. {\em Bottom right:} Spectral Energy Distribution. The red line shows the best MAGPHYS fit of the available photometry (black points), while the blue line shows the corresponding model for the unobscured stellar component. The main output parameters are quoted. 
}\label{fig_id1_d}
\end{figure*}

\begin{figure*}\figurenum{A.5}\begin{center}
\includegraphics[width=0.89\textwidth]{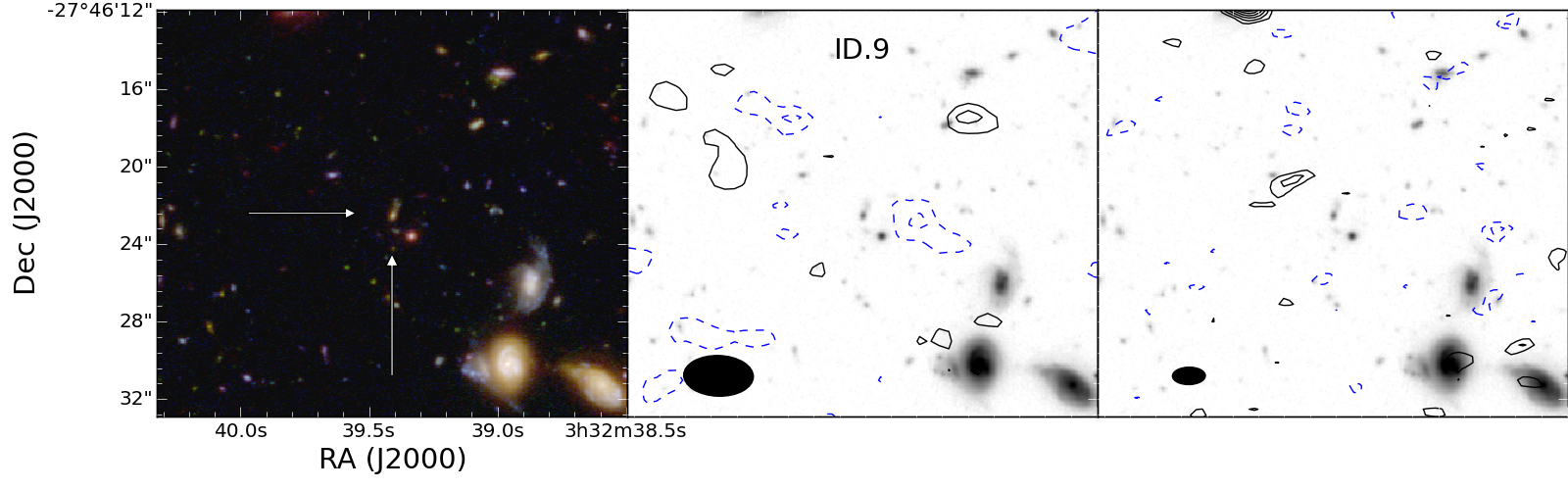}\\
\includegraphics[width=0.49\columnwidth]{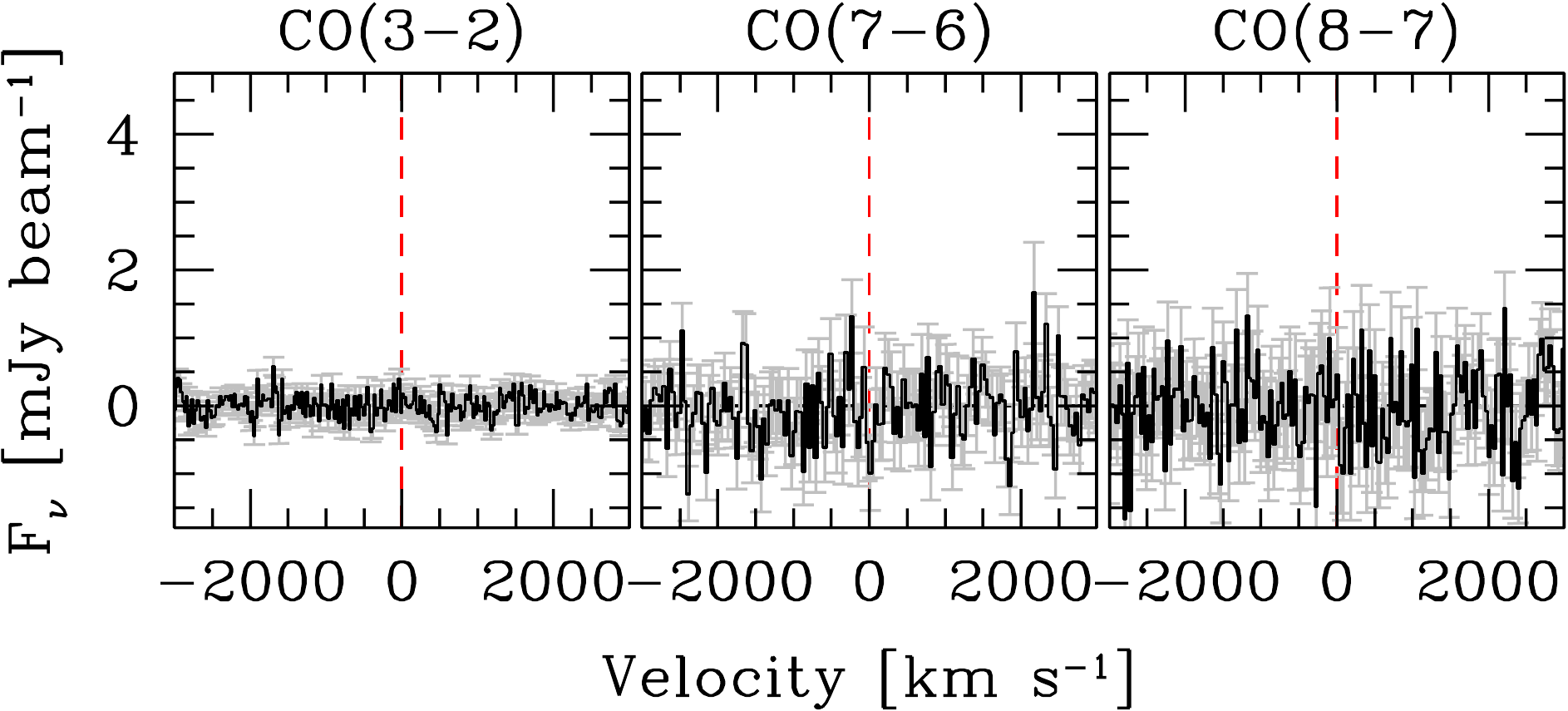}
\includegraphics[width=0.49\columnwidth]{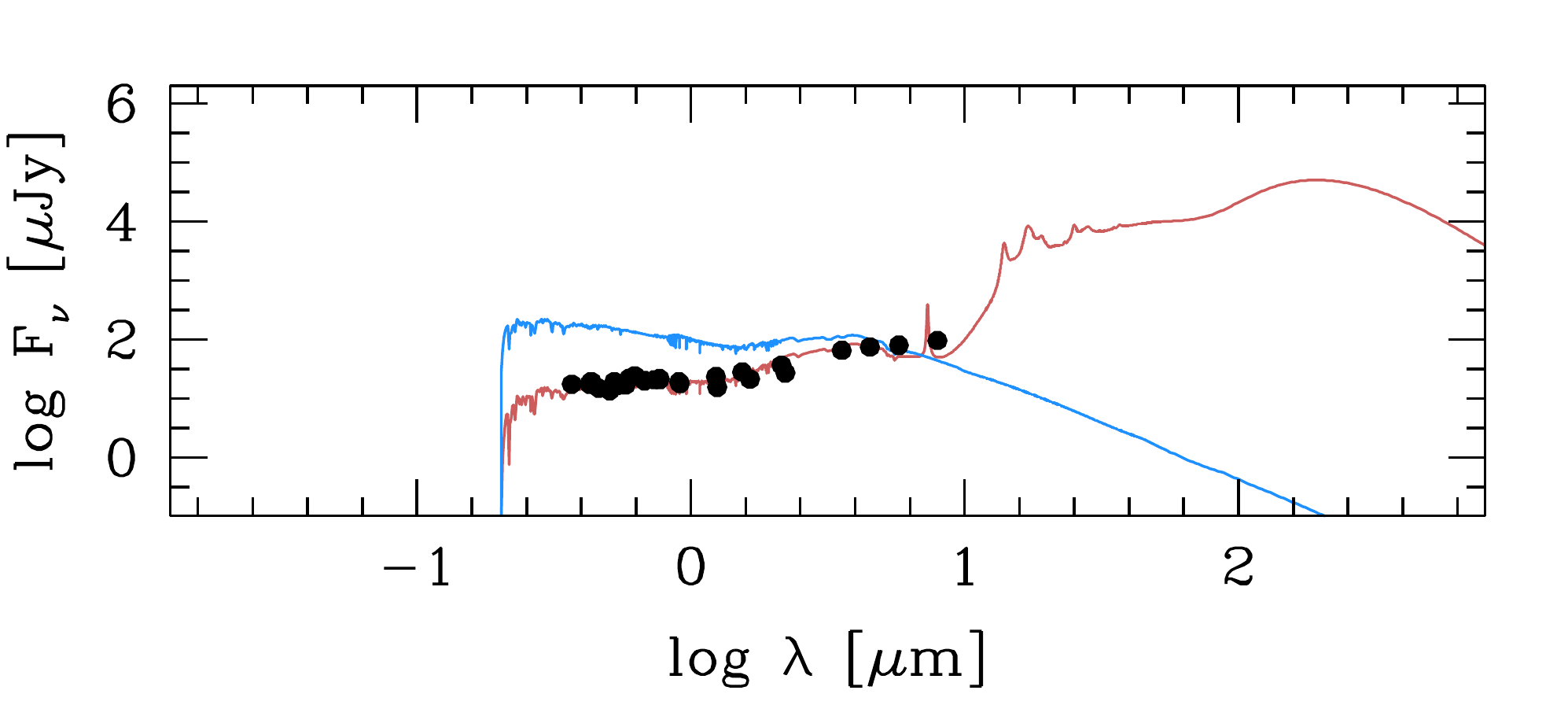}\\
\includegraphics[width=0.89\textwidth]{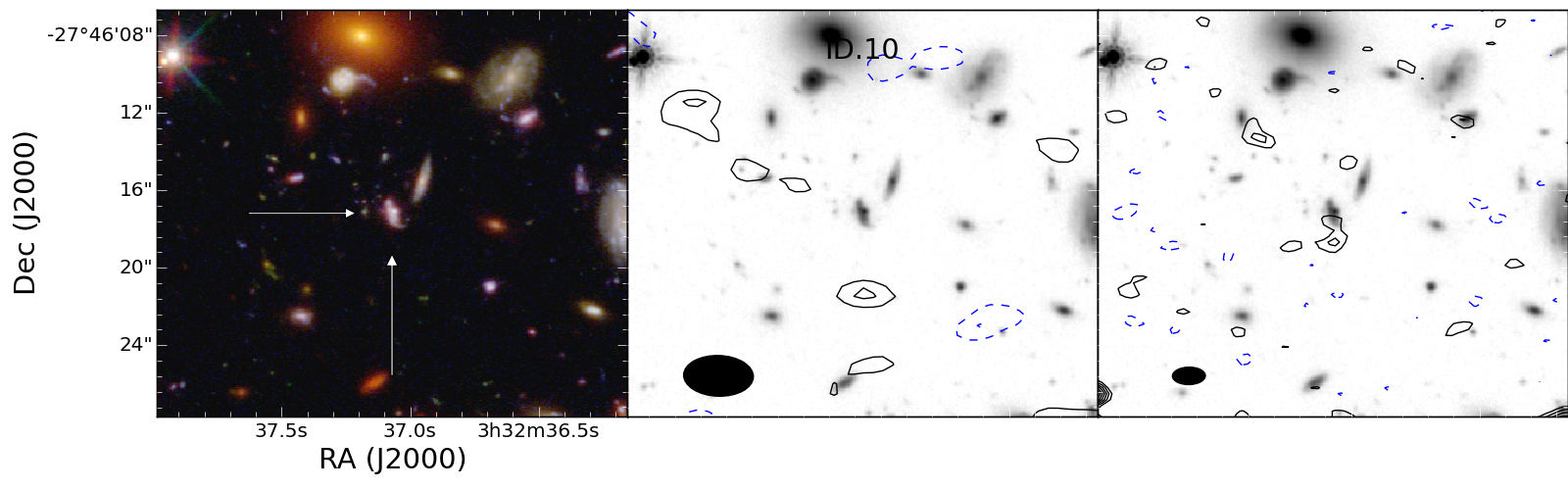}\\
\includegraphics[width=0.49\columnwidth]{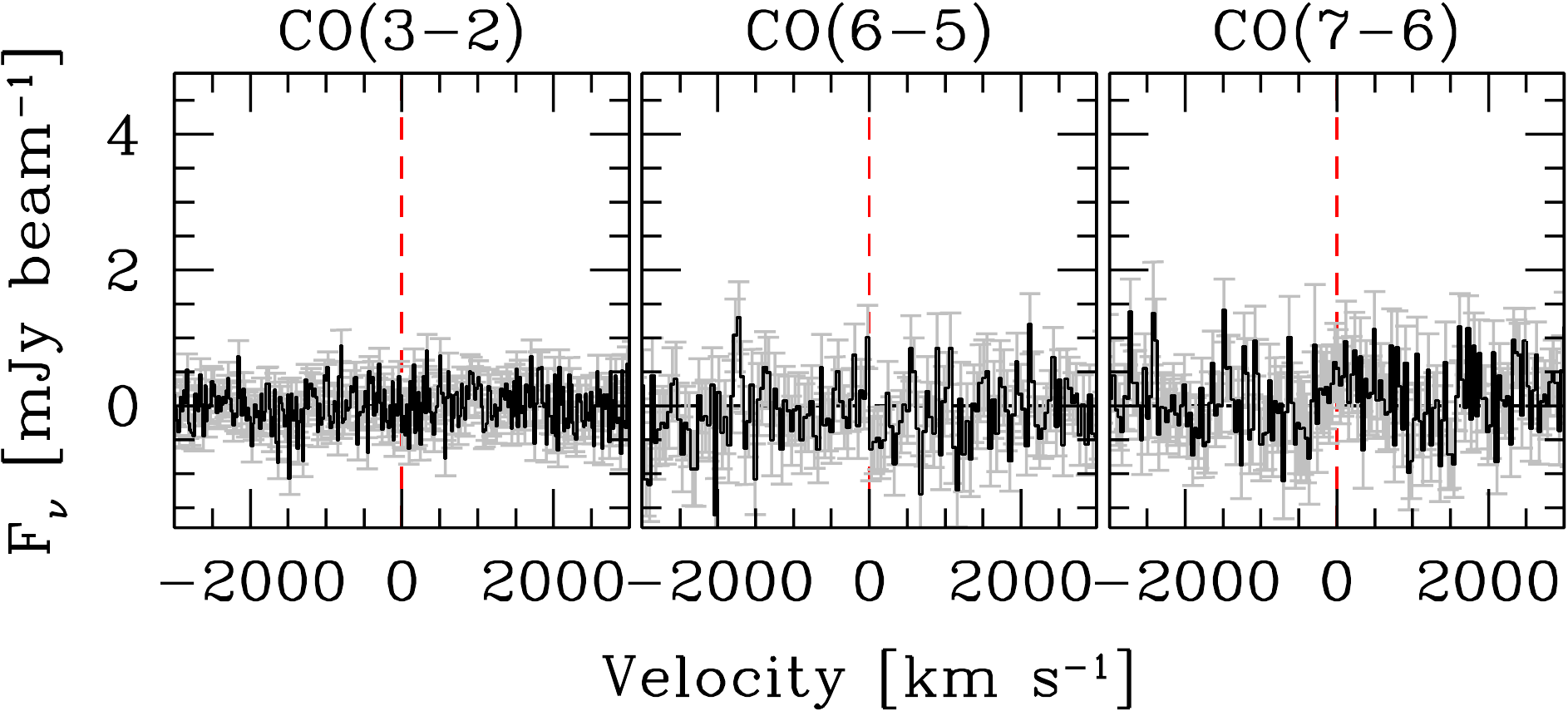}
\includegraphics[width=0.49\columnwidth]{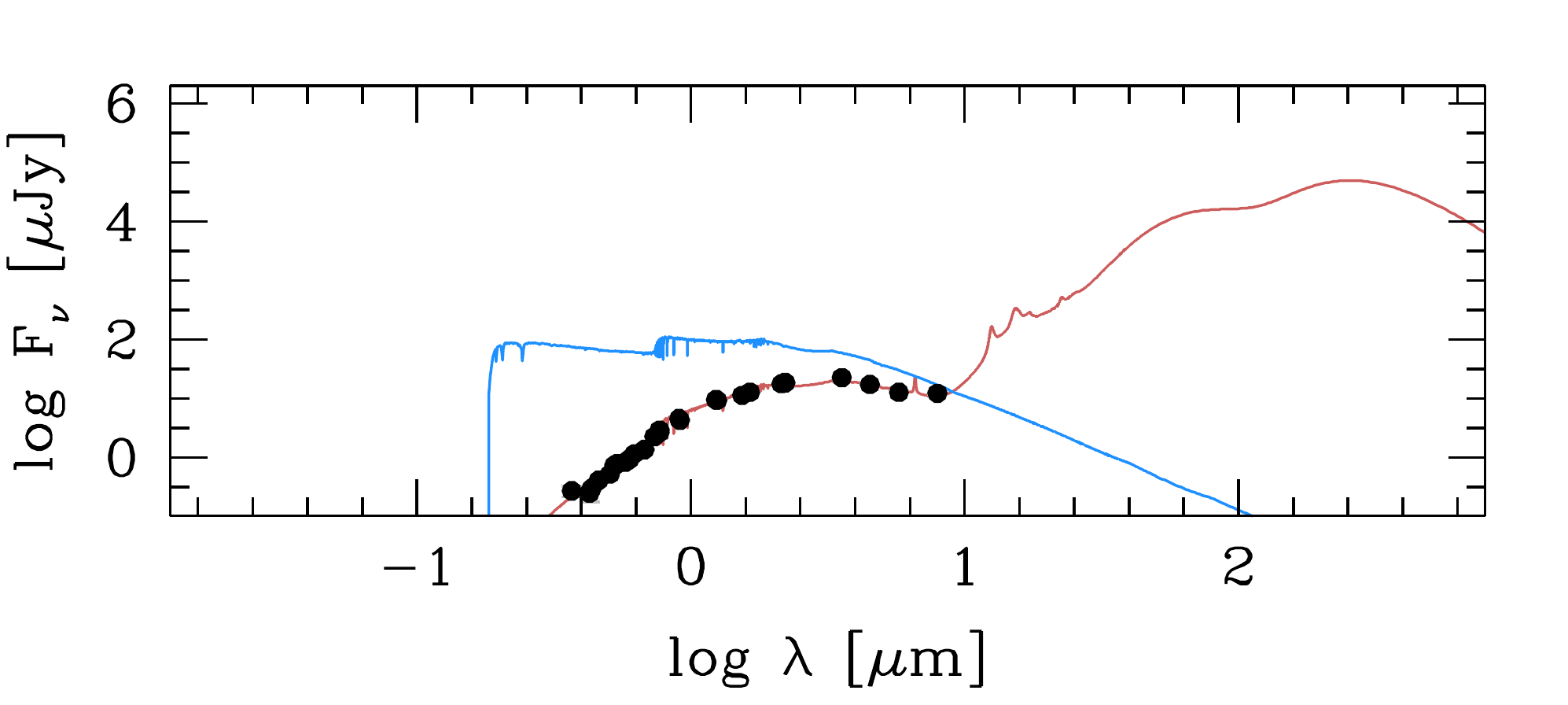}\\
\end{center}
\caption{{\em Top left:} {\em HST} F105W/F775W/F435W RGB image of ID.9 and 10. The postage stamp is $20''\times20''$. {\em Top center:} {\em HST} F125W image of the same field. The map of the lowest-J accessible CO transition (in this case, CO[3-2]) is shown as contours ($\pm 2,3,$\ldots,$20$-$\sigma$ [$\sigma$(ID.9)=0.44\,mJy\,beam$^{-1}$; $\sigma$(ID.10)=0.64\,mJy\,beam$^{-1}$]; solid black lines for the positive isophotes, dashed blue lines for the negative). The synthesized beam is shown as a black ellipse. {\em Top right:} Same as in the center, showing the 1.2mm dust continuum. {\em Bottom left:} Spectra of the CO lines encompassed in our spectral scan. {\em Bottom right:} Spectral Energy Distribution. The red line shows the best MAGPHYS fit of the available photometry (black points), while the blue line shows the corresponding model for the unobscured stellar component. The main output parameters are quoted. 
}\label{fig_id1_e}
\end{figure*}

\begin{figure*}\figurenum{A.6}\begin{center}
\includegraphics[width=0.89\textwidth]{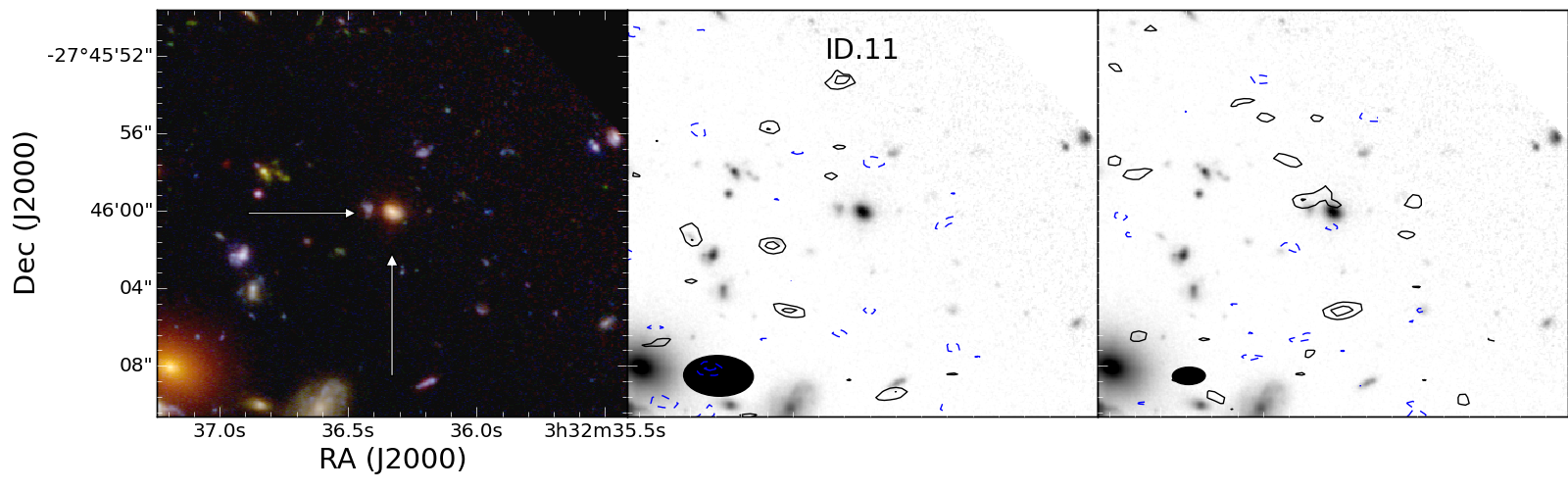}\\
\includegraphics[width=0.49\columnwidth]{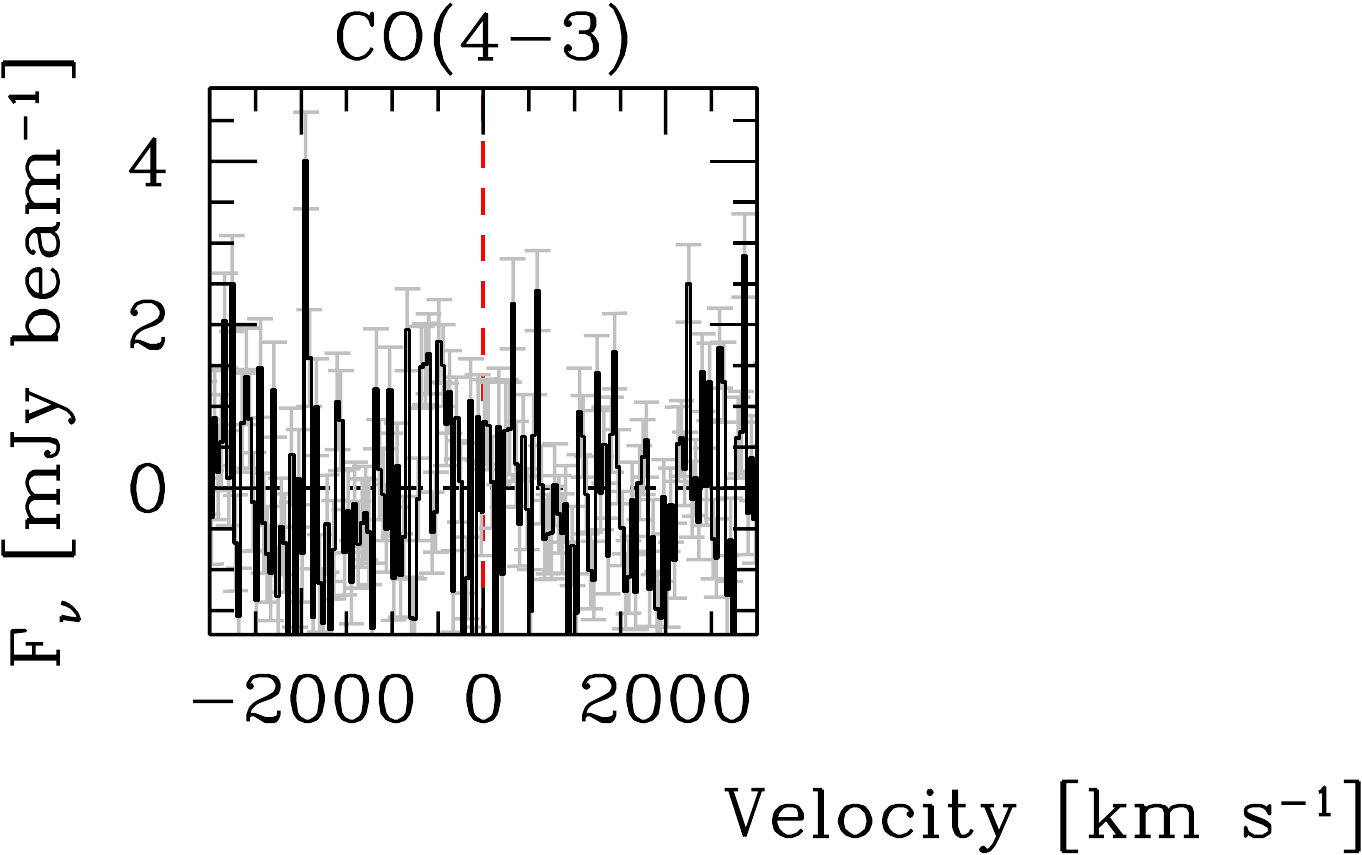}
\includegraphics[width=0.49\columnwidth]{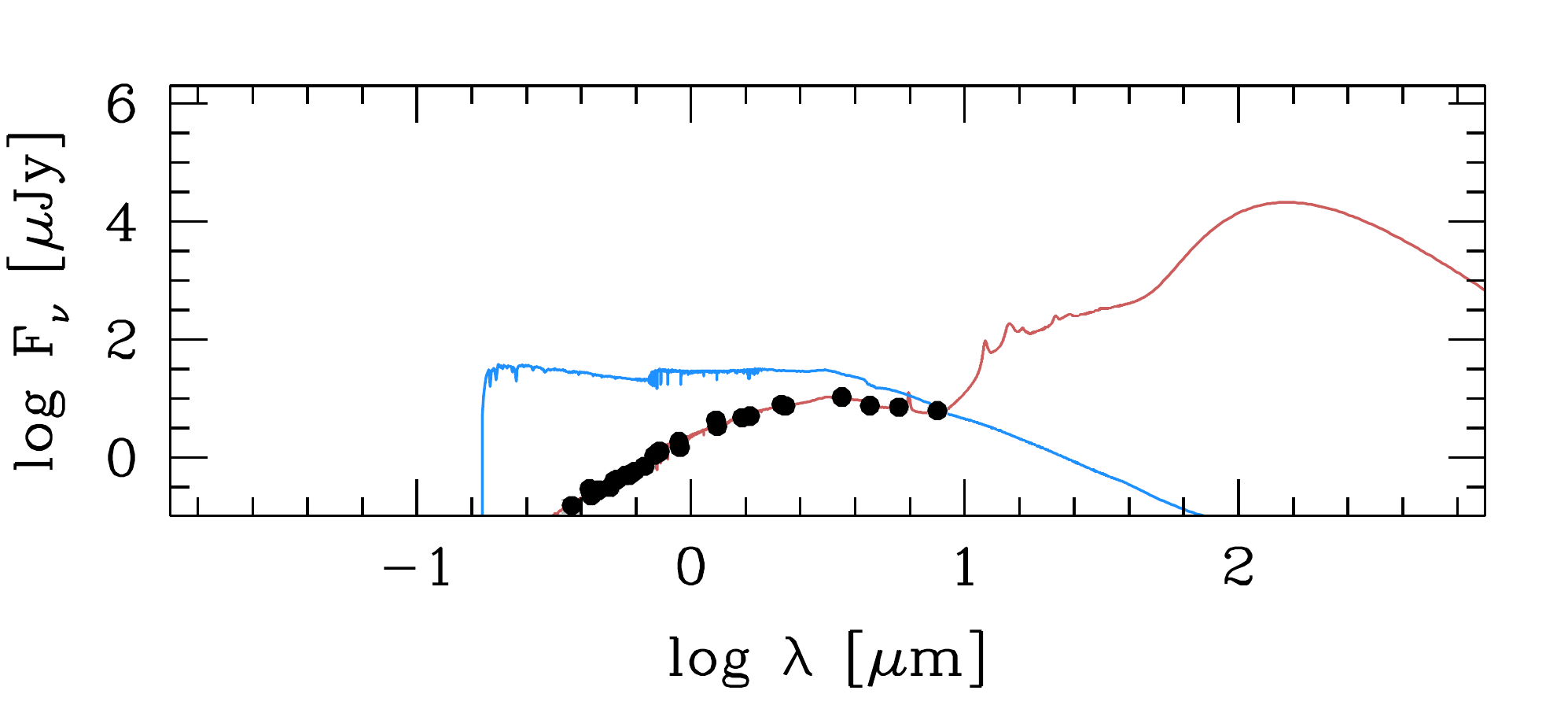}\\
\end{center}
\caption{{\em Top left:} {\em HST} F105W/F775W/F435W RGB image of ID.11. The postage stamp is $20''\times20''$. {\em Top center:} {\em HST} F125W image of the same field. The map of the lowest-J accessible CO transition (in this case, CO[3-2]) is shown as contours ($\pm 2,3,$\ldots,$20$-$\sigma$ [$\sigma$(ID.11)=1.10\,mJy\,beam$^{-1}$]; solid black lines for the positive isophotes, dashed blue lines for the negative). The synthesized beam is shown as a black ellipse. {\em Top right:} Same as in the center, showing the 1.2mm dust continuum. {\em Bottom left:} Spectra of the CO lines encompassed in our spectral scan. {\em Bottom right:} Spectral Energy Distribution. The red line shows the best MAGPHYS fit of the available photometry (black points), while the blue line shows the corresponding model for the unobscured stellar component. The main output parameters are quoted. 
}\label{fig_id1_f}
\end{figure*}

\label{lastpage}

\end{document}